\documentclass[aps,groupedaddress,nofootinbib,floatfix,epfs,preprintnumbers]{revtex4}
\usepackage{amsfonts,amscd,amsmath,amssymb,graphicx,color,float,xcolor}
\usepackage[section]{placeins}
\def\ep{\text{e}}
\def\g{\mathsf{g}}
\def\oh{\frac{1}{2}}
\def\s{\mathsf{s}}
\def\k{\mathsf{k}}

\def\rv{r_v}

\def\r0{r_0}
\def\rvb{r_{\bar v}}

\def\xo{x_{\text{\tiny 0}}}
\def\ro{r_{\text{\tiny 0}}}

\def\vo{{\text v}_{\text{\tiny 1}}}
\def\vp{{\text v}_{\text{\tiny p}}}
\def\vt{{\text v}_\eta}
\def\blt{\lambda_\eta}

\def\vz{{\text v}_{\text{\tiny 0}}}
\def\QQb{\text{\tiny Q}\bar{\text{\tiny Q}}}

\def\5Q{\text{\tiny 5Q}}

\def\3Q{\text{\tiny 3Q}}
\def\2Q{\text{\tiny QQ}}

\def\QQ{\text{\tiny QQ}}
\def\i{\text{\tiny (I)}}
\def\ii{\text{\tiny (II)}}
\begin{document}
\preprint{LMU-ASC 23/25}
\title{$QQ\bar Q\bar Q$ Quark System and Gauge/String Duality}
\author{Oleg Andreev}
 %\affiliation{L.D. Landau Institute for Theoretical Physics, Kosygina 2, 119334 Moscow, Russia}
  %\affiliation{V.A. Steklov Mathematical Institute, Gubkina 8, 119991, Moscow, Russia}
\thanks{Also on leave from L.D. Landau Institute for Theoretical Physics}
\affiliation{Arnold Sommerfeld Center for Theoretical Physics, LMU-M\"unchen, Theresienstrasse 37, 80333 M\"unchen, Germany}
%\date{}
\begin{abstract} 
 We propose a stringy description of a system composed of two heavy quarks and two heavy antiquarks, mimicking that in pure $SU(3)$ gauge theory. We present both analytical and numerical studies of the string configurations for rectangular geometries. As an application, we analyze the two lowest Born-Oppenheimer potentials. Our results suggest that, depending on the geometry and quark ordering, the ground state of the $QQ\bar Q\bar Q$ system may be a hadronic molecule, a tetraquark state, or a superposition of the two. For general geometries, we derive the asymptotic expression for the energy of the tetraquark configuration in the infrared limit and extend this result to multiquark configurations. Here we also demonstrate the universality of the string tension.  
 \end{abstract}
%\pacs{empty}
\maketitle
%_______________________________________________________________________
\vspace{0.5cm}
\section{Introduction}
\renewcommand{\theequation}{1.\arabic{equation}}
\setcounter{equation}{0}

Since the proposal of the quark model by Gell-Mann \cite{GM} and Zweig \cite{Zw}, many exotic hadrons have been discovered \cite{Bram}. However, the long standing question of how quarks are bound inside exotic hadrons still remains open.

The discovery of the $J/\psi$ meson, consisting of the charm quark and the charm antiquark, in 1974 marked the beginning of the modern era of high energy physics \cite{1974}.\footnote{This is often refereed to as the November revolution.} Since then, it has taken almost fifty years to observe two exotic hadrons $T_{c\bar cc \bar c}(6600)$ and $T_{c\bar cc\bar c}(6900)$, composed of four charm quarks and known as fully heavy tetraquark mesons \cite{Tcccc}.

One approach to the fully heavy tetraquark system is as follows. Given the large quark masses, it seems reasonable to apply the Born-Oppenheimer (B-O) approximation, originally developed for use in atomic and molecular physics \cite{bo}.\footnote{For further elaboration on these ideas in the context of QCD, see \cite{braat}.} Within this framework, the corresponding B-O potentials are defined as the energies of stationary configurations of the gluon and light quarks fields in the presence of static heavy quark sources. The hadron spectrum is then determined by solving the Schr\"odinger equation with these potentials.
 
 The calculation of the B-O potentials is strongly influenced by non-perturbative effects and therefore cannot be performed within perturbative QCD. Although lattice gauge theory is one of the basic tools for studying nonperturbative phenomena in QCD and has made significant progress in the study of the fully heavy tetraquark systems \cite{alexa,sug,Bicu-4Q}, the limited results and the need to understand the physics behind computational complexity motivate the use of effective field and string theories. A special class of string models, called AdS/QCD (holographic) models, has received much attention in the last years. The hope is that the gauge/string duality provides new theoretical tools for studying strongly couple gauge theories.\footnote{For the further development of these ideas in the context of QCD, see the book \cite{book-u} and references therein.} In these models the string configurations for tetraquarks were discussed qualitatively in \cite{a3Q2008,coba}. Nevertheless, the existing literature notably lacks a comprehensive discussion on such systems. Bridging these gaps is the main objective of the present paper.
 
The paper continues our study of the $Q\bar Q$ and $QQQ$ heavy quark systems \cite{az1,a3Q2016,a3Q2025}. It is organized as follows. In Sec.II, we briefly review some preliminary results and establish the framework for the reader's convenience. Then in Secs.III and IV, we construct and analyze string configurations in five dimensions that provide a dual description of the lowest B-O potentials in the heavy quark limit. Here, we restrict ourselves to the symmetric case, when the quark sources are located at the vertices of a rectangular. In Sec.V, we examine the two lowest B-O potentials of the system and also discuss the length scales that characterize transitions between different configurations. These length scales are related to various types of string interactions, including string junction annihilation. In Sec.VI, we consider the IR limit for a special class of multiquark string configurations and explores the asymptotic behavior of their energies. We conclude in Sec.VII with several comments on the implications of our findings and discussing directions for future work. Appendix A contains our notation and definitions. To make the paper self-contained, Appendices B and C provide the necessary results on Nambu-Goto strings in five-dimensional space and on the $Q\bar Q$ system. Finally, in Appendices D and E, we discuss the five-dimensional analogs of the tetraquark string configuration. 

%______________________________________________________________
\section{Preliminaries}
\renewcommand{\theequation}{2.\arabic{equation}}
\setcounter{equation}{0}

%______________________________________________________________
\subsection{General procedure}

To study the tetraquark system, we use an approach based on a correlation matrix.\footnote{For more on this in lattice QCD, see \cite{Bicu-4Q}.} In this approach, the diagonal matrix elements are determined by the energies of stationary string configurations, while the off-diagonal ones describe transitions between these configurations. The potentials are determined by the eigenvalues of this matrix.

\begin{figure}[htbp]
\centering
\includegraphics[width=14.5cm]{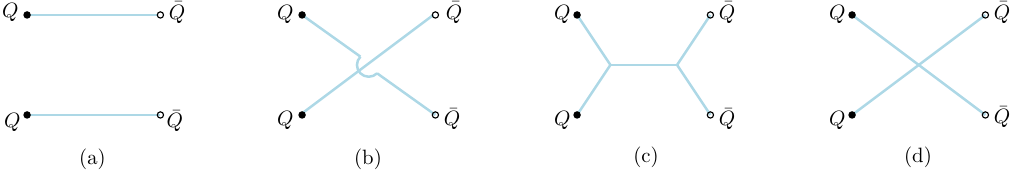}
\caption{{\small Four-dimensional string configurations with type-A ordering: (a)-(b) disconnected configurations, (c) a tetraquark configuration, and (d) a pinched tetraquark configuration.}}
\label{confs4A}
\end{figure}

\begin{figure}[htbp]
\centering
\includegraphics[width=10.5cm]{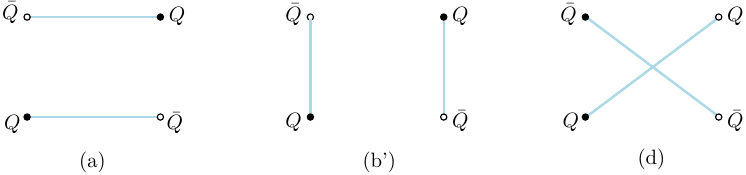}
\caption{{\small Four-dimensional string configurations with type-B ordering: (a)-(b') disconnected configurations and (d) a pinched tetraquark configuration. We retain the label (d) for the pinched tetraquark configuration because its energy equals that of the configuration shown in Figure \ref{confs4A}. Although the tetraquark configuration (c') formally coincides with the pinched configuration in four dimensions, this equivalence no longer holds in five dimensions.}}
\label{confs4B}
\end{figure}

We begin by specializing to the case of pure $SU(3)$ gauge theory and a geometry in which the quark sources are placed at the vertices of a rectangle. There are two possible orderings of these sources: type-A ordering, where the quarks (antiquarks) occupy adjacent vertices, and type-B ordering (bipartite), where each site is connected only to one quark and one antiquark. Let us start with string configurations in four dimensions, where the string picture has long been established \cite{XA,carlson}. The simplest configurations are disconnected (mesonic) ones, as shown in Figures \ref{confs4A} and \ref{confs4B}. Each consists of two quark-antiquark pairs joined by strings. We assume 
%________________________  fig - 1  ___________________________
%_____________________________________________________________
that other configurations are constructed by adding extra string junctions.\footnote{Besides string junctions, other possibilities include excited strings and glueball states, which, however, are not relevant here.} This yields the connected tetraquark configurations, both regular and pinched, shown in the Figures. In the latter case the string junctions coincide.  
%________________________  fig - 2  __________________________
%_____________________________________________________________
In contrast to \cite{XA,carlson}, we include the pinched tetraquark configurations, as their five-dimensional analogs turn out to be relevant for describing the first excited B-O potential. In what follows, we refer to hadronic molecules as being described by the disconnected string configurations, and tetraquarks by the connected ones.

\begin{figure}[htbp]
\centering
\includegraphics[width=10.5cm]{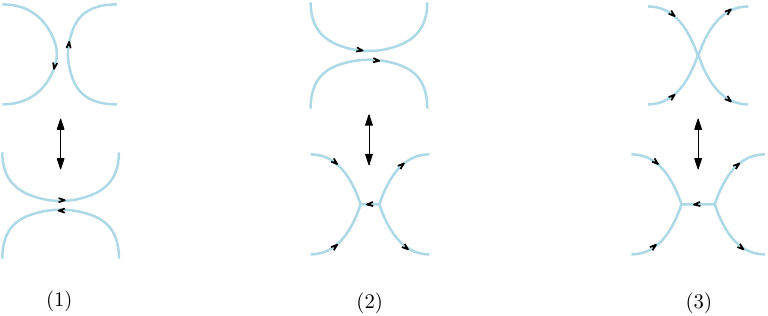}
\caption{{\small Some string interactions: (1) reconnection, (2) junction annihilation, and (3) pinching. }}
\label{stri}
\end{figure}

Transitions between the configurations arise due to string interactions. In Figure \ref{stri}, we sketch three types of interactions that will be discussed in the following sections. These represent only a small part of the broader dynamics of QCD strings. Later, we will introduce the notion of a critical length, which characterizes a transition between two configurations. This helps to deepen our understanding of the physics of QCD strings, the structure of B-O potentials, and, importantly, the nature of multiquark states. 
%________________________  fig - 3  __________________________
%___________________________________________________________

%__________________________________________________________________________________________________
\subsection{A short account of the five-dimensional model}

In our study of the fully heavy tetraquark system, we will use the formalism mainly developed in \cite{a3Q2016}. This formalism is general and can be adapted to any model of AdS/QCD, although we illustrate it by performing calculations in one of the simplest models, the so called soft wall model. In this model, the background five-dimensional geometry is chosen to be a one-parameter deformation of Euclidean $\text{AdS}_5$ space of radius $R$

\begin{equation}\label{metric}
ds^2=\ep^{\s r^2}\frac{R^2}{r^2}\Bigl(dt^2+(dx^i)^2+dr^2\Bigr)
\,.
\end{equation}
Here $r$ denotes the fifth (radial) coordinate, and $\s$ is a deformation parameter. The boundary lies at $r=0$, while the soft wall at $r=1/\sqrt{\s}$. The wall effectively prevents strings from penetrating deep into the bulk. 

Our analysis relies on two key components. The first is a fundamental string governed by the Nambu-Goto action 

\begin{equation}\label{ng}
S_{\text{\tiny NG}}=\frac{1}{2\pi\alpha'}\int d^2\xi\,\sqrt{\gamma^{(2)}}
\,,
\end{equation}
where $\gamma$ is an induced metric, $\alpha'$ is a string parameter, and $\xi^i$ are world-sheet coordinates. The second component is a high-dimensional analogue of the string junction, commonly referred to as the baryon vertex.\footnote{We use this terminology, to distinguish between the four-dimensional string junction and its five-dimensional counterpart.} In the context of AdS/CFT, the vertex is a five brane wrapped on an internal space $\mathbf{X}$ \cite{witten}. From the five-dimensional viewpoint, this object looks point-like. As shown in \cite{a3Q2016}, the action for the baryon vertex written in the static gauge

\begin{equation}\label{baryon-v}
S_{\text{vert}}=\tau_v\int dt \,\frac{\ep^{-2\s r^2}}{r}
\end{equation}
yields the results in very good agreement with lattice QCD calculations of the three-quark potential. Actually, $S_{\text{vert}}$ represents the worldvolume of the brane, assuming $\tau_v={\cal T}_5R\,\text{vol}(\mathbf{X})$ with ${\cal T}_5$ the brane tension.  Unlike AdS/CFT, we treat $\tau_v$ as a free parameter to account for $\alpha'$-corrections as well as the possible effects of other background fields.\footnote{In analogy with AdS/CFT, we expect the presence of Ramond-Ramond background fields on $\mathbf{X}$.} At zero baryon chemical potential, it is natural to suggest the same action for the antibaryon vertex, such that $S_{\bar{\text{vert}}}=S_{\text{vert}}$.

We also need to specify the model parameters. For the purposes of this paper, it is natural to use those from \cite{a3Q2016}, which were obtained by fitting lattice QCD data for the $QQQ$ quark system to the string model under consideration. Accordingly, we set $\g = \frac{R^2}{2\pi\alpha'}=0.176$, $\s = 0.44\,\text{GeV}^2$, and $\k=\frac{\tau_v}{3\g}=-0.083$. These values will be used in all subsequent estimates unless stated otherwise. Crucially, no additional parameters are introduced to describe the tetraquark system.
%___________________________________________________________________  
\section{String configurations in five dimensions for type-A ordering}
\renewcommand{\theequation}{3.\arabic{equation}}
\setcounter{equation}{0}

\begin{figure}[htbp]
\centering
\includegraphics[width=5.5cm]{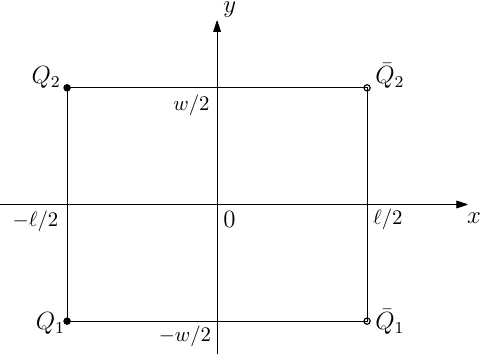}
\caption{{\small A rectangle of length $\ell$ and width $w$. The quarks (antiquarks) are placed at the vertices according to type-A ordering.}}
\label{recta}
\end{figure}

We focus here on the simplest geometry when the quark sources are at the vertices of a rectangle. Without loss of generality, we can place the rectangle in the $xy$-plane with its center at the origin and the $x$- and $y$-axes serving as symmetry axes, as shown in Figure \ref{recta}.
%________________________  fig - 4 __________________________________
%___________________________________________________________________

In the following analysis, we keep the ratio of length to width fixed, so that  
   
\begin{equation}\label{gc}
	\ell=\eta\,w
	\,,
\end{equation}
with $\eta$ a positive real number. We refer to Eq.\eqref{gc} as the geometrical constraint. It reduces the number of parameters by one that significantly simplifies the analysis.  

%______________________________________________________________
\subsection{The disconnected configurations}

We begin with the disconnected configurations, each of which can be interpreted as a hadronic molecule composed of two heavy mesons $Q\bar Q$. We therefore suppose that the energy equals the sum of the rest energies of the quark-antiquark pairs.\footnote{This is, of course, a simplification, as it neglects a binding energy. We return to this point in Sec.V.} In five dimensions, the configuration consists of two disconnected parts, each having the form shown in Figure \ref{conQQb} on the right.
%___________________________________________________________________
\subsubsection{Configuration (a)}

In this case, the quarks and antiquarks are separated by a distance $\ell$, and therefore the energy of configuration (a) is simply

\begin{equation}\label{Ea}
E^{\text{(a)}}=2E_{\QQb}(\ell)
\,,	
\end{equation}
where $E_{\QQb}$ and $\ell$ are given parametrically by Eq.\eqref{EQQb}.

Using the asymptotics for $E_{\QQb}(\ell)$ from Appendix C, we can easily obtain the behavior of $E^{\text{(a)}}$ for small and large $\ell$. Explicitly,

\begin{equation}\label{Ea-small}
E^{\text{(a)}}(\ell)=-\frac{\alpha^{\text{(a)}}}{\ell}+4c+o(1)
\,,\qquad\text{with}\qquad \alpha^{\text{(a)}}=2\alpha_{\QQb}\,,
\end{equation}
and 
\begin{equation}\label{Ea-large}
E^{\text{(a)}}(\ell)=2\sigma\ell
+2C_{\QQb}+o(1)
\,.
\end{equation}
Here $\alpha_{\QQb}$ and $C_{\QQb}$ are given by \eqref{EQQb-small} and \eqref{EQQb-large}, respectively, $c$ is a normalization constant, and $\sigma=\g\ep\s$ is the string tension \cite{az1}.  

\begin{figure}[H]
\centering
\includegraphics[width=8.1cm]{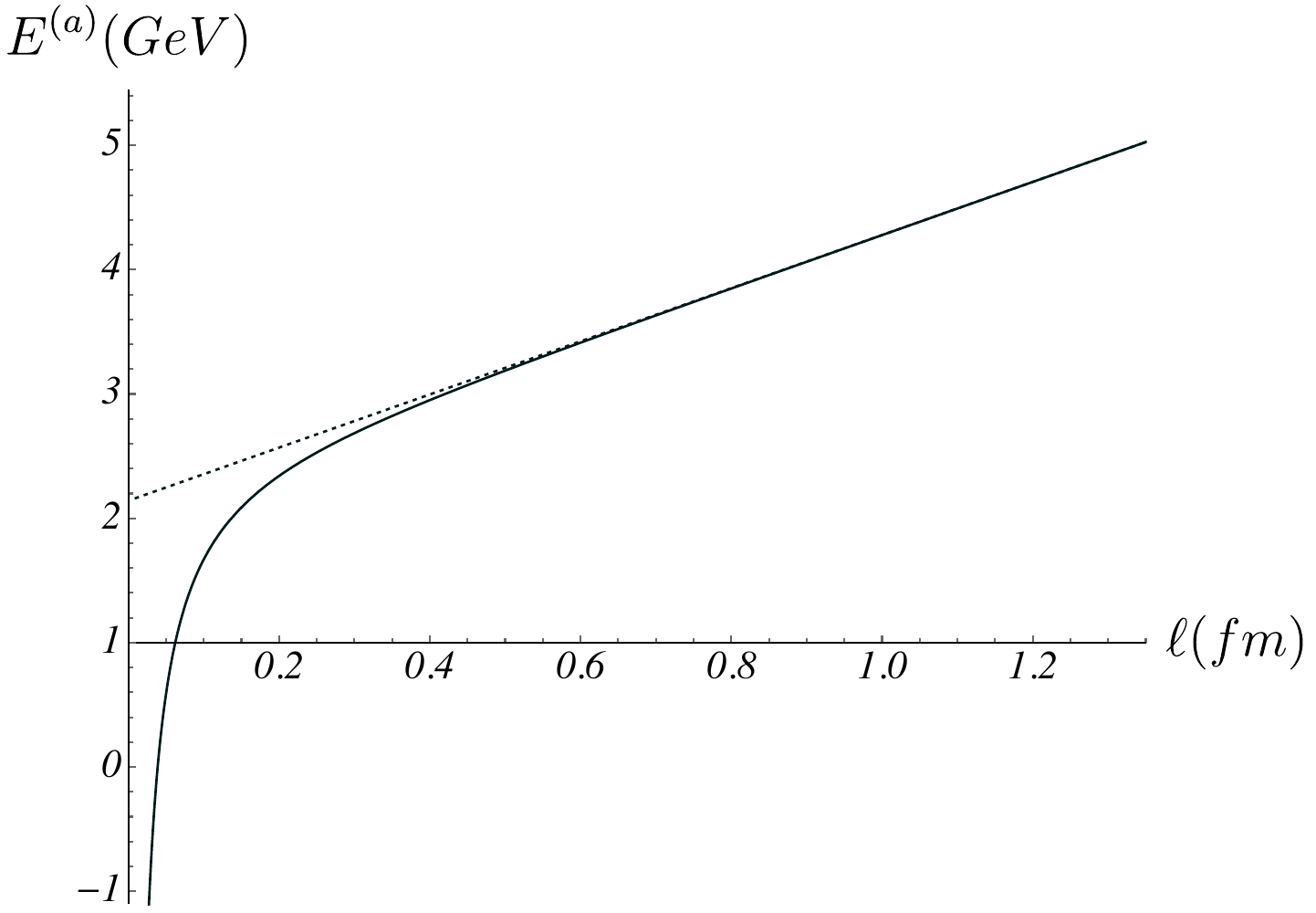}
\caption{{\small $E^{\text{(a)}}$ vs $\ell$. The dotted line represents the large-$\ell$ asymptotics \eqref{Ea-large}.}}
\label{Eal}
\end{figure}

 In Figure \ref{Eal} we plot $E^{\text{(a)}}$ as a function of $\ell$. Notably, this function becomes nearly linear for $\ell\gtrsim 0.6\,\text{fm}$, as  
 %________________________  fig - 5 __________________________________
%______________________________________________________________________
\noindent in the case of the $Q\bar Q$ system.

 %___________________________________________________________________
\subsubsection{Configuration (b)}

Configuration (b) can be treated similarly, but there is one difference. The quarks and antiquarks are now separated by a distance $\sqrt{1+\eta^{-2}}\ell$. Therefore, the energy is 

\begin{equation}\label{Eb}
E^{\text{(b)}}=2E_{\QQb}(\sqrt{1+\eta^{-2}}\,\ell)
\,.
\end{equation}
This implies that its asymptotic expansions can be obtained from those in \eqref{Ea-small} and \eqref{Ea-large} by rescaling $\ell\to\sqrt{1+\eta^{-2}}\ell$. Thus,

\begin{equation}\label{Eb-small}
E^{\text{(b)}}(\ell)=-\frac{\alpha^{\text{(b)}}}{\ell}+4c+o(1)
\,,\qquad\text{with}\qquad 
\alpha^{\text{(b)}}=\frac{2}{\sqrt{1+\eta^{-2}}}\alpha_{\QQb}\,,
\end{equation}
for small $\ell$, and 

\begin{equation}\label{Eb-large}
E^{\text{(b)}}(\ell)=2\sqrt{1+\eta^{-2}}\sigma\ell
+2C_{\QQb}+o(1)
\,
\end{equation}
for large $\ell$. Here, $E^{\text{(b)}}$ is proportional to twice the diagonal length.

We conclude with an important remark. At finite $\eta$, configuration (b) always has a higher energy than configuration (a). The reason is as follows. Both functions ${\cal L}^+(0,\lambda)$ and ${\cal E}^+(0,\lambda)$ increase monotonically in the interval $[0,1]$. For a given $\ell$, the value of $\lambda^{\text{(b)}}$ must be larger than that of $\lambda^{\text{(a)}}$ because $\sqrt{1+\eta^{-2}}>1$. Hence, ${\cal E}^+(0,\lambda^{\text{(b)}})>{\cal E}^+(0,\lambda^{\text{(a)}})$. Consequently, configuration (b) is irrelevant for the ground state potential and contributes only to higher B-O potentials. While this is obvious for large $\ell$, it is less clear for smaller $\ell$.  

%___________________________________________________________________
\subsection{The tetraquark configuration}
 
\begin{figure}[htbp]
\centering
\includegraphics[width=7.15cm]{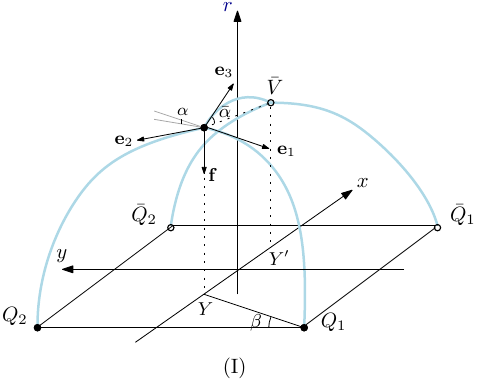}
\hspace{2cm}
 \includegraphics[width=7.15cm]{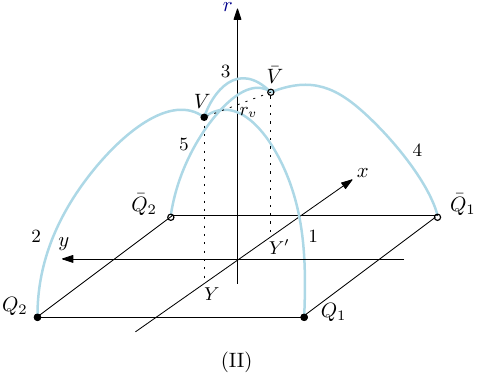}
\caption{{\small The basic tetraquark configurations in five dimensions. The baryon vertices are located at $r=\rv$ in the bulk, with $Y$ and $Y'$ being their projections onto the $x$-axis, and $\beta=\angle YQ_1Q_2$. For the tangent angles (see also Figure \ref{ngs}), we use the symmetry-based abbreviation: $\alpha_{1,2}=\bar\alpha_{1,2}=\alpha$ and $\alpha_3=\bar\alpha$. Left: $\alpha\geq 0$. The arrows indicate the forces acting on $V$, with the gravitational force $\mathbf{f}$ directed downward. Right: $\alpha\leq 0$. The strings are labeled by $1,\dots,5$.}}
\label{tetra5d}
\end{figure}

We now turn to the tetraquark configuration (c) in five dimensions. Its form was already suggested in \cite{a3Q2008}, and here we make it precise. The detailed construction depends nontrivially on $\eta$. In practice, it is convenient to define two basic configurations, shown in Figure \ref{tetra5d}, and to construct the tetraquark configuration from them. 

%________________________  fig - 6 __________________________________
%___________________________________________________________________
 
Before proceeding, let us make a few comments. 

First, the total action governing the tetraquark configuration is the sum of the Nambu-Goto actions for each of the five strings and the actions for the two vertices

\begin{equation}\label{c-action}
	S=\sum_{i=1}^5 S_{\text{\tiny NG}}^{(i)}+2S_{\text{vert}}
	\,.
\end{equation}
In Appendix B, it is shown how to extremize the Nambu-Goto actions to find the solutions that describe the strings. We will explain how to do so for the vertices shortly.

Second, the following simple but useful relations are:

\begin{equation}\label{lw}
\ell=\vert Y\bar Y\vert+2\sin\beta\vert Q_1Y\vert
\,,\qquad
w=2\cos\beta\vert Q_1Y\vert 
\,,
\end{equation}
valid for $0\leq\beta\leq\arctan\eta$. 

Third, at $\beta=\frac{\pi}{6}$, the points $Y$ and $\bar Y$ coincide with the Steiner points of the rectangle. In this case, the length of the Steiner tree is $\vert S\bar S\vert+4\vert Q_1 S\vert =(1+\sqrt{3}\eta^{-1})\ell$.\footnote{Note that the string configurations are not exactly the classical Steiner trees because quarks must connect only to baryon vertices, while antiquarks connect only to antibaryon vertices.} For $\eta\leq\frac{1}{\sqrt{3}}$, the Steiner points collapse to the rectangle center. 

Finally, if the vertices coincide, basic configuration (I) doesn't exist, while configuration (II) reduces to the pinched tetraquark configuration (d). Moreover, we treat the tetraquark configurations (c) and (d) separately because configuration (d) also exists for $\eta>\frac{1}{\sqrt{3}}$, where it differs from configuration (c).
%______________________________________________________________________________
\subsubsection{Strings joining at a baryon vertex}
 
 The string solutions described in Appendix B serve as the building blocks for the basic configurations. These blocks, however, must satisfy certain gluing conditions, which we now describe.

From the physical viewpoint, the gluing conditions express the requirement that the net force vanishes at any vertex where strings meet, as illustrated in Figure \ref{tetra5d} on the left. Translating this into mathematical terms is straightforward: extremizing the total action of the Nambu-Goto strings and baryon vertices with respect to the location of the vertex $V$ yields the force balance equation at that vertex

\begin{equation}\label{fbeV}
\mathbf{e}_1+\mathbf{e}_2+\mathbf{e}_3+\mathbf{f}=0
\,.
\end{equation}
The corresponding string tensions can be read off from the expressions \eqref{e} and \eqref{eb}

\begin{equation}\label{ei}
\begin{split}
	\mathbf{e}_1=-\g w(\rv)\bigl(\sin\beta\cos\alpha,\cos\beta\cos\alpha,&\sin\alpha\bigr)\,, 
	\quad
	\mathbf{e}_2=-\g w(\rv)\bigl(\sin\beta\cos\alpha,-\cos\beta\cos\alpha,\sin\alpha\bigr)\,,\\
\mathbf{e}_3=&\g w(\rv)\bigl(\cos\bar\alpha,0,\sin\bar\alpha\bigr)
\,.
\end{split}
\end{equation}
Note that $\bar\alpha$ is always nonnegative, as explained in Appendix B.

The gravitational force acting on the vertex is given by $\mathbf{f}_r=-\frac{1}{T}\frac{\delta V_{\text{vert}}}{\delta\rv}$, and therefore has only one non-zero component
 
\begin{equation}\label{f}
	\mathbf{f}=\Bigl(0,\,0,\,\tau_v(1+4\s\rv^2)\frac{\ep^{-2\s\rv^2}}{\rv^2}\,\Bigr)
	\,.
\end{equation}
The force balance equation then takes the component form

\begin{equation}\label{fbe}
	2\sin\beta\cos\alpha-\cos\bar\alpha=0\,,\qquad
	2\sin\alpha-\sin\bar\alpha-3\k(1+4v)\ep^{-3v}=0
	\,.
\end{equation}
Obviously, the same analysis applies to the vertex $\bar V$, with identical tangent angles. 

%___________________________________________________________________
\subsubsection{Configuration (I)}

We begin with configuration (I), characterized by nonnegative tangent angles. Using
\eqref{l+} and \eqref{l--}, we get $\vert Q_1Y\vert=\frac{1}{\sqrt{\s}}{\cal L}^+(\alpha,v)$ and $\vert Y\bar Y\vert=\frac{1}{\sqrt{\s}}{\cal L}(\bar\lambda,v)$. Here $\lambda_3$ is denoted by $\bar\lambda$. Thus, the geometrical constraint \eqref{gc} takes the form

\begin{equation}\label{gcp}
{\cal L}(\bar\lambda,v)=2(\eta\cos\beta-\sin\beta) {\cal L}^+(\alpha,v)
\,,
\end{equation}
where the functions ${\cal L}$ and ${\cal L}^+$ are defined in Appendix A, and $\bar\lambda=-\text{ProductLog}(-v\ep^{-v}/\cos\bar\alpha)$, as follows from \eqref{lambda}.

The expression for $\ell$ follows from Eq.\eqref{lw} with the help of \eqref{gc}, while the energy is obtained from Eqs.\eqref{baryon-v}, \eqref{E+} and \eqref{E--}. As a result, we find

\begin{equation}\label{ElI}
\ell^{\,\i}=\frac{2}{\sqrt{\s}}\eta\cos\beta{\cal L}^+(\alpha,v)
\,,\qquad
E^{\,\i}=\g\sqrt{\s}\Bigl(4{\cal E}^+(\alpha,v)+{\cal E}(\bar\lambda,v)+6\k\frac{\ep^{-2v}}{\sqrt{v}}
\Bigr)+4c
\,.
\end{equation}
Here $c$ is the normalization constant. When combined with the geometrical constraint and the force balance equations, this gives the function $E(\ell)$ in parametric form: $\ell=\ell^{\,\i}(v)$ and $E=E^{\,\i}(v)$. 

Let us now consider the small-$\ell$ behavior of $E^{\,\i}$. This limit corresponds to $v\to 0$, as follows from the asymptotic behavior of ${\cal E}^+$. In this case, the force balance equation and the geometric constraint become

\begin{equation}\label{fbe-small}
2\sin\beta\cos\alpha-\cos\bar\alpha=0\,,\qquad
	2\sin\alpha-\sin\bar\alpha -3\k=0
	\,, 
	\end{equation}
\begin{equation}\label{gcI-small}
	\sqrt{\cos\alpha}\,B(\sin^2\hspace{-2pt}\bar\alpha,\tfrac{1}{2},\tfrac{3}{4})
=
(\eta\cos\beta-\sin\beta)
\sqrt{\cos\bar\alpha}\,B(\cos^2\hspace{-2pt}\alpha,\tfrac{3}{4},\tfrac{1}{2})
\,,
\end{equation}
where $B(z;a,b)$ denotes the incomplete beta function. For a given $\eta$, the three angles can be determined from these equations. Taking the limit $v\to 0$ in Eqs.\eqref{ElI} and using Eqs.\eqref{fL+smallx}, \eqref{fE+smallx}, and \eqref{iEx=0}, we arrive at 

\begin{equation}\label{ElI-small}
E^{\,\i}=-
\frac{\alpha^{\i}}{\ell}
+4c+o(1)
\,,\quad\text{with}\quad
\alpha^{\i}=-2\eta\cos\beta\,{\cal L}^+_0(\alpha)
\Bigl(4{\cal E}^+_0(\alpha)+{\cal E}_0(\bar\alpha)
+
6\k\Bigr)\g
\,.
\end{equation}
Here the coefficients ${\cal L}^+_0$, ${\cal E}^+_0$ and ${\cal E}_0$ are defined in Appendix A. The notation ${\cal E}_0(\bar\alpha)$ is used for ${\cal E}_0(\cos\bar\alpha)$. For our parameter values, the factor $4{\cal E}^+_0+{\cal E}_0
+6\k$ is negative, and therefore $\alpha^{\i}$ is positive. 

It is also instructive to discuss the large-$\ell$ behavior. If $\eta$ is finite, the first equation in \eqref{ElI} implies that $\alpha=0$ and $v=1$. In this case, string 3 has $\bar\alpha=0$. However, the second equation in \eqref{fbe} has no solution with $\alpha=\bar\alpha=0$ in the interval $0\leq v\leq 1$. Thus, the large-$\ell$ limit does not exist for finite $\eta$. If $\eta$ is infinite, the limit exists and coincides with the diquark limit. The point is that in this case string 3 becomes infinitely long, while the others remain finite. In practice, it is more convenient to use the geometrical constraint $w=const$ to explore the diquark limit. We will return to this issue in \cite{4Q-2}.

%___________________________________________________________________
\subsubsection{Configuration (II)}

Since this configuration is govern by the same action as configuration (I), the force balance equations \eqref{fbe} still apply. The expressions for the geometrical constraint, the length and the energy can be obtained by replacing ${\cal L}^+$ and ${\cal E}^+$ with ${\cal L}^-$ and ${\cal E}^-$ corresponding to negative $\alpha$. This gives 

\begin{equation}\label{gcm}
{\cal L}(\bar\lambda,v)=2(\eta\cos\beta-\sin\beta) {\cal L}^-(\lambda,v)
\,,
\end{equation}
and 
\begin{equation}\label{ElII}
\ell^{\,\ii}=\frac{2}{\sqrt{\s}}\eta\cos\beta{\cal L}^-(\lambda,v)
\,,\qquad
E^{\,\ii}=\g\sqrt{\s}
\Bigl(4{\cal E}^-(\lambda,v)+{\cal E}(\bar\lambda,v)+6\k\frac{\ep^{-2v}}{\sqrt{v}}
\Bigr)+4c
\,.
\end{equation}
 Here, due to symmetry, we set $\lambda=\lambda_1=\lambda_2=\bar\lambda_1=\bar\lambda_2$. For a given $\eta$, these equations, together with Eqs.\eqref{fbe}, provide a parametric representation of $E$ as a function of $\ell$: $\ell=\ell^{\,\ii}(v)$ and $E=E^{\,\ii}(v)$. 

The small-$\ell$ behavior of $E^{\ii}(\ell)$ can be analyzed in a straightforward way. For $v\to 0$, the force balance equations \eqref{fbe-small} apply to this case as well, while the geometrical constraint becomes

\begin{equation}\label{gcl=02}
\sqrt{\cos\alpha}\,I(\sin^2\hspace{-2pt}\bar\alpha,\tfrac{1}{2},\tfrac{3}{4})
=
(\eta\cos\beta-\sin\beta)
\sqrt{\cos\bar\alpha}\bigl(1+I(\sin^2\hspace{-2pt}\alpha,\tfrac{1}{2},\tfrac{3}{4})\bigr)
\,,
\end{equation}
where $I$ denotes the regularized Beta function. Given $\eta$, the three angles can be determined from Eqs.\eqref{fbe-small} and \eqref{gcl=02}. We also replace ${\cal L}_0^+$ and ${\cal E}_0^+$ with ${\cal L}_0^-$ and ${\cal E}_0^-$ in \eqref{ElI-small}, getting 

\begin{equation}\label{ElII-small}
E^{\,\ii}=-
\frac{\alpha^{\ii}}{\ell}	+4c+o(1)
\,,\quad\text{with}\quad
\alpha^{\ii}=-2\eta\cos\beta\,{\cal L}^-_0(\alpha)
\Bigl(4{\cal E}^-_0(\alpha)+{\cal E}_0(\bar\alpha)
+
6\k\Bigr)\g
\,.
\end{equation}
Here the coefficients ${\cal L}_0^-$ and ${\cal E}^-_0$ are defined in Appendix A. For brevity, the argument is written as $\alpha$ instead of $\cos\alpha$.

Now we turn to the case of large $\ell$. There are two subcases to consider: $\lambda,\bar\lambda\to 1$, where all the strings become infinitely long, and $\lambda\to 1$ with fixed $\bar\lambda<1$, where string 3 remains finite in length. 

We begin with the former case, which leads to two distinct Steiner points and, as a result, to the term $\sigma L_{\text{min}}$ in $E^{\,\ii}$. From \eqref{v-lambda} it follows that in this limit $\cos\alpha=\cos\bar\alpha=v\ep^{1-v}$ and therefore $\alpha=-\bar\alpha$. Using this relation, the second equation in \eqref{fbe} can be readily solved, yielding $\sin\alpha=\k(1+4v)\ep^{-3v}$. Combining both expressions for $\alpha$, we find \footnote{An important point is that this is a general equation. It holds at any vertex where three infinitely long strings meet. For example, in the $QQQ$ system it was obtained in\cite{a3Q2016,a3Q2025}. We return to this issue in Sec.VI.}

\begin{equation}\label{v1}
	1-\k^2(1+4v)^2\ep^{-6v}-v^2\ep^{2(1-v)}=0
	\,,
\end{equation}
whose solution provides the upper bound on $v$. We denote it by $\vo$. Importantly, $\vo$ is independent of $\eta$ and belongs to the interval $[0,1]$. Numerically, $\vo=0.978$. It is noteworthy that in this large-$\ell$ limit the first equation in \eqref{fbe} yields $\sin\beta=\oh$ and thus $\beta=\frac{\pi}{6}$. Hence the points $Y$ and $Y'$ coincide with the Steiner points of the rectangle. 

As $\lambda,\,\bar\lambda\to 1$, the strings become infinitely long. The singularities arise from the functions ${\cal L}^-$ and ${\cal L}$, as seen from Eqs.\eqref{L-y=1} and \eqref{iLy=1}. With this in mind, we keep only the singular terms in $\ell$ and $E^{\,\ii}$. The resulting equations are then 

\begin{equation}\label{EII-large}
\ell=-\sqrt{\frac{3}{\s}}\ln(1-\lambda)\,+O(1)\,,
\qquad
E^{\,\ii}=-\g\ep\sqrt{\s}\bigl(	4\ln(1-\lambda)+\ln(1-\bar\lambda)\bigr)\,+O(1)
\,.
\end{equation}
The relation between $\lambda$ and $\bar\lambda$ follows from the geometrical constraint. Expressing $\bar\lambda$ in terms of $\lambda$, we obtain 

\begin{equation}\label{lambda5}
1-\bar\lambda=(1-\lambda)^{\sqrt{3}\eta-1}
\,.
\end{equation}
Importantly, this relation makes sense only if $\eta>\frac{1}{\sqrt{3}}$. This implies that the Steiner points do not coincide, as expected when an infinite long string terminates on them. Eliminating the parameter from Eqs.\eqref{EII-large} then gives 
 
 \begin{equation}\label{EII-large2}
	E^{\,\ii}=(1+\sqrt{3}\eta^{-1})\sigma\ell\,+O(1)
	\,.
\end{equation}
This is the desired behavior with the minimal length of the string network (Steiner tree) $L_{\min}=(1+\sqrt{3}\eta^{-1})\ell$.

To find the next-to-leading term, consider the difference between $E^{\,\ii}$ and $\sigma L_{\min}$

\begin{equation}\label{EII-large3}
\begin{split}
	E^{\,\ii}-\sigma L_{\min}
	=
	\g &\sqrt{\s}\Bigl(
	4\bigl({\cal E}^-(\lambda,v)-\ep{\cal L}^-(\lambda,v)\bigr)
	+{\cal E}(\bar\lambda,v)-\ep{\cal L}(\bar\lambda,v)
	+6\k\frac{\ep^{-2v}}{\sqrt{v}}\Bigr)\\
	+&
	4c+
	2\g\ep\sqrt{\s}
	\bigl(2-\sin\beta-\sqrt{3}\cos\beta\bigr){\cal L}^-(\lambda,v)
	\,.
\end{split}
\end{equation}
Taking the limit $\lambda,\bar\lambda\to 1$ ($v\to\vo$) and using the formulas \eqref{ILE} and \eqref{JLE}, we get 

\begin{equation}\label{EII-large4}
E^{\,\ii}-\sigma L_{\min}
	=
	\g \sqrt{\s}\Bigl(-4{\cal I}(\vo)-{\cal J}(\vo)
	+6\k\frac{\ep^{-2\vo}}{\sqrt{\vo}}\Bigr)\\
	+
	4c
	\,.
\end{equation}
Here the functions ${\cal I}$ and ${\cal J}$ are defined in Appendix A. The last term in \eqref{EII-large3} vanishes because the factor $2-\sin\beta-\sqrt{3}\cos\beta$ behaves like a power law in $1-\lambda$, whereas ${\cal L}^-$ like a logarithm. Thus,

\begin{equation}\label{EII-large5}
E^{\,\ii}=(1+\sqrt{3}\eta^{-1})\sigma\ell+C^{\,\ii}+o(1)
\,,\qquad\text{with}\qquad
C^{\,\ii}=4c
-
\g \sqrt{\s}\Bigl(4{\cal I}(\vo)+{\cal J}(\vo)
	-6\k\frac{\ep^{-2\vo}}{\sqrt{\vo}}\Bigr)\\
	\,.
\end{equation}
Note that, unlike the linear term, the constant term does not depend on $\eta$ and is, therefore, universal from this viewpoint.

In studying the latter case, we proceed as follows. The angle $\alpha$ is still given by $\cos\alpha=v\ep^{1-v}$, but this no longer holds for the angle $\bar\alpha$, which is now determined by the geometrical constraint. First, since the left hand side of \eqref{gcm} is finite, the factor $(\eta\cos\beta-\sin\beta)$ must vanish in this limit that implies $\tan\beta=\eta$. Then, the first equation in \eqref{fbe} gives $\cos\bar\alpha=\frac{2}{\sqrt{1+\eta^{-2}}}v\ep^{1-v}$. Substituting $\alpha$ and $\bar\alpha$ into the second equation in \eqref{fbe}, one finds that 

\begin{equation}\label{v1eta}
2\sqrt{1-v^2\ep^{2(1-v)}}+\sqrt{	1-\frac{4}{1+\eta^{-2}}v^2\ep^{2(1-v)}}+3\k(1+4v)\ep^{-3v}=0
\,,
\end{equation}
which manifestly depends on $\eta$. Accordingly, its solution is $\eta$-dependent and establishes the upper bound on $v$. We denote this solution in the interval $[0,1]$ as $\vt$. The allowed range of $\eta$ is governed by the two conditions $\bar\lambda=1$ and $\bar\alpha=0$, as discussed in Appendix D. 

Using Eqs.\eqref{L-y=1} and \eqref{E-y=1}, we obtain the singular terms in $\ell$ and $E^{\,\ii}$

\begin{equation}\label{EII-etalarge}
	\ell=-\frac{2}{\sqrt{\s}}
	\frac{\ln(1-\lambda)}{\sqrt{1+\eta^{-2}}}+O(1)
	\,,\qquad
	E^{\,\ii}=-4\g\ep\sqrt{\s}\,\ln(1-\lambda)+O(1)
	\,.
\end{equation}
It follows immediately that 

\begin{equation}\label{EII-etalarge2}
E^{\,\ii}=2\sqrt{1+\eta^{-2}}\sigma\ell+O(1)
\,,
\end{equation}
with the same string tension as before. In this case the length of the string network is equal to twice the length of the diagonal $d=\sqrt{1+\eta^{-2}}\ell$.

To compute the next term in the expansion, consider the difference 

\begin{equation}\label{EII-etalarge3}
	E^{\,\ii}-2\sigma d=
	\g\sqrt{\s}\Bigl(
	4\bigl({\cal E}^-(\lambda,v)-\ep\,{\cal L}^-(\lambda,v)\bigr)
	+{\cal E}(\bar\lambda,v)+2\ep\frac{1-\sqrt{1+\eta^2}\cos\beta}{\eta\cos\beta-\sin\beta}\,{\cal L}(\bar\lambda,v)+6\k\frac{\ep^{-2v}}{\sqrt{v}}
	\Bigr)+4c.
\end{equation} 
Now we take the limit $\lambda\rightarrow 1$ ($v\rightarrow \vt$), getting

\begin{equation}\label{EII-etalarge4}
	E^{\,\ii}-2\sigma d =
	\g\sqrt{\s}\Bigl(
	-4\,{\cal I}^-(\vt)
	+{\cal E}(\blt,\vt)-\frac{2\ep}{\sqrt{1+\eta^{-2}}}\,{\cal L}(\blt,\vt)+6\k\frac{\ep^{-2\vt}}{\sqrt{\vt}}
	\Bigr)+4c,
\end{equation} 
where $\blt=\lim_{v\to\vt}\bar\lambda=-\text{ProductLog}\bigl(-\frac{\sqrt{1+\eta^{-2}}}{2\ep}\bigr)$. Here we used that $\lim_{\beta\rightarrow\arctan\eta}\frac{1-\sqrt{1+\eta^2}\cos\beta}{\eta\cos\beta-\sin\beta}=-\frac{1}{\sqrt{1+\eta^{-2}}}$. Finally, the expansion of $E^{\,\ii}$ reads

\begin{equation}\label{EII-etalarge5}
E^{\,\ii}=2\sqrt{1+\eta^{-2}}\sigma\ell+C^{\,\ii}_\eta+o(1)
\,,\quad\text{with}\quad
C^{\,\ii}_\eta=4c
-
\g \sqrt{\s}\Bigl(4{\cal I}(\vt)
-{\cal E}(\blt,\vt)+\frac{2\ep}{\sqrt{1+\eta^{-2}}}\,{\cal L}(\blt,\vt)
	-6\k\frac{\ep^{-2\vt}}{\sqrt{\vt}}\Bigr)\\
	\,.
\end{equation}
The notable feature here is that the constant term depends explicitly on the ratio $\eta$. 

We have discussed the basic configurations, from which any tetraquark configuration (c) can be constructed. This construction, however, is cumbersome due to its $\eta$-dependence, so we describe it explicitly in Appendix D.

%___________________________________________________________________
\subsection{The pinched tetraquark configuration}

\begin{figure}[H]
\centering
\includegraphics[width=6.5cm]{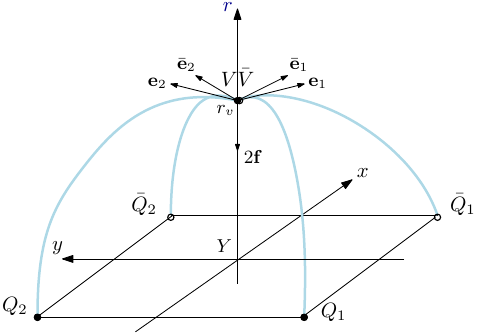}
\caption{{\small A pinched tetraquark configuration in five dimensions. Both baryon vertices are located at $r=\rv$ on the $r$-axis, and $Y$ is at the center of the rectangle. The arrows indicate the forces acting on the vertices. By symmetry, all the tangent angles of the strings are equal.}}
\label{pinched5d}
\end{figure}

In contrast to the tetraquark configuration, the pinched tetraquark configuration exists for all values of $\eta$ and is generally distinct from it.\footnote{For explicit examples, see Sec.V.} Visually, one may think of the pinched configuration as the tetraquark configuration with string 3 pinched into a point, see Figure \ref{pinched5d}. It is governed by the action 

%________________________  fig - 7 __________________________________
%___________________________________________________________________

\begin{equation}\label{d-action}
	S=\sum_{i=1}^4 S_{\text{\tiny NG}}^{(i)}+2S_{\text{vert}}
	\,.
\end{equation}
In this case, the force balance equation at $r=\rv$ takes the form

\begin{equation}\label{fbeV-d}
\mathbf{e}_1+\mathbf{e}_2+\bar{\mathbf{e}}_{1}+\bar{\mathbf{e}}_{2}+2\mathbf{f}=0
\,.
\end{equation}
Its only nontrivial component is the $r$-component, which allows us to explicitly express the tangent angle $\alpha$ in terms of the parameter $v$ 

\begin{equation}\label{fbe-d}
	\sin\alpha=\frac{3}{2}\k(1+4v)\ep^{-3v}
	\,.
\end{equation}
 Note that $\alpha$ is negative for negative $\k$.

A quick way to get some formulas is to set $\bar\alpha=0$ and $\tan\beta=\eta$ in the corresponding formulas of subsection B. So, the length and the energy are given by 

\begin{equation}\label{El-d}
\ell=\frac{2}{\sqrt{\s}} \frac{{\cal L}^-(\lambda,v)}{\sqrt{1+\eta^{-2}}}
\,,\quad
E^{\text{(d)}}=2\g\sqrt{\s}\Bigl(2{\cal E}^-(\lambda,v)+3\k\frac{\ep^{-2v}}{\sqrt{v}}\Bigr)+4c
\,,	
\end{equation}
where $v$ varies from $0$ to $\vp$, defined shortly. For a fixed $\eta$, these parametric equations determine the function $E^{\text{(d)}}(\ell)$.

The small-$\ell$ behavior of $E^{\text{(d)}}$ can be read off from Eq.\eqref{ElII-small}. Thus,

\begin{equation}\label{Eld-small}
E^{\text{\,(d)}}=-
\frac{\alpha^{\text{(d)}}}{\ell}	+4c+o(1)
\,,\quad\text{with}\quad
\alpha^{\text{(d)}}=
-\frac{4}{\sqrt{1+\eta^{-2}}}{\cal L}^-_0(\alpha)
\bigl(2{\cal E}^-_0(\alpha)+3\k\bigr)\g
\,.
\end{equation}
The factor $2{\cal E}^-_0(\alpha)+3\k$ is negative for our parameter values, so the coefficient $\alpha^{\text{(d)}}$ is positive. 

Next, let us consider the large-$\ell$ behavior. This corresponds to the limit $\lambda\to 1$, where the tangent angle is given by $\cos\alpha=v\ep^{1-v}$. Combining this with \eqref{fbe-d}, we get the equation

\begin{equation}\label{v1p}
	1-v^2\ep^{2(1-v)}-\frac{9}{4}\k^2(1+4v)^2\ep^{-6v}=0
	\,,
\end{equation}
whose solution $\vp$ in the interval $[0,1]$ gives the upper bound on $v$. Numerically, $\vp=0.967$. 

From the foregoing, it is clear that at leading order in $\ell$, $E^{\text{(d)}}$ is proportional to twice the diagonal length

\begin{equation}\label{Eld-large}
	E^{\text{(d)}}=2\sqrt{1+\eta^{-2}}\sigma\ell\,+O(1)
	\,,
\end{equation}
as $E^{\,\ii}$ in Eq.\eqref{EII-etalarge2}. To compute the constant term, consider the difference between $E^{\text{(d)}}$ and the linear term  

\begin{equation}\label{Eld-large3}
E^{\text{(d)}}-2\sigma d=	
2\g\sqrt{\s}\Bigl(2{\cal E}^-(\lambda,v)-2\ep{\cal L}^-(\lambda,v)
+3\k\frac{\ep^{-2v}}{\sqrt{v}}\Bigr)+4c
\,.
\end{equation}
After taking the limit $\lambda\rightarrow 1$, the right hand side becomes $2\g\sqrt{\s}\Bigl(-2{\cal I}(\vp)+3\k\frac{\ep^{-2\vp}}{\sqrt{\vp}}\Bigr)+4c$, where we have used \eqref{ILE}. Thus, we arrive at 

\begin{equation}\label{Eld-large5}
E^{\text{(d)}}=2\sqrt{1+\eta^{-2}}\sigma\ell+C^{\text{(d)}}+o(1)\,,
\qquad\text{with}\qquad
C^{\text{(d)}}=4c-2\g\sqrt{\s}\Bigl(2{\cal I}(\vp)-3\k\frac{\ep^{-2\vp}}{\sqrt{\vp}}\Bigr)
\,.
\end{equation}
An important feature of this expression is that the constant term $C^{\text{(d)}}$ is independent of $\eta$. 

%___________________________________________________________________  
\section{String configurations in five dimensions for type-B ordering}
\renewcommand{\theequation}{4.\arabic{equation}}
\setcounter{equation}{0}

The above analysis can be straightforwardly extended to the string configurations with type- B ordering. As before, the quarks (antiquarks) are placed at the vertices of the rectangle shown in Figure \ref{recta}, but with $Q_2$ and $\bar Q_2$ interchanged. Obviously, nothing happens with configurations (a) and (d). The main subtlety arises for the analog of configuration (c). While it is puzzling in four dimensions, in five dimensions a possible resolution is to align string 3 along the radial direction. 
%______________________________________________________________
\subsection{The disconnected configuration (b')}

We start with configuration (b'). Here, the quarks and antiquarks are separated by a distance $w$, and the energy is given by 

\begin{equation}\label{Ebp}
E^{\text{(b')}}=2E_{\QQb}(\eta^{-1}\,\ell)
\,.
\end{equation}
The function $E_{\QQb}$ is defined parametrically by $\ell= \frac{2\eta}{\sqrt{\s}}{\cal L}^+(0,\lambda)$ and $E_{\QQb}=2\g\sqrt{\s}\,{\cal E}^+(0,\lambda)+2c$.

The asymptotic expansions follow directly from the expressions \eqref{Ea-small} and \eqref{Ea-large} after rescaling $\ell\to\frac{\ell}{\eta}$. Thus, for small $\ell$ one finds 

\begin{equation}\label{Ebp-small}
E^{\text{(b')}}(\ell)=-\frac{\alpha^{\text{(b')}}}{\ell}+4c+o(1)
\,,\qquad\text{with}\qquad 
\alpha^{\text{(b')}}=2\eta\,\alpha_{\QQb}\,,
\end{equation}
and for large $\ell$ 

\begin{equation}\label{Ebp-large}
E^{\text{(b')}}(\ell)=\frac{2}{\eta}\sigma\ell
+2C_{\QQb}+o(1)
\,.
\end{equation}

%___________________________________________________________________
\subsection{The tetraquark configuration (c')}

\begin{figure}[htbp]
\centering
\includegraphics[width=7cm]{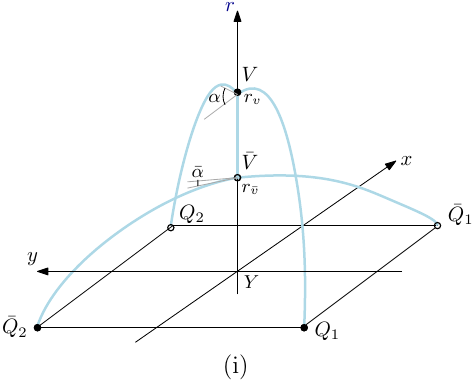}
\hspace{1.5cm}
\includegraphics[width=7cm]{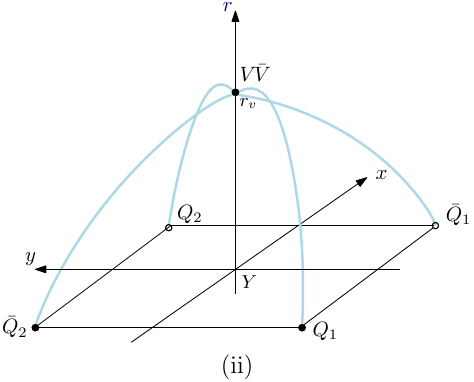}
\caption{{\small Basic configurations for type-B ordering in five dimensions. The baryon vertices lie on the $r$-axis, with $Y$ located at the origin. Because the configuration is symmetric under a permutation of the quarks (antiquarks), we use the shorthand $\alpha_{1,2}=\alpha$ and $\bar\alpha_{1,2}=\bar\alpha$ for the tangent angles. Left: $\rv>\rvb$. Right: $\rv=\rvb$. Shown here is a basic configuration (ii), where $\bar\alpha\geq 0$.}}
\label{tetrap5d}
\end{figure}

The problem of constructing the tetraquark configuration for type-B ordering is slightly tricky. One way to do so is to imagine "cutting" the strings attached to the quarks in configuration (c), while keeping those attached to the antiquarks intact, and then reconnecting them to the quarks reordered according to type-B ordering. The two resulting configurations are shown in Figure \ref{tetrap5d}.\footnote{A third configuration is obtained from configuration (ii) by changing the signs of the tangent angles of the strings attached to the antiquarks.}  An unusual feature is that string 3 is stretched along the $r$-axis.  
%________________________  fig - 8 __________________________________
%___________________________________________________________________
When projected onto the $xy$-plane, it becomes indistinguishable from configuration (d) in Figure \eqref{confs4B}. This makes the tetraquark configuration difficult to visualize in four dimensions.

This tetraquark configuration may be analyzed along the lines of Sec.III by introducing basic configurations and then constructing the tetraquark configuration from them. The total action is again given by the sums of the Nambu-Goto actions and the vertex contributions, as in \eqref{c-action} and \eqref{d-action}. 
	
%___________________________________________________________________
\subsubsection{Configuration (i)}

The first basic configuration is shown on the left of Figure \ref{tetrap5d}. We must extremize the corresponding action with respect to $x(r)$, which describes the string profiles, and with respect to $\rv$ and $\rvb$, which  specify the position of the vertices. In Appendix B, we outline how this is done for the strings. Varying $\rv$ and $\rvb$ results in the force balance equations of the form \eqref{fbeV} at $V$ and $\bar V$. Due to symmetry, only the $r$-component is non-zero. Using \eqref{e} and \eqref{eb} together with \eqref{f}, we get 

\begin{equation}\label{fbeW}
	2\sin\alpha+1-3\k(1+4v)\ep^{-3v}=0
	\,,
\qquad
	2\sin\bar\alpha-1-3\k(1+4\bar v)\ep^{-3\bar v}=0
	\,
\end{equation}
which allow us to express the tangent angles in terms of $v$ and $\bar v$, where  $\bar v=\s\rvb^2$. For $\k=-0.083$, $\alpha$ is negative in the interval $[0,1]$ while $\bar\alpha$ is positive.

The form of the expression for $\ell$ is the same as that in the case of configuration (d). For the energy, however, one must take into account string 3 and the different tangent angle of the strings attached to the antiquarks. Putting all together with the help of \eqref{Evertb}, we find 

\begin{equation}\label{El-cpI}
 \ell^{\text{\,(i)}}=\frac{2}{\sqrt{\s}}\,\frac{{\cal L}^-(\lambda,v)}{\sqrt{1+\eta^{-2}}}
\,,\quad
E^{\text{\,(i)}}=\g\sqrt{\s}
\Bigl(2{\cal E}^-(\lambda,v)+2{\cal E}^+(\bar\alpha,\bar v)
+
{\cal Q}(v)-{\cal Q}(\bar v)+
3\k\frac{\ep^{-2v}}{\sqrt{v}}
+
3\k\frac{\ep^{-2\bar v}}{\sqrt{\bar v}}
\Bigr)
+4c
\,.
\end{equation}
This is not the whole story. A geometrical constraint arises from the relation $\vert Q_1Y\vert=\vert\bar Q_1Y\vert$, which yields

\begin{equation}\label{gc-pI}
	{\cal L}^-(\lambda,v)={\cal L}^+(\bar\alpha,\bar v)
	\,.
\end{equation}
Using this, one can express $\bar v$ in terms of $v$ and write the energy in parametric form as $E=E^{\text{\,(i)}}(v)$ and $\ell=\ell^{\text{\,(i)}}(v)$. 

A numerical analysis shows that, for any given $v$ (except $v=0$), there is no solution to Eqs.\eqref{fbeW} and \eqref{gc-pI} satisfying $\bar v\leq v$. Thus, the basic configuration (i) does not exist for our parameter values. 
%_______________________________________________________________________
\subsubsection{Configuration (ii)}

The second basic configuration is shown on the right of the Figure. It may be viewed as configuration (i) with string 3 pinched into a point, so that $\bar v=v$ but $\alpha\not=\bar\alpha$. The latter distinguishes it from configuration (d).

In this case, the geometrical constraint \eqref{gc-pI} simplifies to 

\begin{equation}\label{gc-pII}
	{\cal L}^-(\lambda,v)={\cal L}^+(\bar\alpha,v)
	\,.
\end{equation}
The force balance equation takes the form of \eqref{fbeV-d}, with only the $r$-component non-zero. A simple way to obtain it is to add the equations in \eqref{fbeW} and then to set $\bar v=v$. This gives

\begin{equation}\label{fbeW-II}
	\sin\alpha+\sin\bar\alpha-3\k(1+4v)\ep^{-3v}=0
	\,.
\end{equation}
The expression for the length in terms of $v$ is the same as in \eqref{El-cpI},  but that for the energy becomes

\begin{equation}\label{El-cpII}
E^{\text{\,(ii)}}=2\g\sqrt{\s}
\Bigl({\cal E}^-(\lambda,v)+{\cal E}^+(\bar\alpha,v)
+
3\k\frac{\ep^{-2v}}{\sqrt{v}}
\Bigr)
+4c
\,.
\end{equation}

For a given $v$, one may determine the tangent angles from the geometrical constraint and the force balance equation. Numerics, however, shows that there are no solutions with $\alpha\not=\bar\alpha$. Thus, the basic configuration (ii) also does not exist. 

%_______________________________________________________________________
\subsubsection{Configuration (iii)}

This basic configuration has the same structure as (ii), but with all the tangent angles negative.\footnote{It may be thought of as an asymmetric version of configuration (d) with $\alpha\not =\bar\alpha$.} Hence all the formulas can be obtained by replacing ${\cal L}^+$ and ${\cal E}^+$ with ${\cal L}^-$ and ${\cal E}^-$, corresponding to negative $\bar\alpha$. In this way, from \eqref{gc-pII} we get 

\begin{equation}\label{gc-pIII}
	{\cal L}^-(\lambda,v)={\cal L}^-(\bar\lambda,v)
	\,.
\end{equation}
Note that $\bar\lambda\not=\lambda$ because $\bar\alpha\not=\alpha$. The above constraint becomes trivial at $\bar\alpha=\alpha$, as in configuration (d). Similarly, for the energy, we have 

\begin{equation}\label{El-cpIII}
E^{\text{\,(iii)}}=2\g\sqrt{\s}
\Bigl({\cal E}^-(\lambda,v)+{\cal E}^-(\bar\lambda,v)
+
3\k\frac{\ep^{-2v}}{\sqrt{v}}
\Bigr)
+4c
\,.
\end{equation}
Meanwhile, the force balance equation \eqref{fbeW-II} remains unchanged. 

In principle, the geometrical constraint and force balance equation allow one to express the tangent angles in terms of $v$. A simple numerical analysis, however, show that for a given $v$ the only solution has $\alpha=\bar\alpha$. In other words, the result reduces to configuration (d). Thus, the basic configuration (iii) also does not exist. We conclude that for our parameter set configuration (c') does not exist either. One might suspect that this conclusion holds for other geometries as well, but this is not the case: configuration (c') does exist when the diagonals are unequal. A concrete example is presented in Appendix E. This marks an important difference between four- and five-dimensional string models.

We can summarize all this by saying that for our parameter values, only three types of configurations exist in the case of type-B ordering.
%___________________________________________________________________  
\section{The potentials of the $QQ\bar Q\bar Q$ system}
\renewcommand{\theequation}{5.\arabic{equation}}
\setcounter{equation}{0}

We can gain insight into the lowest Born-Oppenheimer potentials by following the approach used in lattice QCD. Consider a model Hamiltonian, which in the general case is an $n\times n$ matrix

\begin{equation}\label{H}
{\cal H}(\ell)=
\begin{pmatrix}
\,E_1 & {} & \Theta_{ij} \, \\
\,{} & \ddots & {} \,\\
\,\Theta_{ij} & {} & E_n \,
\end{pmatrix}
\,,
\end{equation}
where the diagonal elements represent the energies of string configurations, and the off-diagonal elements describe the strength of mixing between them.\footnote{Importantly, the binding energies of disconnected configurations are partially encoded in the $\Theta$'s.} The lowest B-O potentials are given by the smallest eigenvalues of ${\cal H}$. Note that if $\Theta_{ij}$ vanishes for the states $i$ and $j$, then there is no mixing between them, and the plots of $E_i$ and $E_j$ do not intersect. Although in principle the Hamiltonian can be determined from a correlation matrix in lattice QCD, it remains challenging to compute the off-diagonal elements within the effective string model. Because of this, it is not possible to visualize the exact shape of the potentials. Nevertheless, valuable insight can be gained by treating $\Theta_{ij}$ as free parameters, for example in terms of the approximate magnitudes of the $\Theta$ values near the transition points. In this case, we may assume that the $\Theta$'s are approximately constant, with values of about $40$-$60\,\text{MeV}$, as in \cite{Bicu-4Q,bulava}. 

As a special case of \eqref{H}, consider $n=2$. The corresponding potentials are given by 

\begin{equation}\label{H2}
V_{0,1}=\oh\bigl(E_1+E_2\bigr)\mp\sqrt{\frac{1}{4}\bigl(E_1-E_2 \bigr)^2+\Theta^2_{12}}
\,.
\end{equation}
Here $V_0=\min\{E_1,E_2\}$ and $V_1=\max\{E_1,E_2\}$.
This formula is useful for plotting the potentials in the vicinity of intersection points. 
%___________________________________________________________________
\subsection{The case of type-A ordering}

Once the string configurations are constructed in Sec.III, we will use them to analyze the two lowest B-O potentials, denoted as $V_0$ and $V_1$. However, the complexity of configuration (c) makes the following analysis somewhat cumbersome.  
%___________________________________________________________________
\subsubsection{$0<\eta<\frac{1}{\sqrt{3}}$}

In this range, the tetraquark configuration (c) does not exist that considerably simplifies the analysis. The ground state potential is provided by configuration (a) and the first excited one by configuration (d), namely $V_0=E^{\text{(a)}}$ and $V_1=E^{\text{(d)}}$. Thus, if the $\Theta$'s are sufficiently small, we expect the ground state to correspond to a hadronic molecule, as the potential $V_0$ is described by the disconnected configuration, whereas the first excited potential gives rise to tetraquark states.

\begin{figure}[htbp]
\centering
\includegraphics[width=8.25cm]{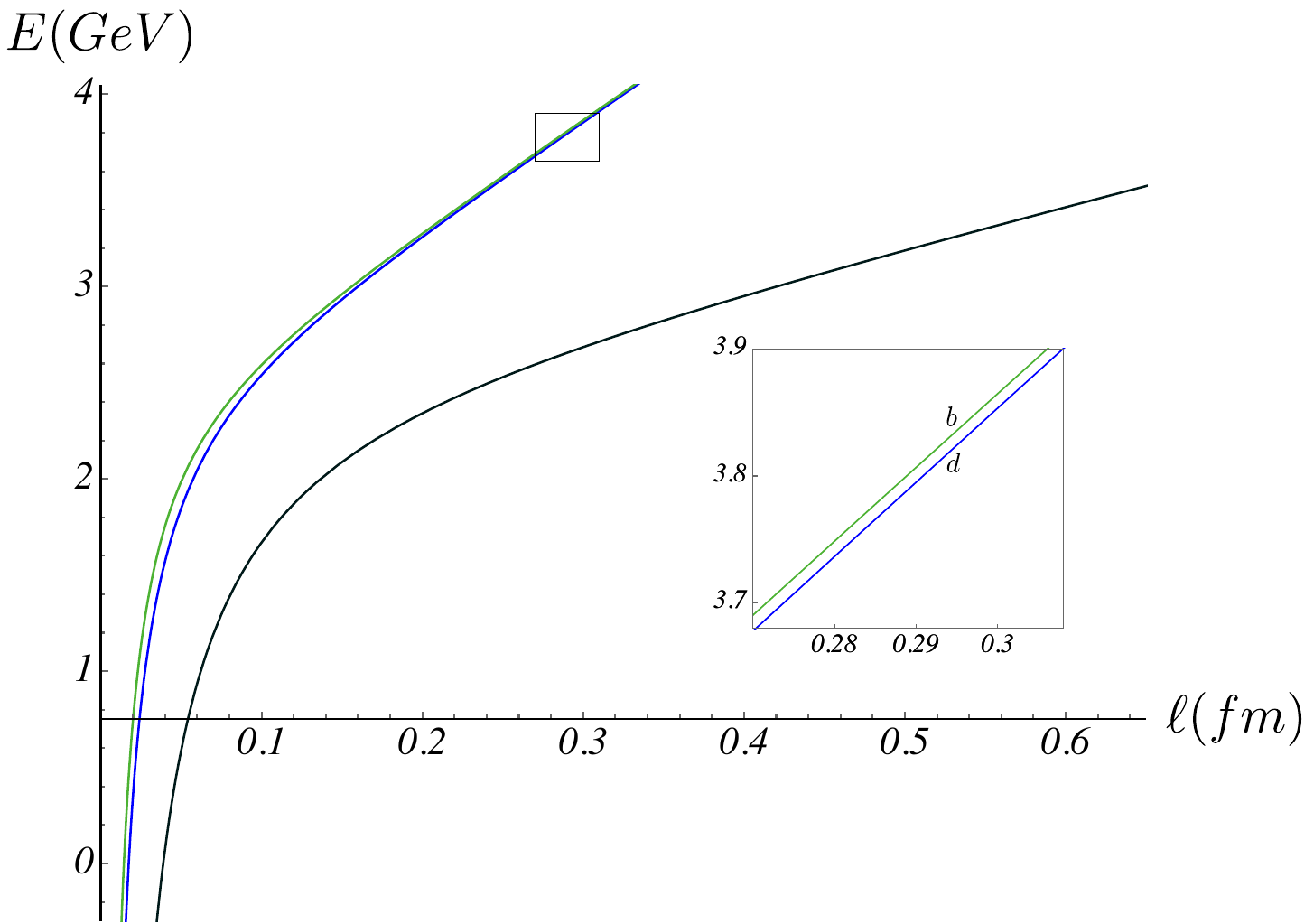}
\hspace{1cm}
\includegraphics[width=8.25cm]{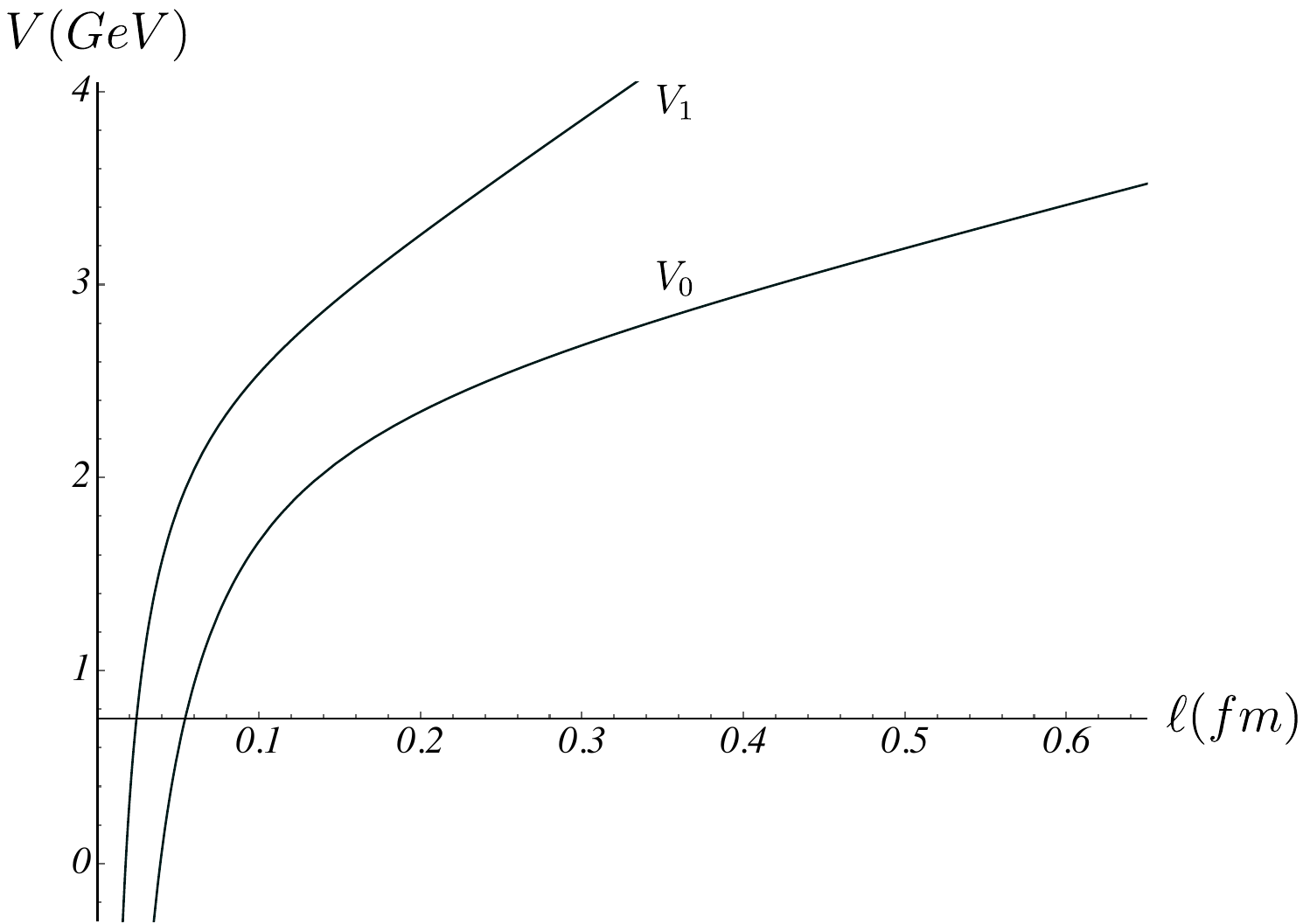}
\caption{{\small Various plots at $\eta=0.4$. Left: The $E$'s vs $\ell$. The black, green, and blue curves correspond respectively to configurations (a), (b), and (d), here and below. Right: The two lowest potentials. The energies of irrelevant configurations are omitted, here and below.}}
\label{all04}
\end{figure}

  As an illustration, Figure \ref{all04} shows the results of numerical computations for $\eta=0.4$. The plots of $E^{\text{(b)}}$ and $E^{\text{(d)}}$ 
%________________________  fig - 9 __________________________________
%___________________________________________________________________
\noindent do not intersect. Although the linear asymptotics \eqref{Eb-large} and \eqref{Eld-large5} have the same slopes, their intercepts differ by about $8\,\text{MeV}$.\footnote{This is true for any $\eta$.} 
%____________________________________________________________________
\subsubsection{$\eta_{\textup{\tiny p}}<\eta\leq 0.5840$}

As in Appendix D, we skip the very narrow range $\frac{1}{\sqrt{3}}\leq\eta\leq\eta_{\text{\tiny p}}$, where the construction of configuration (c) is puzzling, and proceed to the next interval. Here the lower bound $\eta_{\text{\tiny p}}$ is defined by \eqref{etap}. The ground state potential is still provided by configuration (a), while the first excited potential by two configurations (c) and (d). Thus, $V_0=E^{\text{(a)}}$ and $V_1=\min\{E^{\text{(c)}}, E^{\text{(d)}}\}$, where configuration (d) dominates at small $\ell$ and configuration (c) at large. This does not change the physical interpretation compared to the previous range: the ground state may still correspond to a hadronic molecule, depending on the strength of mixing.
%________________________  fig - 10 __________________________________
\begin{figure}[]
\centering
\includegraphics[width=8.25cm]{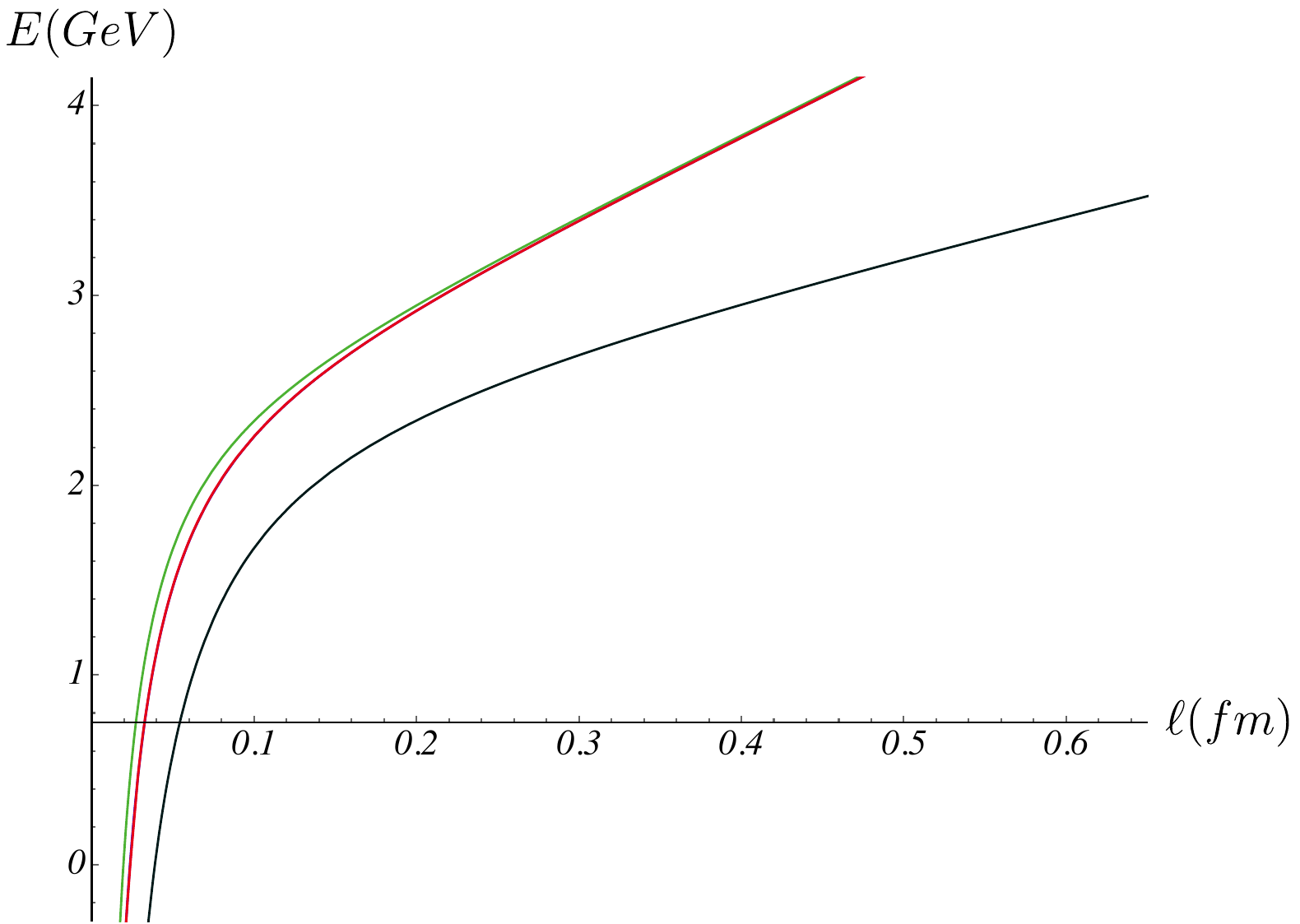}
\hspace{1cm}
\includegraphics[width=8.15cm]{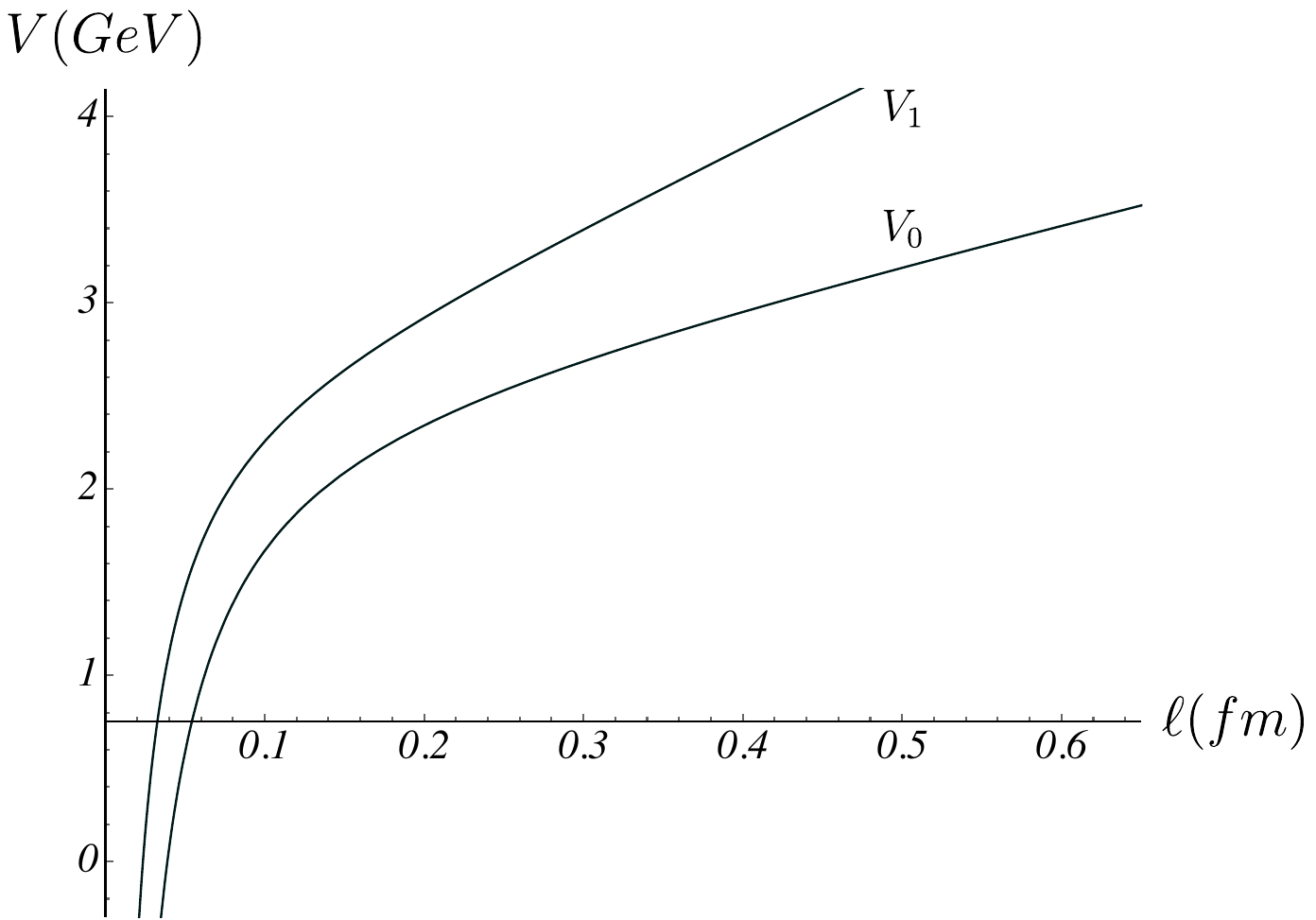}
\caption{{\small Various plots at $\eta=0.58$ Left: The $E$'s vs $\ell$. The red curve corresponds to configuration (c), here and below. Right: Shown here are the two lowest potentials. The approximate formula \eqref{H2} with $\Theta_{12}=50\,\text{MeV}$ is used to plot the potentials near the transition points, here and below.}}
\label{all058}
\end{figure}
%___________________________________________________________________

As the next example, consider $\eta=0.58$. The results are presented in Figure \ref{all058}. The plots of $E^{\text{(c)}}$ and $E^{\text{(d)}}$ are indistinguishable because the difference between them is extremely small. At small $\ell$, $E^{\text{(c)}}$ is larger than $E^{\text{(d)}}$, with the Coulomb coefficients satisfying $\alpha^{\text{(c)}}/\alpha^{\text{(d)}}=1-1.5\times 10^{-3}$, whereas at large $\ell$ the situation is reversed, with the slope ratio $0.5(\sqrt{3}+\eta)/\sqrt{1+\eta^2}=1-2\times 10^{-6}$. The transition between these configurations can be interpreted as pinching (see Figure \ref{stri}). To make this more quantitative, we define a critical length by equating the energies

\begin{equation}\label{lc}
E^{\text{(c)}}(\ell_c)=E^{\text{(d)}}(\ell_c)
\,,
\end{equation}
and similarly for other transitions. At sufficiently large $\ell$, one may estimate the critical length using the linear asymptotics \eqref{EII-large5} and \eqref{Eld-large5}. For $\eta=0.58$ this yields $\ell_c=13.6\,\text{fm}$. Before proceeding further, we note that the upper bound of this range corresponds to $\ell_c=0$, where the Coulomb coefficients become equal.

%____________________________________________________________________
\subsubsection{$0.5840<\eta<1.17937$}

This range is easier to analyze. The ground state potential is given by configuration (a) and the first excited potential by configuration (c). Thus, $V_0=E^{\text{(a)}}$ and $V_1=E^{\text{(c)}}$. As in the previous ranges, we expect the ground state to be a hadronic molecule. 

\begin{figure}[H]
\centering
\includegraphics[width=8.25cm]{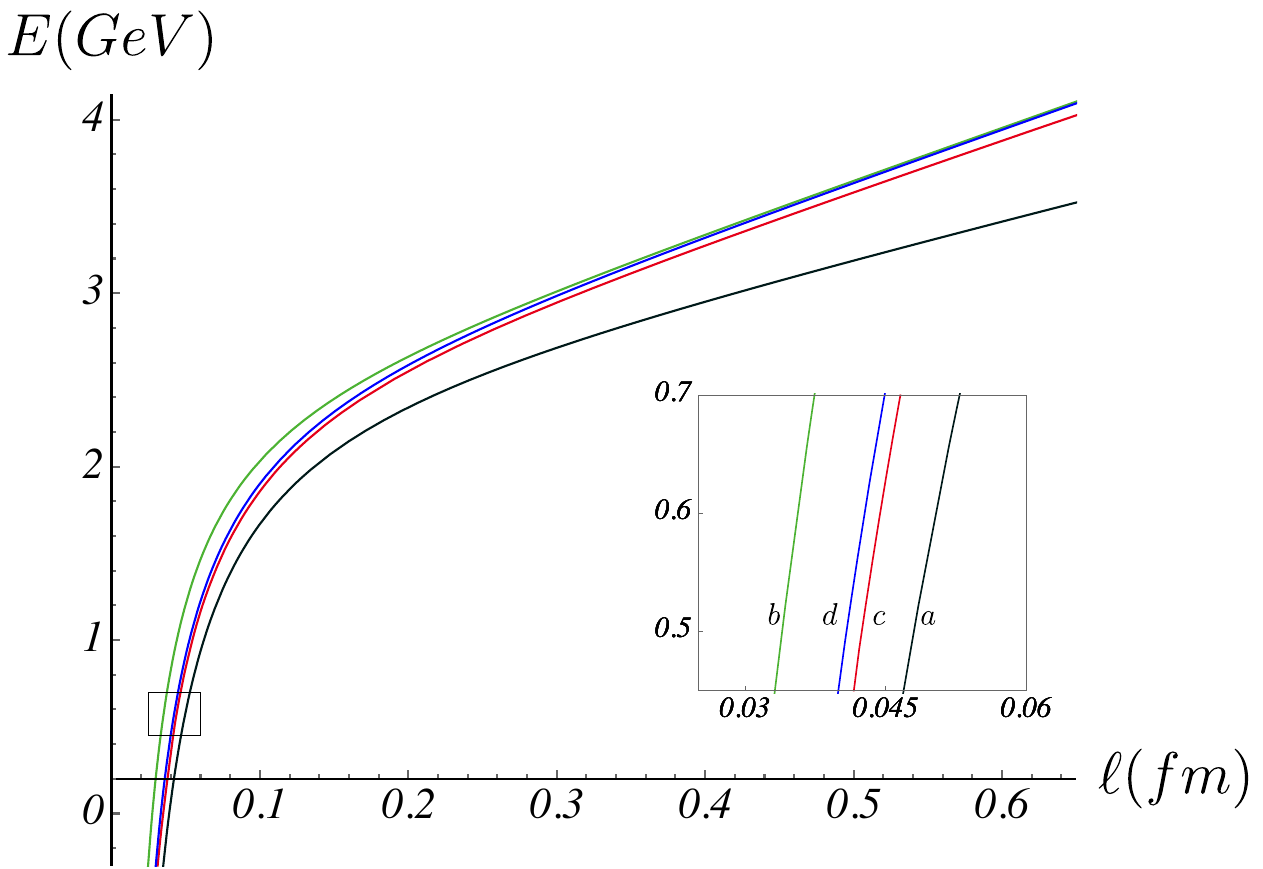}
\hspace{1.25cm}
\includegraphics[width=8cm]{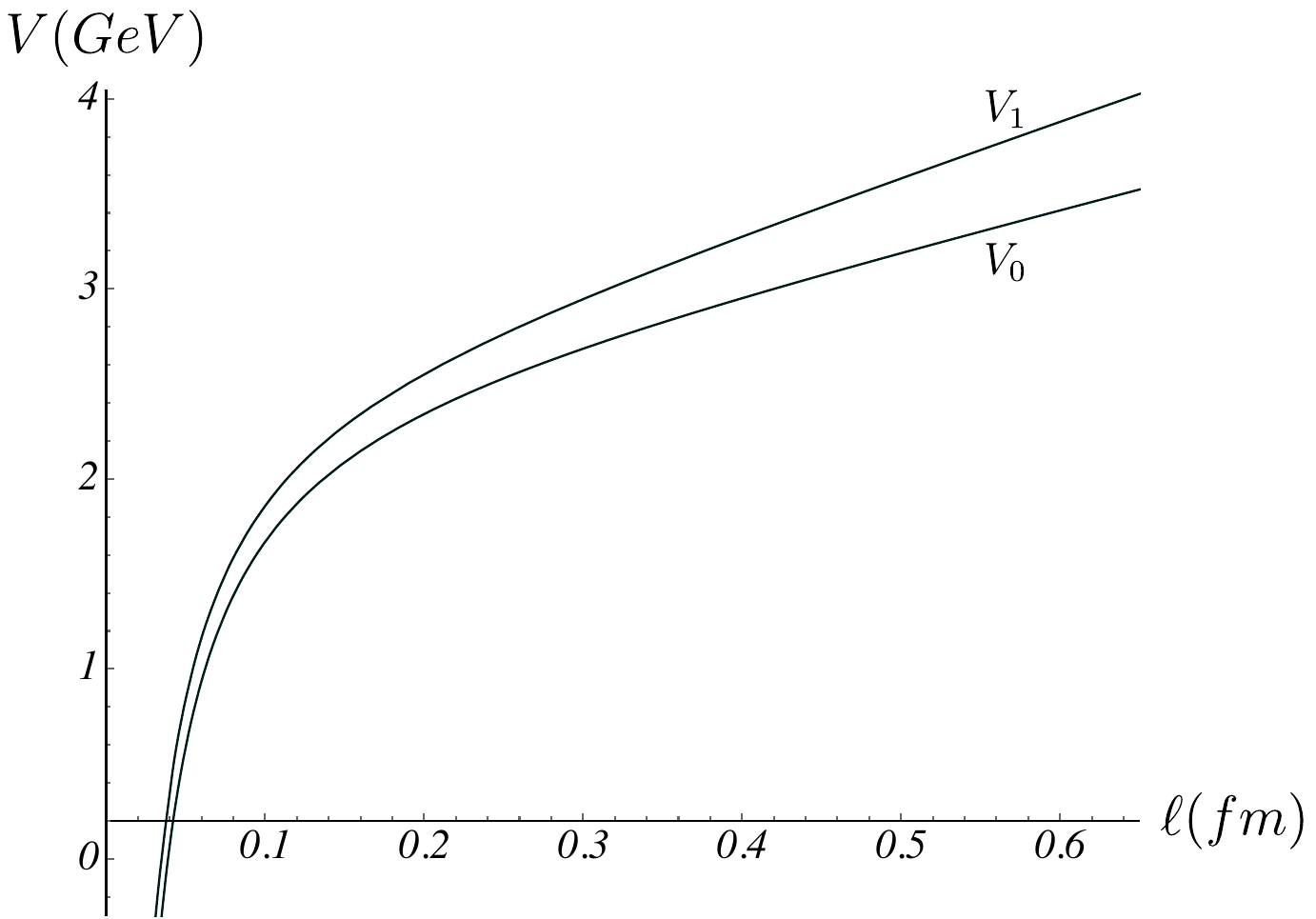}
\caption{{\small Various plot at $\eta=1$. Left: The $E$'s vs $\ell$. Right: The two lowest potentials.}}
\label{all1}
\end{figure}

To illustrate these statements, let us choose $\eta=1$. Figure \ref{all1} shows the resulting plots.
%________________________  fig - 11 __________________________________
%___________________________________________________________________
 
%____________________________________________________________________
\subsubsection{$1.17937 \leq\eta\leq 1.434792$}

At the lower bound of this range, the Coulomb coefficients of $E^{\text{(a)}}$ and $E^{\text{(c)}}$ coincide, implying an intersection point between them at $\ell=0$. The coordinate of the intersection point increases with $\eta$. This implies that a transition occurs between these configurations, which can be interpreted as the process of string junction annihilation. As before, we define the critical length by equating the energies, $E^{\text{(a)}}=E^{\text{(c)}}$. At the upper bound, the Coulomb coefficients of $E^{\text{(a)}}$ and $E^{\text{(d)}}$ coincide, indicating another intersection at $\ell=0$. The potentials are therefore described by two configurations (a) and (c) as $V_0=\min\{E^{\text{(a)}},E^{\text{(c)}}\}$ and $V_1=\max\{E^{\text{(a)}},E^{\text{(c)}}\}$. This suggests that the ground state is no longer a pure state but a mixed state of a hadronic molecule and a tetraquark state.  

\begin{figure}[htbp]
\centering
\includegraphics[width=8.25cm]{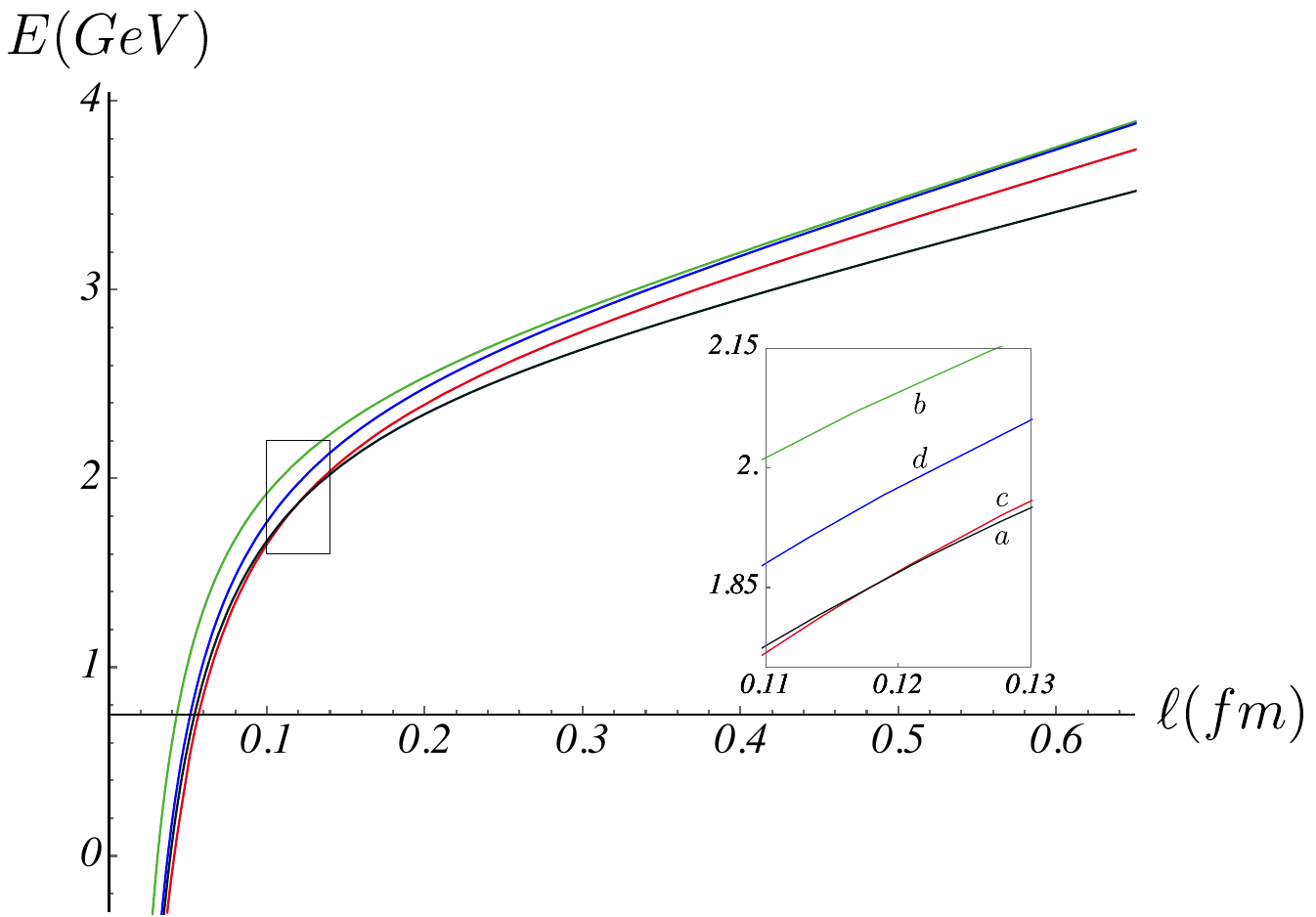}
\hspace{1.25cm}
\includegraphics[width=8cm]{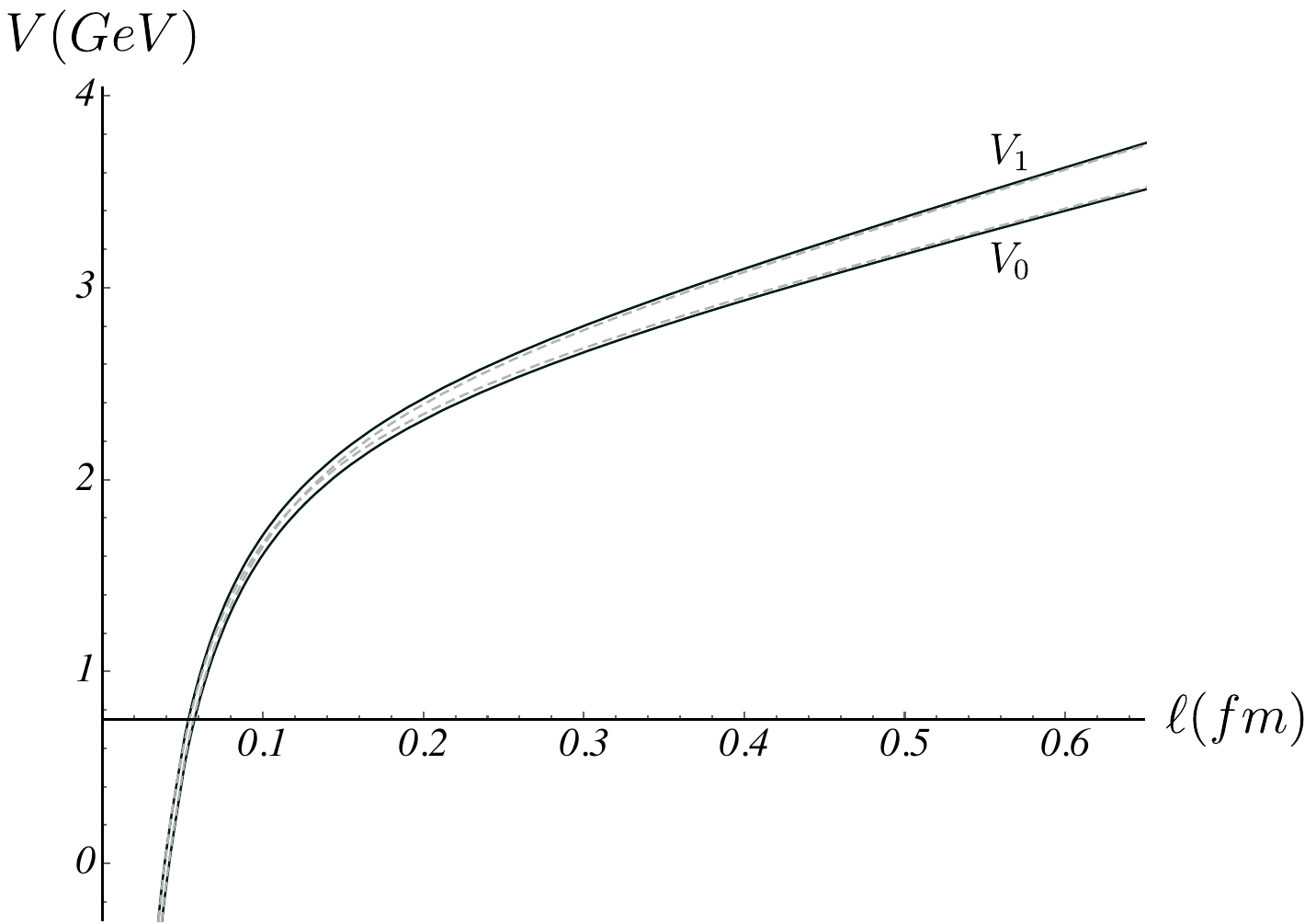}
\caption{{\small Various plot at $\eta=1.3$. Left: The $E$'s vs $\ell$. Right: The two lowest potentials. The dashed curves indicate the plots of the relevant $E$'s, here and below.}}
\label{all13}
\end{figure}

As an illustration, Figure \ref{all13} shows the results of numerical computations for $\eta=1.3$. Here the critical length is 
%________________________  fig - 12   __________________________________
%___________________________________________________________________
$\ell_c=0.118\,\text{fm}$. For larger $\ell$, configuration (c) has a higher energy, while for smaller $\ell$ the energy of configuration (a) is higher.
%___________________________________________________________________
\subsubsection{$1.434792\leq\eta\leq\sqrt{3}$}

\begin{figure}[H]
\centering
\includegraphics[width=8.25cm]{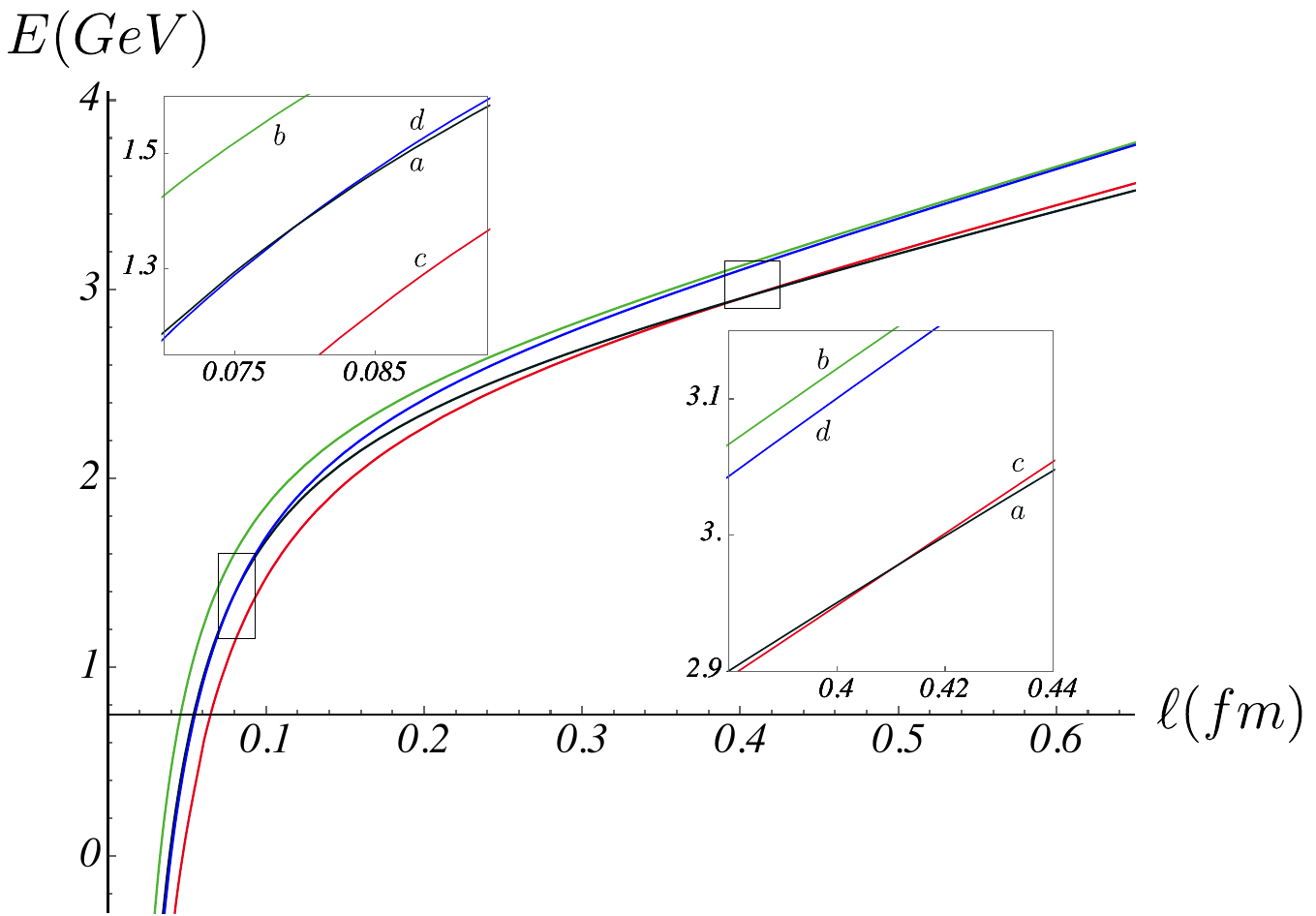}
\hspace{1.25cm}
\includegraphics[width=8cm]{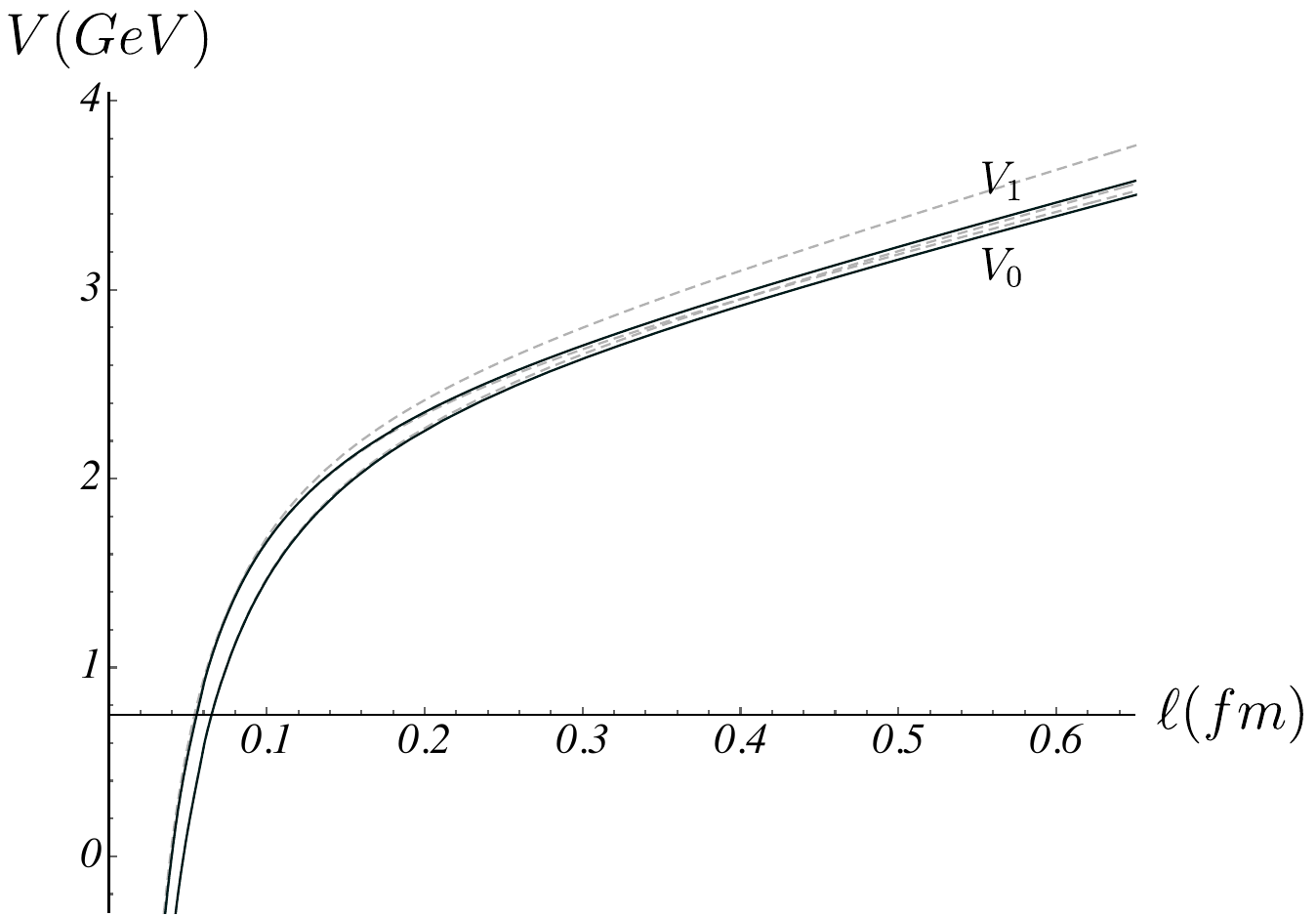}
\caption{{\small Various plots at $\eta=1.6$. Left: The $E$'s vs $\ell$. Right: Sketched here are the two lowest potentials.}}
\label{all16}
\end{figure}

As $\eta$ increases, the intersection points between $E^{\text{(a)}}$ and $E^{\text{(d)}}$ as well as between $E^{\text{(a)}}$ and $E^{\text{(c)}}$ shift to larger values of $\ell$. At the upper bound, the slopes of $E^{\text{(a)}}$ and $E^{\text{(c)}}$ become equal and, as a consequence, their plots no longer intersect at finite $\ell$. This leads to a rather entangled pattern of the energies, as illustrated in Figure \ref{all16} on the left. The ground state potential is described by configurations (a) and (c), while the first excited potential involves configurations (a), (c), and (d). Explicitly, $V_0=\min\{E^{\text{(a)}},E^{\text{(c)}}\}$ and $V_1=\min\{E^{\text{(a)}}, E^{\text{(c)}}, E^{\text{(d)}}\}$.\footnote{The minimum in $V_0$ must be taken first.} Hence, the ground state is expected to be a mixed state, involving both hadronic and tetraquark components. 

As an example, consider $\eta=1.6$. The resulting plots are presented in Figure \ref{all16}. The plot of $E^{\text{(a)}}$ intersects with 
%________________________  fig -__________________________________
%___________________________________________________________________
\noindent $E^{\text{(c)}}$ and $E^{\text{(d)}}$. We define the corresponding critical length by equating the energies. So, we have $\ell_c=0.411\,\text{fm}$ and $\ell_c=0.080\,\text{fm}$, respectively. Both transitions can be interpreted as string junction annihilation.   
 
%__________________________________________________________________
\begin{center}
{${\it 6}.\quad\sqrt{3}<\eta$}
\end{center}

In this range, $E^{\text{(c)}}$ does not intersect with the others, while $E^{\text{(a)}}$ and $E^{\text{(d)}}$ intersect with each other. Thus, the ground state potential is given simply by $V_0=E^{\text{(c)}}$ and the first excited one by $V_1=\min\{E^{\text{(a)}}, E^{\text{(d)}}\}$. Because of this, we expect the ground state to correspond to a tetraquark state. 

\begin{figure}[htbp]
\centering
\includegraphics[width=8cm]{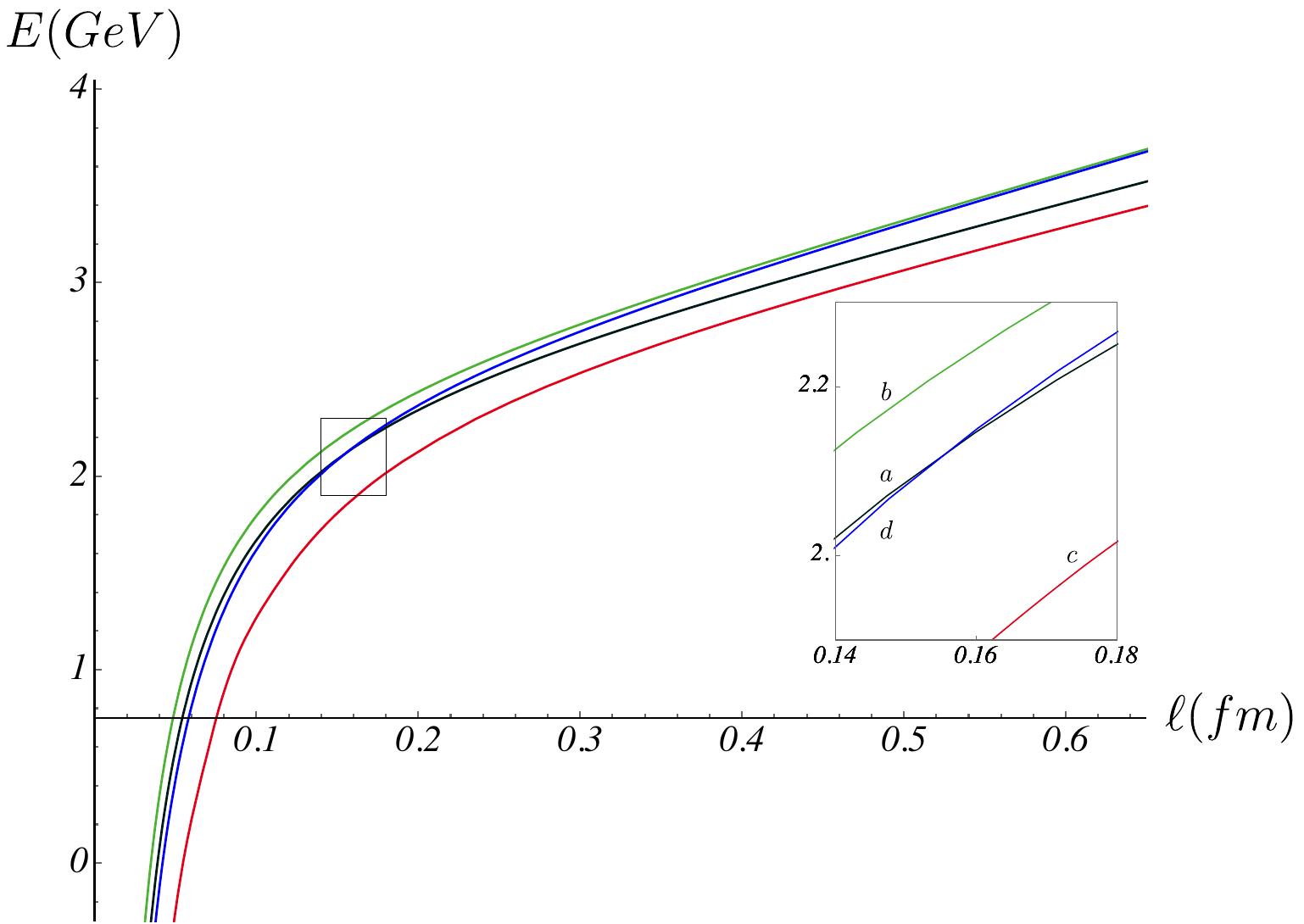}
\hspace{1.5cm}
\includegraphics[width=8cm]{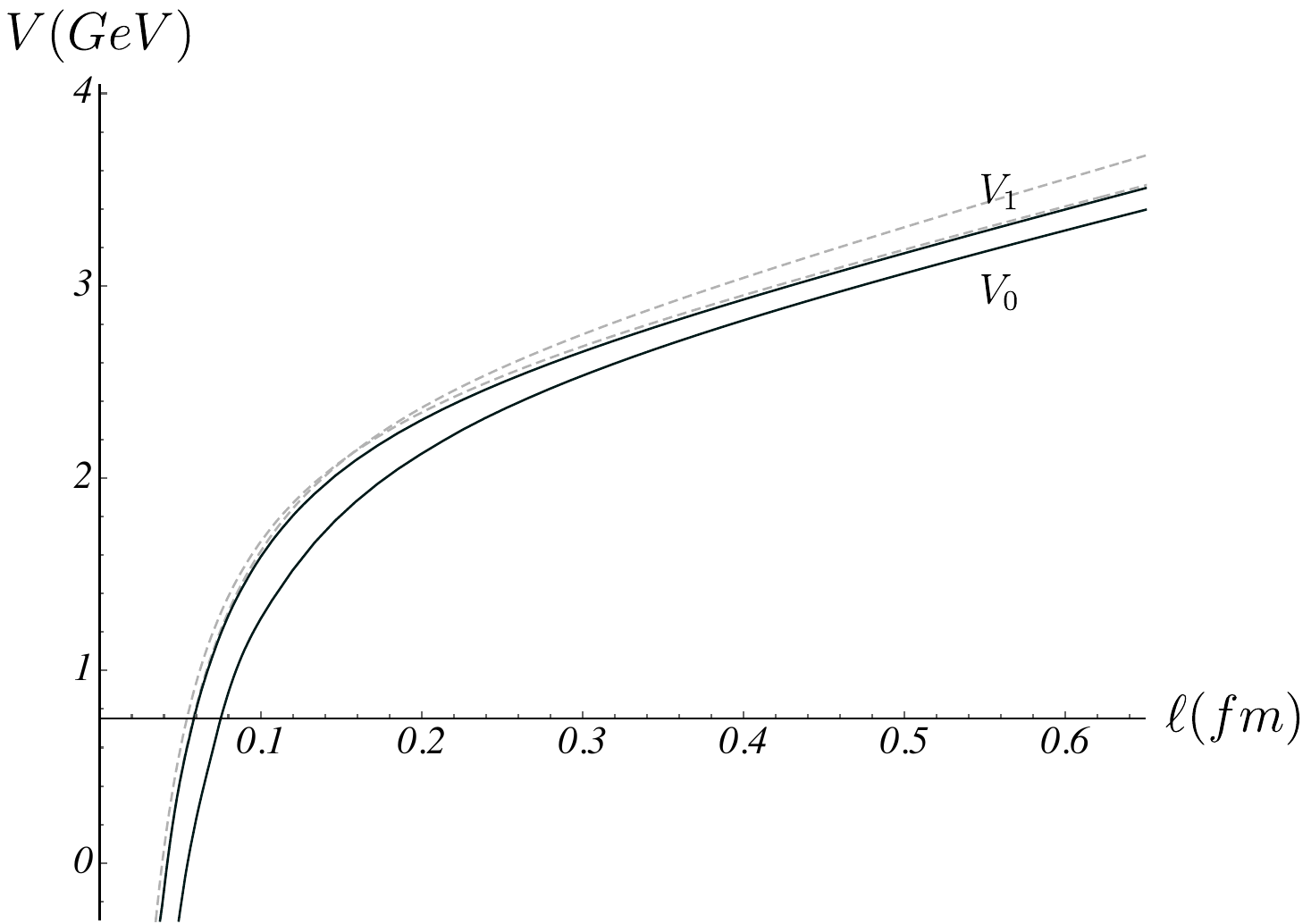}
\caption{{\small Various plots at $\eta=2$. Left: The $E$'s vs $\ell$. Right: The two lowest potentials.}}
\label{all2}
\end{figure}

We illustrate this with the example $\eta=2$. Figure \ref{all2} shows the corresponding numerical results. The transition 
%________________________  fig - 14 __________________________________
%___________________________________________________________________
between configurations (a) and (d) occurs at $\ell_c=0.156\,\text{fm}$ and can be interpreted as string junction annihilation. Note that $\ell_c$ increases with $\eta$ and tends to infinity in the diquark limit ($\eta\to\infty$).

%__________________________________________________________________
\subsection{Type-B ordering}

As we saw in Sec.IV, the tetraquark configuration does not exist for this ordering of the quark sources.\footnote{Note that this does not exclude the existence of tetraquark states, since the pinched tetraquark configuration still exists for type-B ordering.} This, together with an additional symmetry, constitutes a special feature that significantly simplifies the following analysis. On the other hand, the analog of configuration (b), namely configuration (b'), now becomes relevant for the potential $V_1$. 

%______________________________________________________________________
\subsubsection{$0<\eta\leq 0.6969651$}

A numerical analysis shows that in this range, the ground state potential is described by a single configuration, (a), whereas the first excited potential involves two configurations, (b') and (d). In the latter case, the tetraquark (pinched) configuration dominates at small $\ell$, while configuration (a) dominates at large $\ell$. The transition between them occurs due to string junction annihilation. So, we have $V_0=E^{(\mathrm{a})}$ and $V_1=\min\{E^{(\text{b'})},E^{(\text{d})}\}$, and we expect the ground state to be a hadronic molecule.
%________________________  fig - 15 __________________________________
\begin{figure}[]
\centering
\includegraphics[width=8cm]{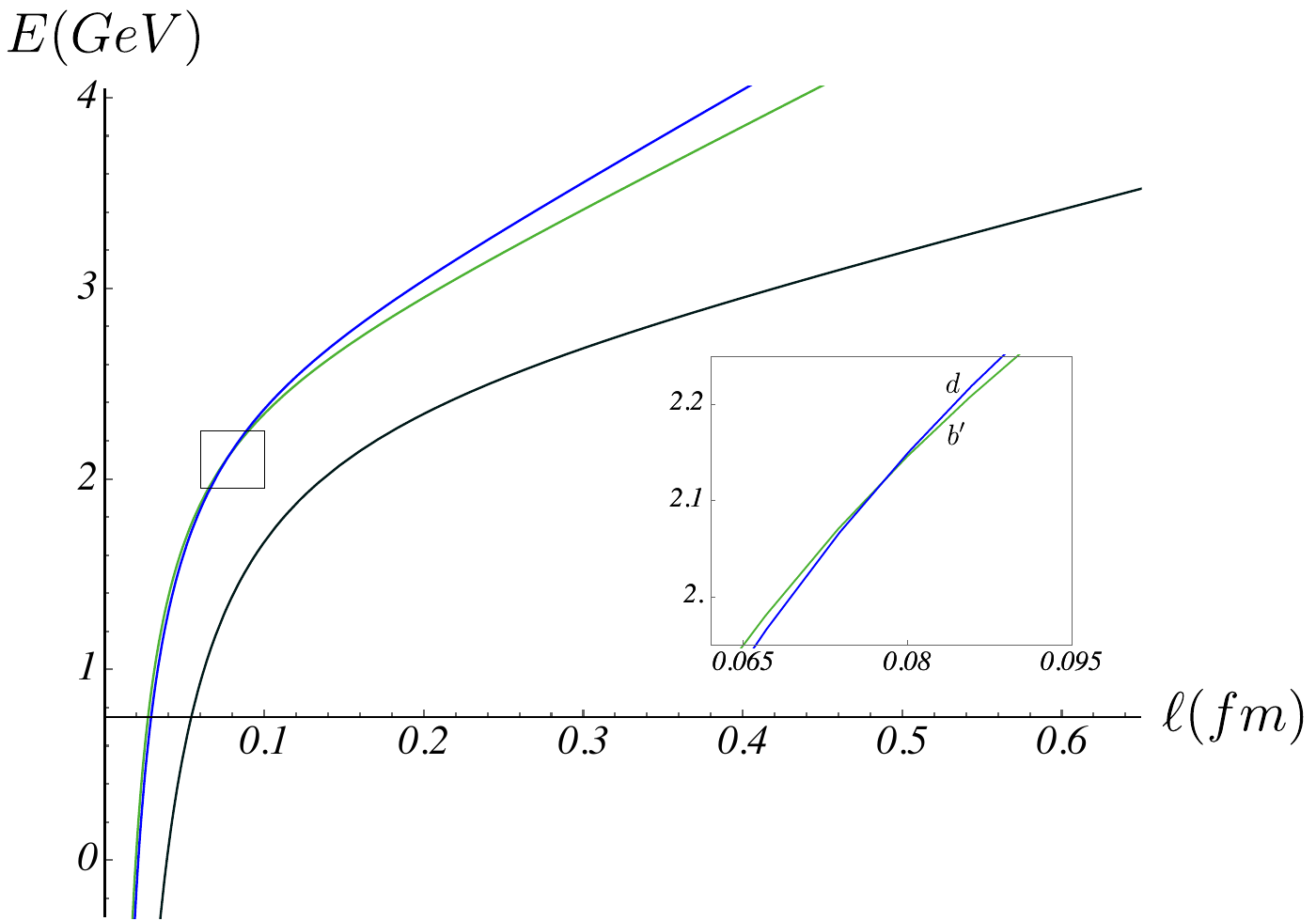}
\hspace{1.5cm}
\includegraphics[width=8cm]{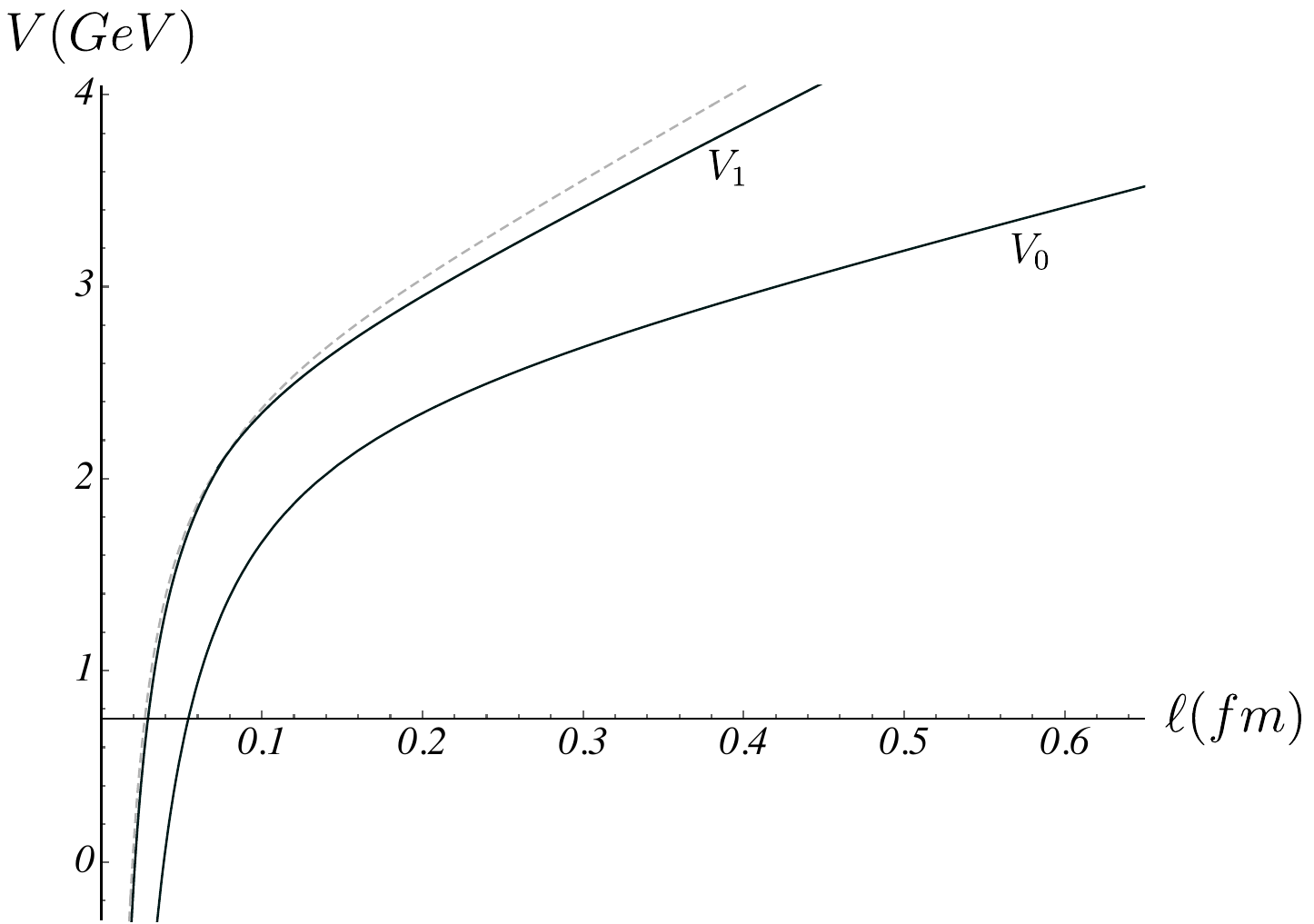}
\caption{{\small Various plots at $\eta=0.5$. Left: The $E$'s vs $\ell$. Right: The two lowest potentials.}}
\label{all05}
\end{figure}
%___________________________________________________________________

To illustrate this, let us choose $\eta=0.5$. The corresponding plots are shown in Figure \ref{all05}. The critical length characterizing the transition between configuration (b') and (d) is given by $\ell_c=0.078\,\text{fm}$. 

%_______________________________________________________
\subsubsection{$0.6969651<\eta <1$}

As $\eta$ increases, the critical length $\ell_c$ decreases and eventually vanishes at $\eta=0.6969651$, where the Coulomb coefficients $\alpha^{\text{(b')}}$ and $\alpha^{\text{(d)}}$ become equal. This defines the lower bound. For larger values of $\eta$, $E^{\text{(b')}}$ and $E^{\text{(d)}}$  no longer intersect. Hence, $V_0=E^{(\text{a})}$ and $V_1=E^{(\text{b'})}$. In this case, the ground state is also expected to correspond to a hadronic molecule. 

\begin{figure}[H]
\centering
\includegraphics[width=8cm]{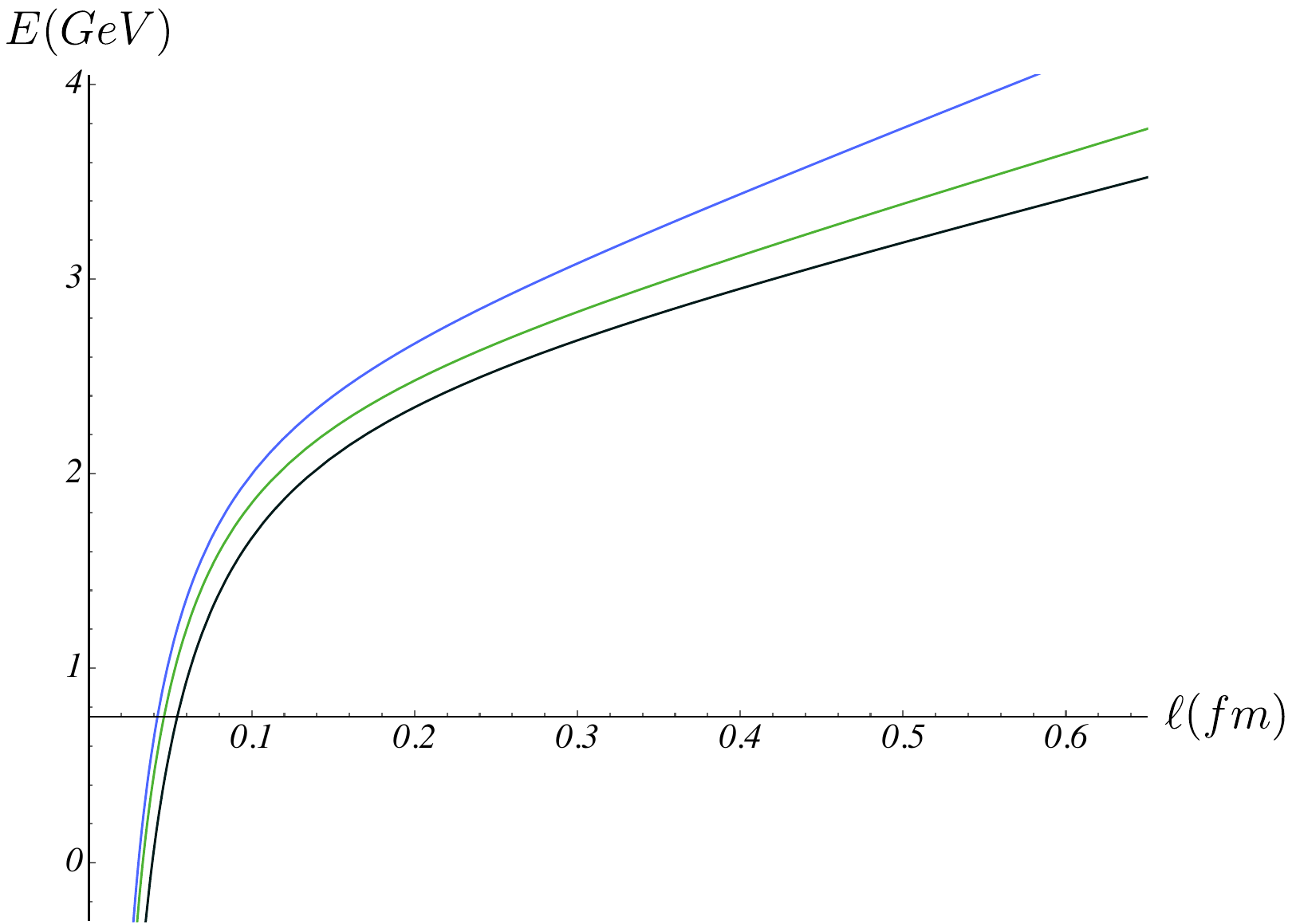}
\hspace{1.5cm}
\includegraphics[width=8cm]{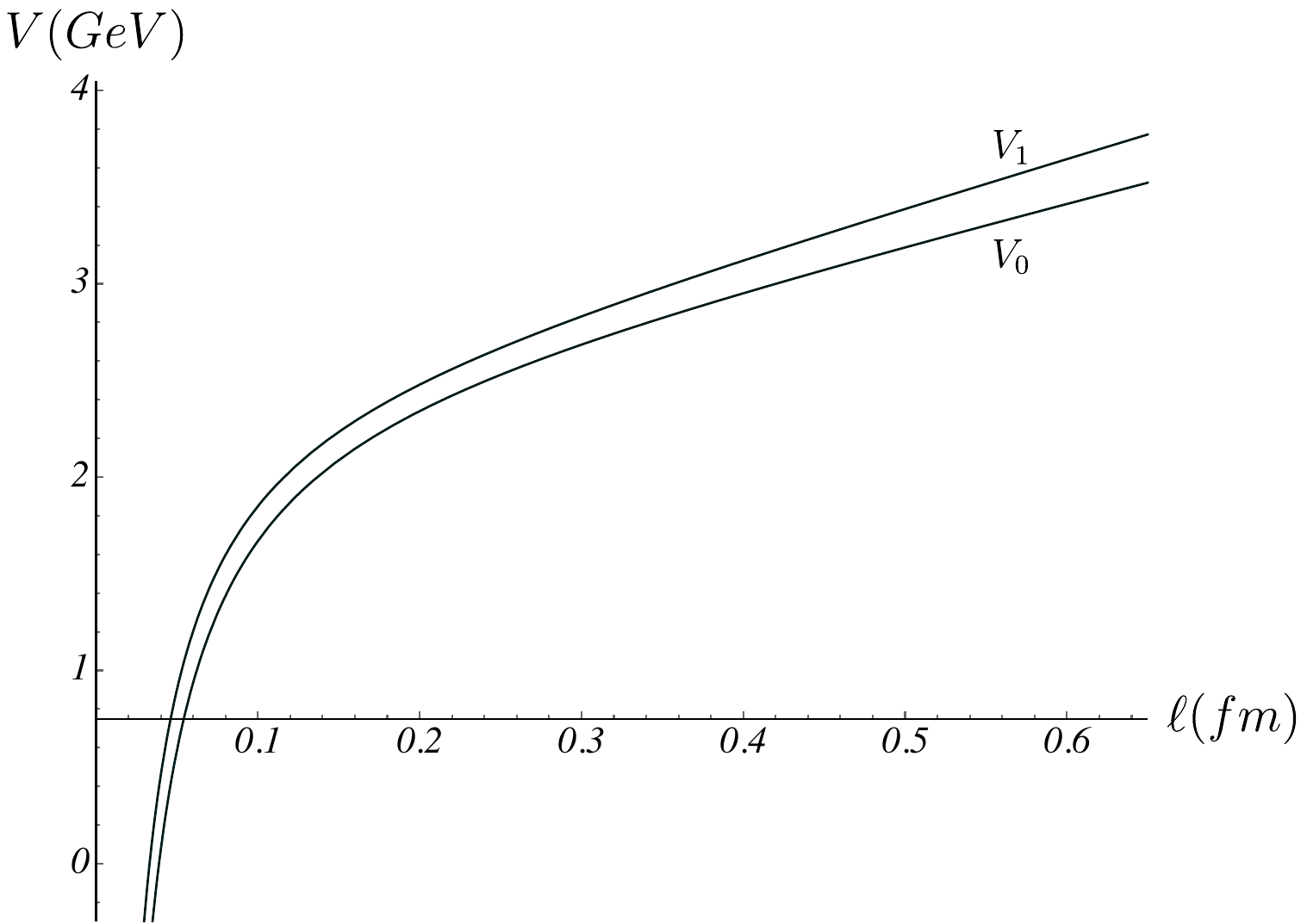}
\caption{{\small Various plots at $\eta=0.85$. Left: The $E$'s vs $\ell$. Right: The two lowest potentials.}}
\label{all085}
\end{figure}

As a typical example, consider $\eta=0.85$. Figure \ref{all085} shows the resulting plots. 

%________________________  fig - 16 __________________________________
%___________________________________________________________________

%___________________________________________________________________________________
\subsubsection{$1\leq\eta$}
At $\eta=1$, the rectangle becomes a square, and hence configurations (a) and (b') are indistinguishable. The potentials are then simply given by $V_0=E^{(\text{a})}$ and $V_1=E^{(\text{d})}$. Thus, the ground state remains a hadronic molecule. 

The behavior of the potentials for larger values of $\eta$ can be understood purely on symmetry grounds. The key point is that for type-B ordering there is an additional symmetry which exchanges the length and the width. It acts on $\eta$ and $\ell$ as $\eta\rightarrow \eta^{-1}$ and $\ell\rightarrow \eta^{-1}\ell$, with a fixed point at $\eta=1$. In terms of configurations, the symmetry exchanges the meson configurations $E^{\text{(a)}}$ and $E^{\text{(b')}}$.\footnote{In the literature, such an exchange is referred to as a flip-flop of strings \cite{flip-flop}. It is, in fact, a special case of string reconnection (see Figure \ref{stri}). Although the flip-flop is hidden when the constraint $\ell/w=const$ is imposed, it becomes manifest under the constraint $w=const$ and has been observed in lattice simulations (see, e.g., \cite{Bicu-rev} and reference therein). We will return to this issue in \cite{4Q-2}.}

%________________________  fig - 17 __________________________________
\begin{figure}[htbp]
\centering
\includegraphics[width=8cm]{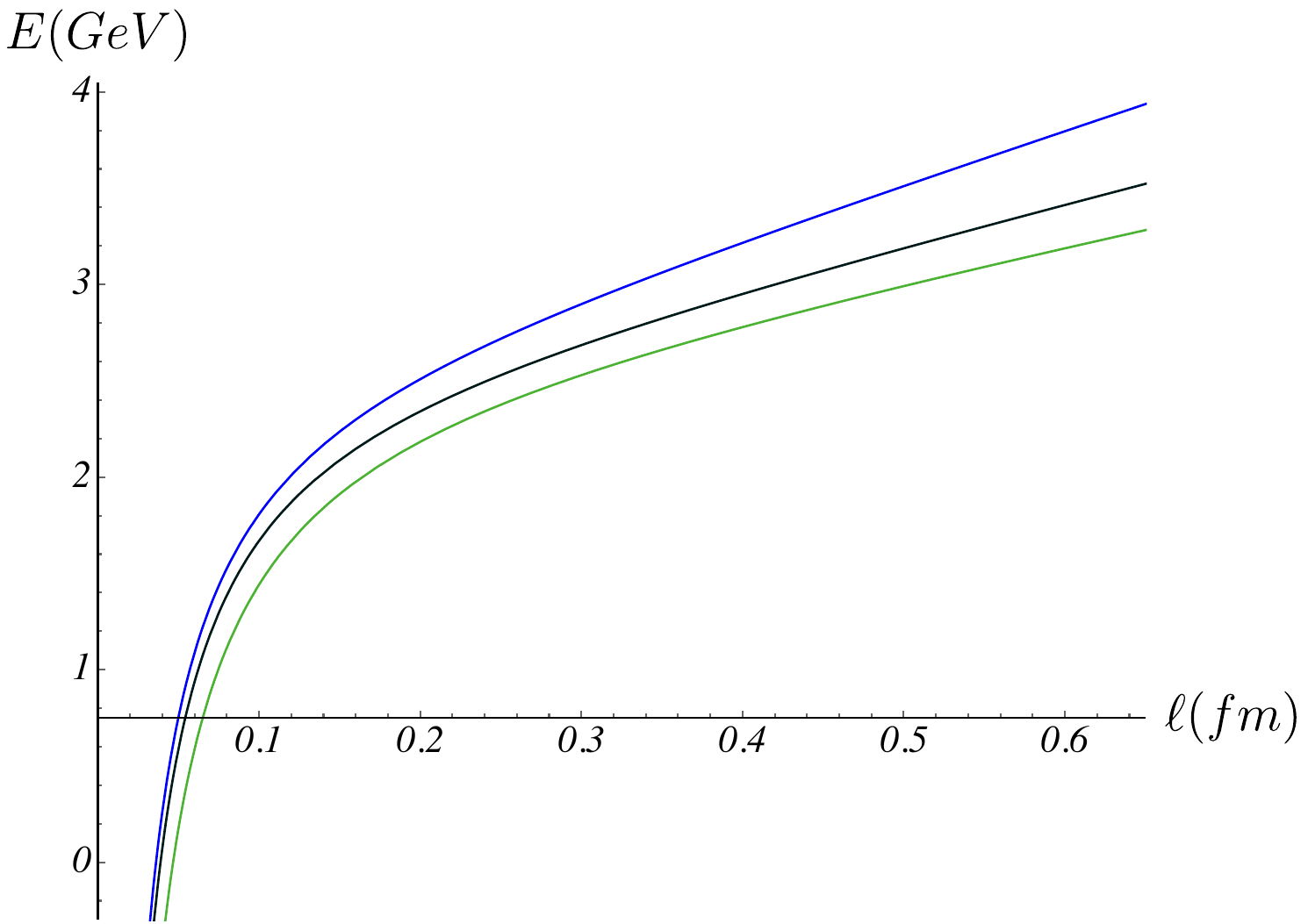}
\hspace{1.5cm}
\includegraphics[width=8cm]{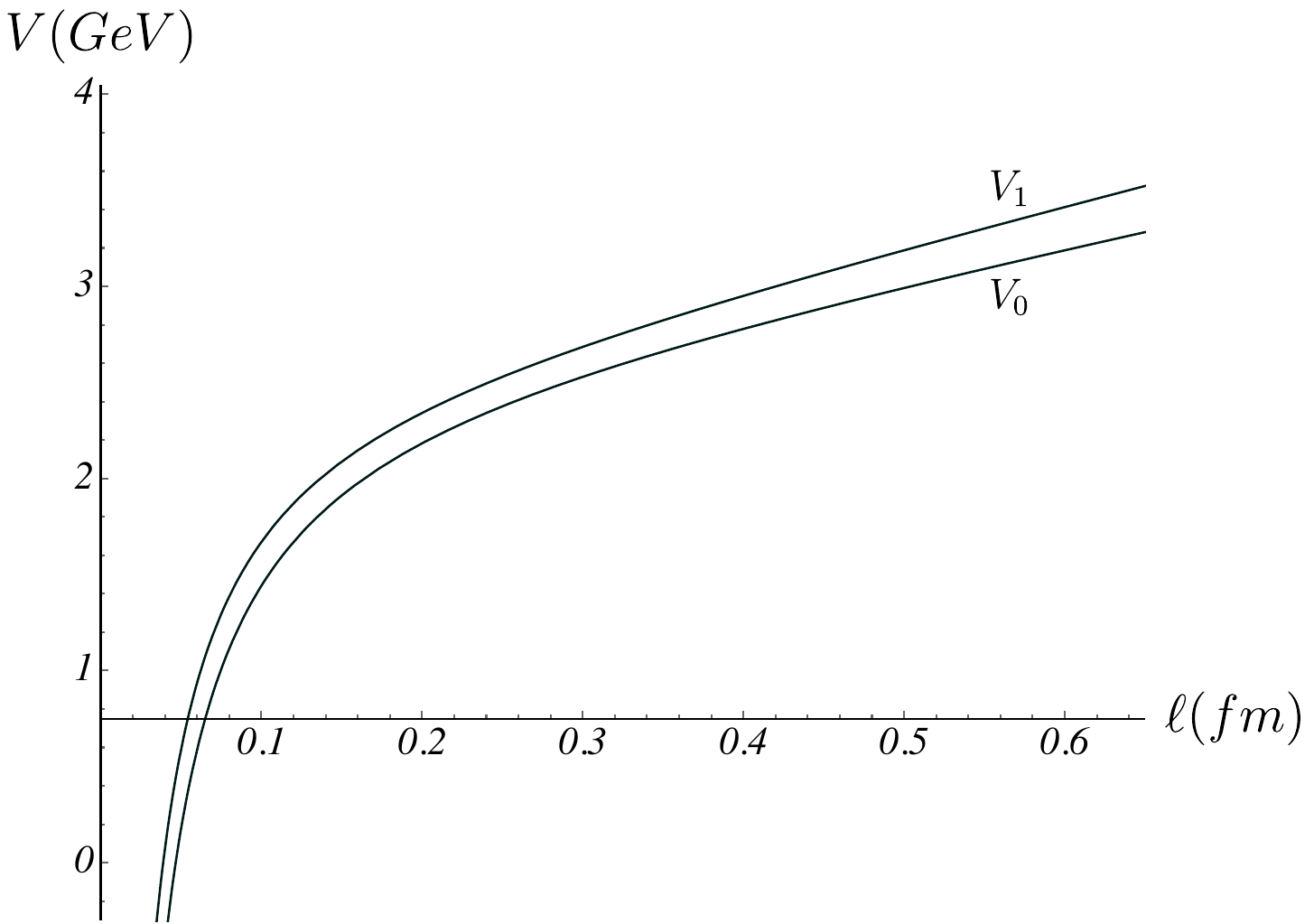}
\caption{{\small Various plots at $\eta=1.2$. Left: The $E$'s vs $\ell$. Right: The two lowest  potentials.}}
\label{all12}
\end{figure}
%___________________________________________________________________

In this description, the range $0.6969651<\eta <1$ is mapped to $1<\eta<1.43479$. The corresponding potentials are simply $V_0=E^{(\text{b'})}$ and $V_1=E^{(\text{a})}$. Therefore, as before, the ground state is expected to be a hadronic molecule. For illustration, the resulting plots for $\eta=1.2$ are shown in Figure \ref{all12}. 

\begin{figure}[htpb]
\centering
\includegraphics[width=8cm]{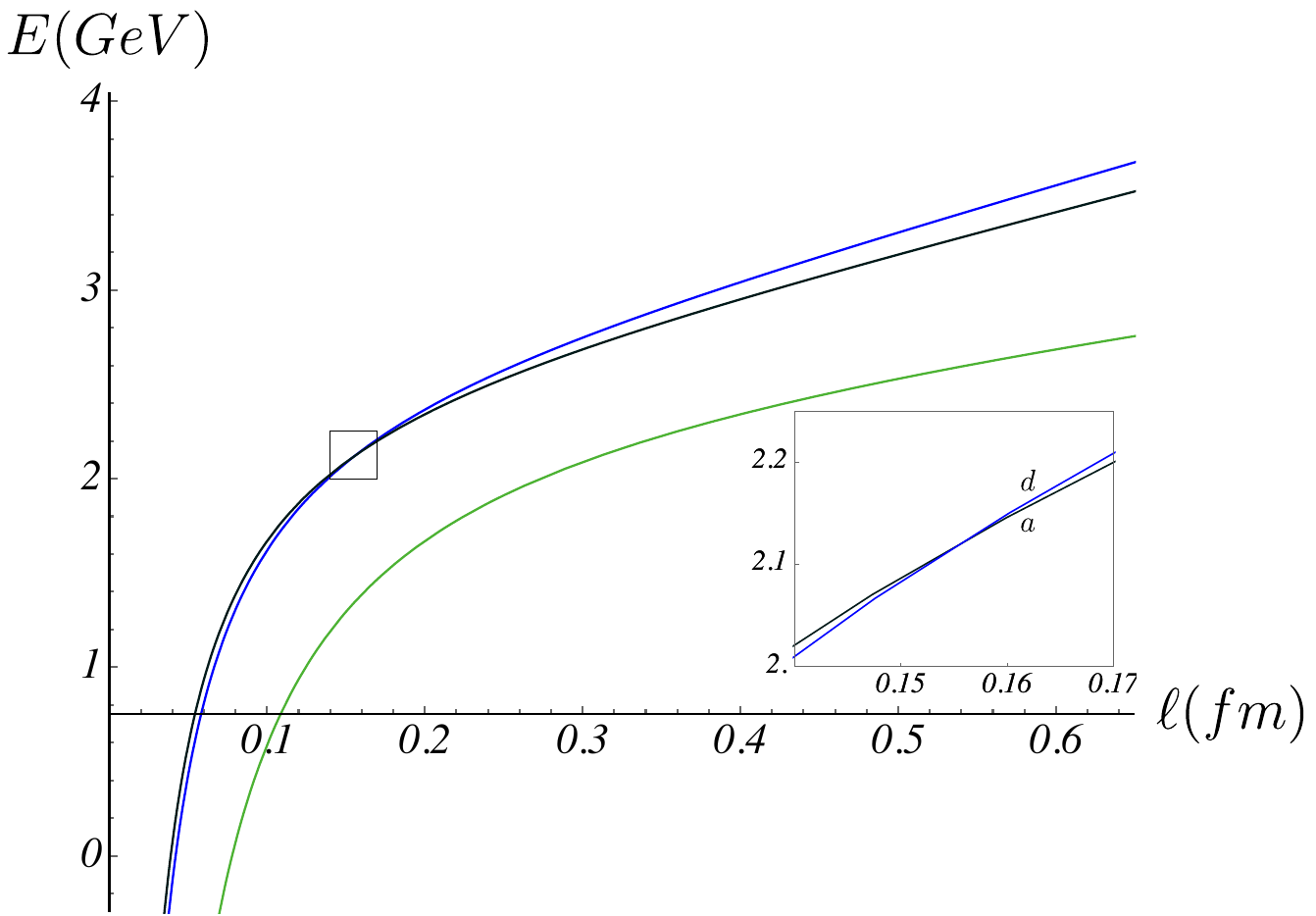}
\hspace{1.5cm}
\includegraphics[width=8cm]{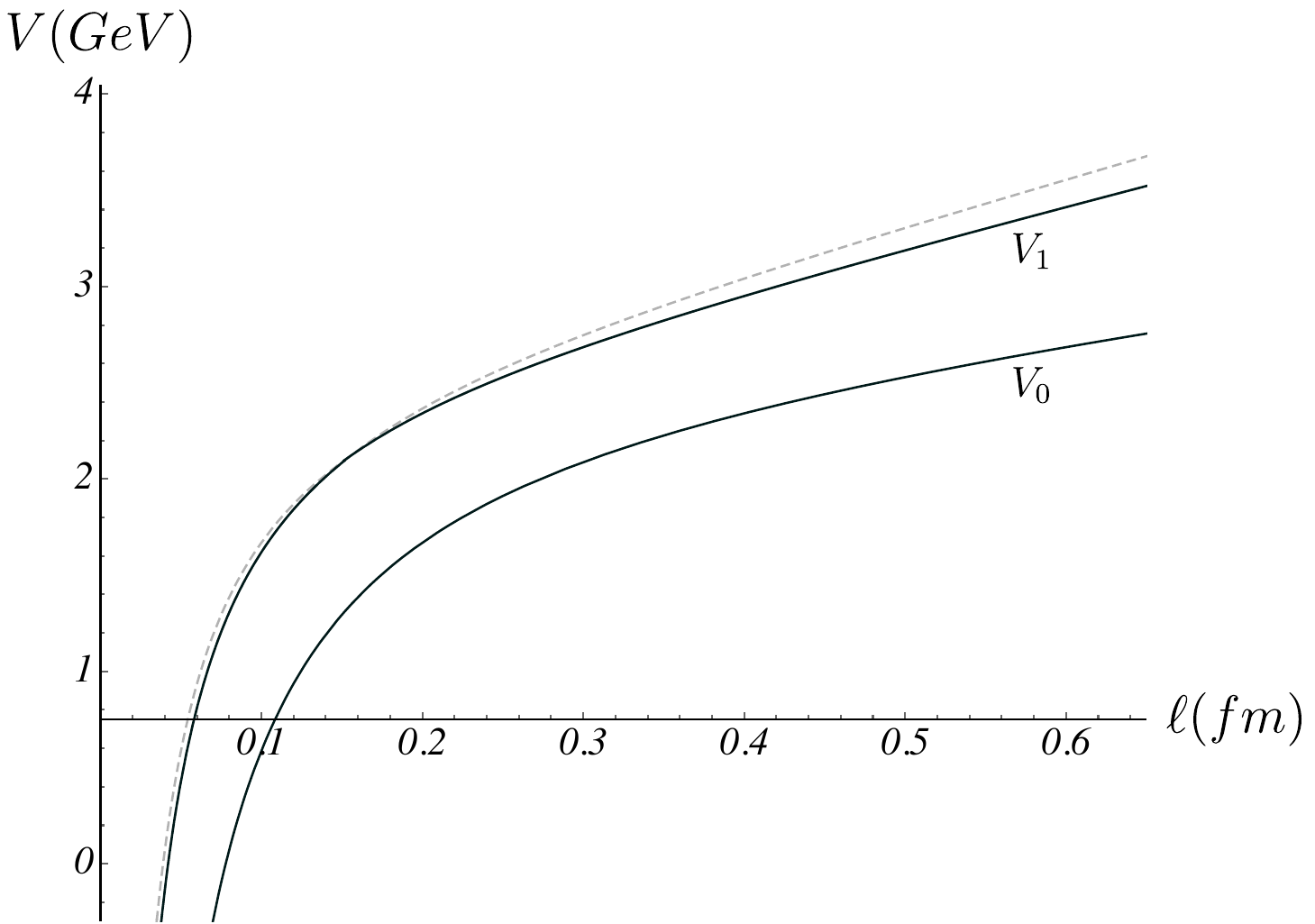}
\caption{{\small Various plots at $\eta=2$. Left: The $E$'s vs $\ell$. Right: The two lowest potentials.}}
\label{all2B}
\end{figure}

Similarly, the range $\eta<0.6969651$ is mapped to the range $1.43479<\eta$. The potentials are then written as $V_0=E^{\text{(b')}}$ and $V_1=\min\{E^{\text{(a)}},E^{\text{(d)}}\}$. For $V_1$, the tetraquark configuration dominates at small $\ell$, as in the range $\eta<0.6969651$. The ground state is again expected to correspond to a hadronic molecule. As an example, we choose $\eta=2$. The corresponding plots are shown in Figure \ref{all2B}. The transition between configurations (a) and (d) occurs at $\ell_c=0.156\,\text{fm}$. 
%________________________  fig - 18 __________________________________
%____________________________________________________________________________________________ 

%____________________________________________________________________________________________
\section{the IR limit of some string configurations and Steiner trees}
\renewcommand{\theequation}{6.\arabic{equation}}
\setcounter{equation}{0}

Consider now the general case in which $N$ quark sources are placed at arbitrary points on the boundary of the five-dimensional space. We impose several restrictions on a string configuration connecting these sources. First, we assume that the configuration is connected. Second, it is tree, meaning that it contains no string loops. We also assume that in the IR limit, when the quark sources are infinitely separated, its projection onto the boundary forms a regular Steiner tree whose Steiner points do not coincide either with each other or with the sources. The physical meaning of the last assumption is straightforward: in such a limit, all strings become infinitely long, so that baryon vertices are infinitely separated from one another and from the quark sources. 

Within the present model, the asymptotic expansion of the energy can be easily obtained. The leading term comes from the strings. Summing over them gives  

\begin{equation}\label{EN0}
E_{\text{\tiny NQ}}=\sigma L_{\text{min}}+O(1)
\,,\quad
\end{equation}
where $L_{\text{min}}$ is the length of the Steiner tree. Importantly, the string tension is universal: it is the same for all the strings, as follows from the asymptotic formulas \eqref{El-large4} and \eqref{Ebulklarge}. This is the well-known asymptotic behavior. Our goal is to determine the leading correction to it.\footnote{We will not discuss the leading term in detail here. There exists a vast literature on the Steiner tree problem. See, e.g., the book \cite{IT} and references therein.} 

Before getting to arbitrary $N$, recall that for $N=3$ the desired correction was computed in \cite{a3Q2016, a3Q2025}. Explicitly,

\begin{equation}\label{C3Q}
C_{\3Q}=3c
-
\g \sqrt{\s}\Bigl(3{\cal I}(\vo)-3\k\frac{\ep^{-2\vo}}{\sqrt{\vo}}\Bigr)
	\,.
	\end{equation}
Here $\vo$ is a solution to Eq.\eqref{v1}, and $c$ is the normalization constant, as before. Two important remarks are in order. First, \eqref{C3Q} is meaningful only if none of the triangle's angles formed by the quarks exceeds $\frac{2}{3}\pi$. In other words, the Steiner tree is regular.\footnote{In this case, the Fermat point of the triangle does not coincide with any of its vertices.} Second, the correction is universal in the sense that it does not depend on the triangle's angles. 

The case $N=4$ may be analyzed in a similar way. The corresponding string configuration resembles those of Figure \ref{tetra5d}, but with the quark sources placed at arbitrary points on the boundary. Consider the force balance equation \eqref{fbeV} at $V$. It is convenient to decompose the vectors $\mathbf{e}$ as $\mathbf{e}=(\vec\ep, \ep_r)$, where $\ep_r$ is the radial component, and the remaining $x$, $y$, and $z$ components are grouped into $\vec\ep$. Then $\vert\vec{\ep}\vert=\g w(r_v)\cos\alpha$ and $\ep_r=-\g w(r_v)\sin\alpha$, as follows from \eqref{e} and \eqref{eb}. The force balance equation can then be written in component form as \footnote{String 3 is internal, as it is stretched between the baryon vertices. Its tangent angle is always positive (see Appendix B).}

\begin{equation}\label{fbeg}
\vec\ep_1+\vec\ep_2+\vec\ep_3=0\,,\qquad
\sin\alpha_1+\sin\alpha_2-\sin\alpha_3=3\k(1+4v)\ep^{-3v}
\,.	
\end{equation}
As explained in Appendices A and B, string 1 becomes infinitely long as its $\lambda$ parameter approaches $1$. In this IR limit, its tangent angle at the baryon vertex is negative and given by $\cos\alpha_1=v\ep^{1-v}$ (see Eq.\eqref{v-lambda}). The same argument applies to the two remaining strings, yielding $\alpha_1=\alpha_2=-\alpha_3$. Consequently, all the vectors $\vec{\ep}$ have the same length. This implies that the angles between them must be $\frac{2}{3}\pi$ to satisfy the first equation in \eqref{fbeg}, and therefore the point $Y$ coincides with the Steiner point $S$. Moreover, from the second equation it follows that $\sin\alpha_1=\k(1+4v)\ep^{-3v}$. Combining this with the above expression for $\cos\alpha_1$ leads directly to Eq.\eqref{v1}, whose solution is $\vo$. Clearly, the same argument can be applied to the vertex $\bar V$ to show that its projection $\bar Y$ coincides with the second Steiner point $\bar S$. The important consistency condition is that both vertices are equidistant from the boundary, namely at $r=\sqrt{\vo/\s}$. The asymptotic expression for the energy of the four external strings (ending on the quark sources) is given by \eqref{El-large3}, while that for the internal string (ending only on the vertices) is given by \eqref{Ebulklarge}. So after summing over the strings and adding the vertex contributions, we arrive at 

\begin{equation}\label{E4}
E_{\text{\tiny 4Q}}=\sigma L_{\text{min}}
+
C_{\text{\tiny 4Q}}
+o(1)
\,,\quad\text{with}\quad 
C_{\text{\tiny 4Q}}=4c
-
\g \sqrt{\s}\Bigl(4{\cal I}(\vo)+{\cal J}(\vo)
	-6\k\frac{\ep^{-2\vo}}{\sqrt{\vo}}\Bigr)
	\,,
\end{equation}
which includes the universal constant term. The expression \eqref{EII-large5} obtained in Sec.III for the rectangular geometry is a special case of \eqref{E4}.

\begin{figure}[htbp]
\centering
\includegraphics[width=3.25cm]{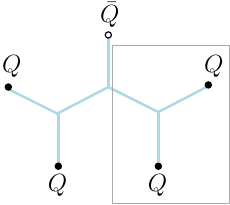}
\caption{{\small A schematic illustration of a regular Steiner tree for five points corresponding to the quark positions. It can be obtained from the $N=4$ tree by cutting one external string together with the quark source at its endpoint and then attaching a giant diquark, the object shown in the box.}}
\label{SQ5}
\end{figure}

Technically, the case $N=5$ differs from the previous one only by the numbers of strings and vertices. Using essentially the same arguments, one can show that in the IR limit the projections of the vertices on the boundary correspond to the Steiner points, and all the vertices lie at the same distance from the boundary. A typical Steiner tree is sketched in Figure \ref{SQ5}. In this case, summing over the strings and the vertices gives 

%________________________  fig - 19 _____________________________________________________
%________________________________________________________________________________________

\begin{equation}\label{E5}
E_{\text{\tiny 5Q}}=\sigma L_{\text{min}}
+
C_{\text{\tiny 5Q}}
+o(1)
\,,\quad\text{with}\quad 
C_{\text{\tiny 5Q}}=5c
-
\g \sqrt{\s}\Bigl(5{\cal I}(\vo)+2{\cal J}(\vo)
	-9\k\frac{\ep^{-2\vo}}{\sqrt{\vo}}\Bigr)
	\,,
\end{equation}
again with a universal constant term. Note that $C_{\text{\tiny 5Q}}$ is obtained from $C_{\text{\tiny 4Q}}$ by adding the contributions of one external string, one internal string, and one vertex, according to the procedure illustrated in the Figure.

We are now in a position to write down the formula for arbitrary $N$: 
 
\begin{equation}\label{EN}
E_{\text{\tiny NQ}}=\sigma L_{\text{min}}
+
C_{\text{\tiny NQ}}
+o(1)
\,,\quad\text{with}\quad 
C_{\text{\tiny NQ}}=Nc
-
\g \sqrt{\s}\Bigl(N{\cal I}(\vo)+(N-3){\cal J}(\vo)
	-3(N-2)\k\frac{\ep^{-2\vo}}{\sqrt{\vo}}\Bigr)
	\,,\quad N\geq 2\,.
\end{equation} 
This is the main result of this Section. Note that the above formula applies to the $Q\bar Q$ system, where $C_{\text{\tiny 2Q}}$ reduces to $C_{\QQb}$ defined in \eqref{EQQb-large}. However, it should be stressed that the result is valid for a specific class of string configurations rather than for the ground state potentials of the multiquark systems. 

We conclude our discussion with a few remarks. 
 
(1) In \cite{a3Q2016} it was noted that the constant terms appearing in the expansions of the energy of the $QQQ$ system at small and large quark separations are different. More precisely, the constant term in the IR limit is smaller than that in the UV. The same feature holds for the string configurations considered above, with $Nc$ being the constant term in the UV. It is noteworthy that the difference between these constants is well-defined (scheme-independent), and can be straightforwardly estimated using \eqref{EN}. For a few small values of $N$, we get $3c-C_{\3Q}=267\,\text{MeV}$, $4c-C_{\text{\tiny 4Q}}=359\,\text{MeV}$, and $5c-C_{\text{\tiny 5Q}}=450\,\text{MeV}$.  

(2) If the normalization constant $c$ is chosen to be sufficiently large so that the first term dominates in $C_{\text{\tiny NQ}}$, then 

\begin{equation}\label{CN-lattice}
\frac{C_{\text{\tiny NQ}}}{N}\simeq const
\,,\quad N\geq 2\,.
\end{equation}
For $N\leq 5$, such a relation has been observed on the lattice \cite{Bicu-rev}. Since the parameter values used here were obtained by fitting the lattice data for the three-quark potential \cite{a3Q2016}, it is worth making some estimates. We obtain $\frac{C_{\QQb}}{2}=0.535\,\text{GeV}$, $\frac{C_{\text{\tiny 3Q}}}{3}=\frac{C_{\text{\tiny 4Q}}}{4}=\frac{C_{\text{\tiny 5Q}}}{5}=0.533\,\text{GeV}$, and $\frac{C_{\text{\tiny NQ}}}{N}=0.531\,\text{GeV}$ as $N\to\infty$.

(3) Another special choice is to set $c=\g\sqrt{\s}I_0$. In this case, the following relations hold \footnote{The relation $C_{\text{\tiny 4Q}}=2C_{\3Q}$ was suggested by Z. Komargodski.}

\begin{equation}\label{CN-unitary}
C_{\QQb}=0\,,\qquad
C_{\3Q}=3\g\sqrt{\s}\Bigl(\k\frac{\ep^{-2\vo}}{\sqrt{\vo}}-\oh{\cal J}(\vo)\Bigr)
\,,\qquad
C_{\text{\tiny NQ}}=(N-2)C_{\3Q}
\,.
\end{equation}
With our parameter values, we find $c=88\,\text{MeV}$ and $C_{\3Q}=-4\,\text{MeV}$. 

(4) There are other IR limits in which some of the strings remain of finite length or even shrink to points. In such limits, the corresponding Steiner points coincide with each other or with given points (quark positions), as discussed in Sec.III and in \cite{a3Q2025}. These limits are more complex and generally do not yield universal constant terms.   

%_____________________________________________________________________________________ 
\section{Conclusions}
\renewcommand{\theequation}{7.\arabic{equation}}
\setcounter{equation}{0}

We conclude our discussion of the fully tetraquark system with several remarks.

(1) The results of Sec.V suggest that, depending on the geometry and quark ordering, the ground state of the $QQ\bar Q\bar Q$ system may be a hadronic molecule, a tetraquark state, or a superposition of the two.

(2) In general, the tetraquark and pinched tetraquark configurations are distinct and may coexist. In terms of string interactions in ten dimensions, a transition between them can be interpreted as the creation or decay of a brane-antibrane bound state. For the type-B ordering, configurations (c') and (d) are indistinguishable in four dimensions but become distinguishable in five dimensions, as explained in Appendix E. This is puzzling and warrants further study. Hopefully, this will shed light on the five-dimensional origin of QCD strings \cite{polyakov}.  

 (3) For the class of the string configurations discussed in Sec.VI, we can deduce from \eqref{EN} that  

\begin{equation}\label{CN-universal}
\frac{C_{\text{\tiny NQ}}}{N}-\frac{C_{\QQb}}{2}
=
3 \frac{N-2}{N}\biggl( \frac{C_{\3Q}}{3}-\frac{C_{\QQb}}{2}\biggr)
\,,\qquad N\geq 3\,.
\end{equation}
This is a general relation which is well-defined (i.e., scheme independent). It is obviously satisfied if $\frac{C_{\text{\tiny NQ}}}{N}=const$, and \eqref{CN-unitary} is its special case. It will be interesting to test this prediction through high precision computer simulations.  

 (4) In general, for the rectangular geometry the potentials are complicated functions of both length and width. In this work we imposed the constraint $\ell=\eta w$ to simplify the analysis. Another option, used in lattice QCD, is to fix $w=const$. This is convenient for studying the diquark limit ($\eta\to\infty$) but is less suitable for exploring the UV and IR limits. We plan to return to this issue in future work  \cite{4Q-2}.

(5) In the light of the recent discovery of the $T_{c\bar cc \bar c}(6600)$ and $T_{c\bar cc\bar c}(6900)$ mesons at the LHC, understanding the properties of the tetraquark systems has become a matter of primary importance. The $QQ\bar Q\bar Q$ system exhibits a rich and complex landscape of physics, with many open questions yet to be answered. Advancing our theoretical understanding and connecting it to hadron spectroscopy will require a concerted effort from the high-energy physics community. We hope this study provides useful insights and motivation for future research.

%______________________________________________________________________
\begin{acknowledgments}
We would like to thank Zohar Komargodski for discussions concerning this topic. This work was conducted as part of Program FFWR-2024-0011 at the Landau Institute.
 \end{acknowledgments}

%_____________________________________________________________________
 \appendix
 \section{Notation and useful formulas}
%\label{notation}
\renewcommand{\theequation}{A.\arabic{equation}}
\setcounter{equation}{0}

Throughout this paper, we denote heavy quarks (antiquarks) by $Q(\bar Q)$ and baryon (antibaryon) vertices by $V(\bar V)$. Heavy quark sources are always located on the boundary of the five-dimensional space at $r=0$, while vertices are located in its interior at $r=\rv(\rvb)$. For convenience, we introduce the dimensionless variables $v=\s\rv^2$ and $\bar v=\s\rvb^2$, which take values in the interval $[0,1]$ and quantify the proximity of these objects to the soft wall, located at $1$ in such units.

To present the resulting formulas in a compact form, we make use of the basic functions introduced in \cite{astbr3Q}, together with several newly defined ones:

(i) The function ${\cal L}^+$ is defined as  

\begin{equation}\label{fL+}
{\cal L}^+(\alpha,x)=\cos\alpha\sqrt{x}\int^1_0 du\, u^2\, \ep^{x (1-u^2)}
\Bigl[1-\cos^2{}\hspace{-1mm}\alpha\, u^4\ep^{2x(1-u^2)}\Bigr]^{-\frac{1}{2}}
\,,
\qquad
0\leq\alpha\leq\frac{\pi}{2}\,,
\qquad 
0\leq x\leq 1
\,.
\end{equation}
This non-negative function vanishes at $\alpha=\frac{\pi}{2}$ or $x=0$. For small $x$, if $\alpha$ tends to $\alpha_0$, it behaves as 

\begin{equation}\label{fL+smallx}
{\cal L}^+(\alpha,x)=\sqrt{x}\bigl({\cal L}^+_0(\alpha_0)+O(x)\bigr)
\,,
\qquad\text{with}\qquad
{\cal L}^+_0(\alpha_0)=\frac{1}{4}
\cos^{-\oh}\hspace{-.9mm}\alpha_0\,B\bigl(\cos^2\hspace{-.9mm}\alpha_0;\tfrac{3}{4},\tfrac{1}{2}\bigr)
\,.
\end{equation}
Here $B(z;a,b)$ denotes the incomplete beta function. At $(0,1)$, ${\cal L}^+$ develops a logarithmic singularity

\begin{equation}\label{fL+x1}
{\cal L}^+(\alpha,x)=-\oh\ln(1-x)+O(1)
\,,\quad
\text{as} \quad \alpha\rightarrow 0\,.
\end{equation}

(ii) The ${\cal L}^-$ function is given by 

\begin{equation}\label{fL-}
{\cal L}^-(y,x)=\sqrt{y}
\biggl(\,
\int^1_0 du\, u^2\, \ep^{y(1-u^2)}
\Bigl[1-u^4\,\ep^{2y(1-u^2)}\Bigr]^{-\frac{1}{2}}
+
\int^1_
{\sqrt{\frac{x}{y}}} 
du\, u^2\, \ep^{y(1-u^2)}
\Bigl[1-u^4\,\ep^{2y(1-u^2)}\Bigr]^{-\frac{1}{2}}
\,\biggr)
\,,
\quad
0\leq x\leq y\leq 1
\,.
\end{equation}
This function is non-negative and vanishes at the origin. For $y=x/\rho$, with non-zero $\rho$ as $x\rightarrow 0$, its small-$\ell$ behavior is

\begin{equation}\label{L-y=0}
	{\cal L}^-(y,x)=\sqrt{x}\bigl({\cal L}^-_0(\rho)+O(x)\bigr)
	\,,\qquad\text{with}\qquad
	{\cal L}^-_0(\rho)=\frac{1}{4}\rho^{-\oh}{\cal B}\bigl(1-\rho^2;\tfrac{1}{2},\tfrac{3}{4}\bigr) 
\,.
\end{equation}
Here ${\cal B}(z;a,b)=B(a,b)+B(z;a,b)$. At $y=1$ it exhibits a logarithmic singularity

\begin{equation}\label{L-y=1}
{\cal L}^-(y,x)=-\ln(1-y)+O(1)\,,
\quad
\text{at fixed}\,\,\,x
\,.
\end{equation}
The ${\cal L}^{\pm}$ functions are related by ${\cal L}^+(0,x)={\cal L}^-(x,x)$. 

(iii) The ${\cal L}$ function is defined as

\begin{equation}\label{iL}
{\cal L}(y,x)=2\sqrt{y}
\int^1_
{\sqrt{\frac{x}{y}}} 
du\, u^2\, \ep^{y(1-u^2)}
\Bigl[1-u^4\,\ep^{2y(1-u^2)}\Bigr]^{-\frac{1}{2}}
\,,
\qquad 
0<x\leq y
\,.
\end{equation}
It is nonnegative and vanishes at $y=0$ and $x=y$. For $y=x/\rho$, with nonzero $\rho$ as $x\rightarrow 0$, it behaves as

\begin{equation}\label{iLy=0}
{\cal L}(y,x)
=
\sqrt{x}
\bigl({\cal L}_0(\rho)+O(1)\bigr)\,,
\qquad\text{with}\qquad
	{\cal L}_0(\rho)=\frac{1}{2}\rho^{-\oh}B\bigl(1-\rho^2;\tfrac{1}{2},\tfrac{3}{4}\bigr)
\,.	
\end{equation}
It also develops a logarithmic singularity at $y=1$

\begin{equation}\label{iLy=1}
{\cal L}(y,x)=-\ln(1-y)+O(1)\,,
\quad
\text{at fixed}\,\,\, x
\,.
\end{equation}
The ${\cal L}$ functions satisfy

\begin{equation}\label{LL-}
	{\cal L}(y,x)={\cal L}^-(y,x)-{\cal L}^+(\alpha,x)
	\,,
	\end{equation}
valid if $\frac{\ep^y}{y}=\cos\alpha\frac{\ep^{x}}{x}$.

(iv) The function ${\cal E}^+$ is given by  

\begin{equation}\label{fE+}
{\cal E}^+(\alpha,x)=
\frac{1}{\sqrt{x}}
\int^1_0\,\frac{du}{u^2}\,\biggl(\ep^{x u^2}
\Bigl[
1-\cos^2{}\hspace{-1mm}\alpha\,u^4\ep^{2x (1-u^2)}
\Bigr]^{-\frac{1}{2}}-1-u^2\biggr)
\,,
\qquad
0\leq\alpha\leq\frac{\pi}{2}\,,
\qquad 
0\leq x\leq 1
\,.
\end{equation}
It is singular at $x=0$ and at $(0,1)$. If $\alpha\to\alpha_0$ as $x\to 0$, then 

\begin{equation}\label{fE+smallx}
{\cal E}^+(\alpha,x)=\frac{1}{\sqrt{x}}\bigl({\cal E}^+_0(\alpha_0)+O(x)\bigr)
\,,
\qquad\text{with}\qquad
{\cal E}_0^+(\alpha_0)=
\frac{1}{4}
\cos^{\oh}\hspace{-.9mm}\alpha_0\,B\bigl(\cos^2\hspace{-.9mm}\alpha_0;-\tfrac{1}{4},\tfrac{1}{2}\bigr)
\,.
\end{equation}
Near $x=1$, $\alpha\to 0$, it behaves as 

\begin{equation}\label{E-x=1}
	{\cal E}^+(\alpha,x)=-\oh\ep\ln(1-x)+O(1)
	\,.
\end{equation}

(v) The ${\cal E}^-$ function is defined as 

\begin{equation}\label{fE-}
{\cal E}^-(y,x)=\frac{1}{\sqrt{y}}
\biggl(
\int^1_0\,\frac{du}{u^2}\,
\Bigl(\ep^{y u^2}\Bigl[1-u^4\,\ep^{2y(1-u^2)}\Bigr]^{-\frac{1}{2}}
-1-u^2\Bigr)
+
\int^1_{\sqrt{\frac{x}{y}}}\,\frac{du}{u^2}\,\ep^{y u^2}
\Bigl[1-u^4\,\ep^{2y(1-u^2)}\Bigr]^{-\frac{1}{2}}
\biggr) 
\,,
\,\,\,
0\leq x\leq y\leq 1
\,.
\end{equation}
This function is singular at $(0,0)$ and at $y=1$. If $y=x/\rho$ with nonzero $\rho$ as $x\to 0$, then

\begin{equation}\label{E-y=0}
	{\cal E}^-(y,x)=\frac{1}{\sqrt{x}}\Bigl({\cal E}^-_0(\rho)+O(x)\Bigr)
	\,,\qquad\text{with}\qquad
	{\cal E}^-_0(\rho)=\frac{1}{4}\rho^{\oh}{\cal B}\bigl(1-\rho^2;\tfrac{1}{2},-\tfrac{1}{4}\bigr)
\,.
\end{equation}
Near $y=1$ with fixed $x$, it is singular

\begin{equation}\label{E-y=1}
	{\cal E}^-(y,x)=-\ep\ln(1-y)+O(1)
	\,.
\end{equation}
Note that the relation ${\cal E}^+(0,x)={\cal E}^-(x,x)$ holds.

(vi) The ${\cal E}$ function is defined by 

\begin{equation}\label{ifE-}
{\cal E}(y,x)=
\frac{2}{\sqrt{y}}
\int^1_{\sqrt{\frac{x}{y}}}\,\frac{du}{u^2}\,
\ep^{y u^2}
\Bigl[1-u^4\ep^{2y(1-u^2)}
\Bigr]^{-\frac{1}{2}}
\,,
\qquad
0<x\leq y
\,.
\end{equation}
It is singular at the origin. For $y=x/\rho$ with nonzero $\rho$ as $x\to 0$, it behaves as

\begin{equation}\label{iEx=0}
	{\cal E}(y,x)=\frac{1}{\sqrt{x}}\Bigl({\cal E}_0(\rho)+o(x)\Bigr)
	\,,\qquad\text{with}\qquad
	{\cal E}_0(\rho)=\frac{1}{2}\rho^{\oh}B\bigl(1-\rho^2;\tfrac{1}{2},-\tfrac{1}{4}\bigr)\,.
\end{equation}
It also has a logarithmic singularity at $y=1$

\begin{equation}\label{Ey=1}
	{\cal E}(y,x)=-\ep\ln(1-y)+O(1)
	\,,\quad\text{with}\,\,\,\text{fixed}\,\,\,x\,.
\end{equation}
For the ${\cal E}$ functions, the analog of \eqref{LL-} is 

\begin{equation}\label{EE-}
	{\cal E}(y,x)
	={\cal E}^-(y,x)
	-
	{\cal E}^+(\alpha,x)\,,
	\end{equation}
which is also valid if $\frac{\ep^y}{y}=\cos\alpha\frac{\ep^{x}}{x}$.

(vii) The ${\cal Q}$ function is given by 

\begin{equation}\label{Q}
{\cal Q}(x)=\sqrt{\pi}\text{erfi}(\sqrt{x})-\frac{\ep^x}{\sqrt{x}}
\,,
\qquad
0<x\leq 1
\,,
\end{equation}
where $\text{erfi}(x)$ stands for the imaginary error function. This is a special case of ${\cal E}^+$ with $\alpha=\tfrac{\pi}{2}$. For small $x$,

 \begin{equation}\label{Q0}
{\cal Q}(x)=-\frac{1}{\sqrt{x}}+\sqrt{x}+O(x^{\frac{3}{2}})
\,.
\end{equation}

(viii) The ${\cal I}$ function is given by 

\begin{equation}\label{I}
	{\cal I}(x)=
	I_0
	-
	\int_{\sqrt{x}}^1\frac{du}{u^2}\ep^{u^2}\Bigl[1-u^4\ep^{2(1-u^2)}\Bigr]^{\frac{1}{2}}
	\,,
\quad\text{with}\quad 
I_0=\int_0^1\frac{du}{u^2}\Bigl(1+u^2-\ep^{u^2}\Bigl[1-u^4\ep^{2(1-u^2)}\Bigr]^{\frac{1}{2}}\Bigr)
\,,
\quad
0< x\leq 1
\,.
\end{equation}
In particular, ${\cal I}(1)=I_0$, and numerically $I_0=0.751$. This function is related to the ${\cal L}^-$ and $ {\cal E}^-$ functions as 

\begin{equation}\label{ILE}
	{\cal I}(x)=\ep\,{\cal L}^-(y,x)-{\cal E}^-(y,x)
	\quad\text{as}\,\,\, y\rightarrow 1\,\,\text{at}
	\,\,\,\text{fixed}\,\,\,x\,.
	\end{equation}

(ix) The ${\cal J}$ function is defined by 

\begin{equation}\label{iI}
	{\cal J}(x)
	=2\bigl({\cal I}(x)-I_0\bigr)
	=
	-2
	\int_{\sqrt{x}}^1\frac{du}{u^2}\ep^{u^2}\Bigl[1-u^4\ep^{2(1-u^2)}\Bigr]^{\frac{1}{2}}
	\,,
\quad
0< x\leq 1
\end{equation}
Analogous to ${\cal I}$, it  satisfies the relation 

\begin{equation}\label{JLE}
	{\cal J}(x)=\ep\,{\cal L}(y,x)-{\cal E}(y,x)
	\quad\text{as}\,\,\, y\rightarrow 1\,\,\text{at}
	\,\,\,\text{fixed}\,\,\,x\,.
	\end{equation}

%_______________________________________________________________________________
\section{Static Nambu-Goto strings with fixed endpoints}
%\label{notation}
\renewcommand{\theequation}{B.\arabic{equation}}
\setcounter{equation}{0}

The aim here is to describe some facts about static Nambu-Goto strings in the curved geometry \eqref{metric}, which are essential for constructing the string configurations discussed in Secs.III and IV. Some of this material is not new and can be found in \cite{a3Q2016}, whose notation we largely adopt.

%___________________________________________________________________
\renewcommand \thesubsubsection {\arabic{subsubsection}}
\subsubsection{A static string with one endpoint on the boundary}

\begin{figure}[H]
\centering
\includegraphics[width=5.25cm]{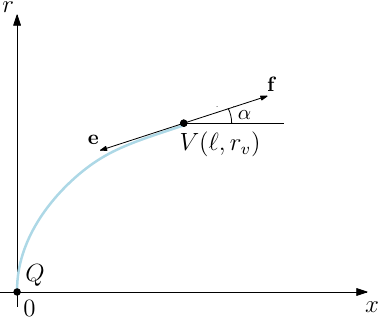}
\hspace{3cm}
\includegraphics[width=5.25cm]{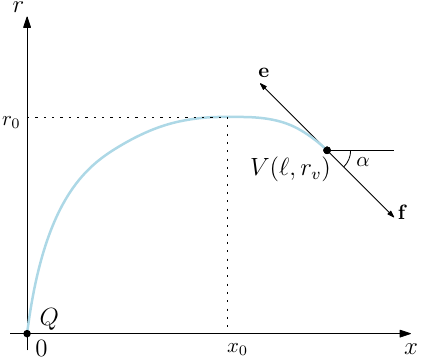}
\caption{{\small A string stretched between two points, $Q(0,0)$ and $V(\ell,\rv)$, where $\rv<1/\sqrt{\s}$. The tangent angle at $V$ is denoted by $\alpha$, and the arrows indicate the forces acting on this point. The left and right panels show the cases for $\alpha\geq 0$ and $\alpha\leq 0$, respectively.}}
\label{ngs}
\end{figure}

Consider a static string stretched between two fixed points in the $xr$-plane, as shown in Figure \ref{ngs}. 
%________________________  f - 13 __________________________________
%_________________________________________________________________

In static gauge, where $\xi^1=t$ and $\xi^2=x$, the boundary conditions for $r(x)$ describing the string profile are 

\begin{equation}\label{string-bc}
r(0)=0
\,,\qquad
r(\ell)=\rv
\,.
\end{equation}
The Nambu-Goto action \eqref{ng} takes the form

\begin{equation}\label{NG2}
S=T\g\int_{0}^{\ell} dx\,w(r)\sqrt{1+(\partial_x r)^2}
\,,
\qquad
w(r)=\frac{\ep^{\s r^2}}{r^2}
\,.
\end{equation}
For convenience, we use the shorthand notation $\g=\frac{R^2}{2\pi\alpha'}$, $T=\int dt$, and $\partial_x r=\frac{\partial r}{\partial x}$. Since the integrand does not depend explicitly on $x$, the equation of motion admits the first integral

\begin{equation}\label{Int}
I=\frac{w(r)}{\sqrt{1+(\partial_x r)^2}}\,.
\end{equation}
At point $V$, it can be written as

\begin{equation}\label{I-PB}
I=w(\rv)\cos\alpha
\,,
\end{equation}
where $\tan\alpha=\partial_xr\vert_{x=\ell}$ and $\alpha\in[-\frac{\pi}{2},\frac{\pi}{2}]$.

In this paper, point $Q$ is associated with an infinitely heavy quark, while point $V$ with a baryon vertex. It is therefore natural to consider the forces acting on $V$ in order to maintain equilibrium. The force balance equation is simply 

\begin{equation}\label{fbeNG}
\mathbf{e}+\mathbf{f}=0
\,,
\end{equation}
where $\mathbf{e}$ is the string tension and $\mathbf{f}$ is an external force (see the Figure). It is straightforward to compute the $r$-component of $\mathbf{e}$, which arises from the boundary term in the variation of the action. Indeed, with $\delta r\vert_{x=\ell}=\delta\rv$, we find that 

\begin{equation}\label{NG3}
\delta S=T\g \frac{w(r)\partial_x r}{\sqrt{1+(\partial_x r)^2}}\delta\rv
\,
\end{equation}
which implies $\mathbf{e}_r=-T^{-1}\delta S/\delta\rv=-\g w(\rv)\sin\alpha$. Similarly, in static gauge $\xi^1=t$ and $\xi^2=r$, the boundary term provides the $x$-component: $\mathbf{e}_x=-\g w(\rv)\cos\alpha$.\footnote{Alternatively, one can choose a gauge in which the boundary terms provide both components \cite{a3Q2016}.} Putting both components together, we have 

\begin{equation}\label{e}
\mathbf{e}=-\g w(\rv)(\cos\alpha,\sin\alpha)
\,.
\end{equation}
 A simple but important observation is that the magnitude of the tension is determined by a radial coordinate, specifically $\lvert\mathbf{e}\rvert=\g w(r)$. 
 
 In general, the tangent angle $\alpha$ may be positive or negative. For $\alpha\geq 0$, the function $r(x)$ describing a string profile  monotonically increases on the interval $[0,\ell]$. Conversely, for $\alpha<0$, the situation is more intricate: $r(x)$ increases on $[0,x_{\text{\tiny 0}}]$ and decreases on $[x_{\text{\tiny 0}},\ell]$, reaching a local maximum at $x=\xo$. Both cases are depicted in Figure \ref{ngs}. 

 Let us examine these cases more systematically, starting with $\alpha\geq 0$. First, we express $I$ in terms of $\alpha$ and $\rv$. This yields the differential equation $w(\rv)\cos\alpha=w(r)/\sqrt{1+(\partial_x r)^2}$, which can be integrated over $x$ and $r$. Using the boundary conditions \eqref{string-bc}, we obtain 

 \begin{equation}\label{l+}
\ell
=
\cos\alpha\sqrt{\frac{v}{\s}}\int^1_0 du\, u^2\, \ep^{v (1-u^2)}
\Bigl(1-\cos^2{}\hspace{-1mm}\alpha\, u^4\ep^{2v(1-u^2)}\Bigr)^{-\frac{1}{2}}
=
\frac{1}{\sqrt{\s}}{\cal L}^+(\alpha,v)
\,,
\end{equation}
where $v=\s\rv^2$, and the function ${\cal L}^+$ is defined in Appendix A.

To compute the string energy, we reduce the integral over $x$ in $S$ to an integral over $r$ using the differential equation. Since the resulting expression diverges at $r=0$, we regularize it by imposing a short-distance cutoff $\epsilon$. This gives

\begin{equation}\label{E+reg}
E_{R}=\frac{S_R}{T}=
\g\sqrt{\frac{\s}{v}}\int^1_{\sqrt{\tfrac{\s}{v}}\epsilon}\,\frac{du}{u^2}\,\ep^{v u^2}
\Bigl[1-\cos^2{}\hspace{-1mm}\alpha\,u^4\,\ep^{2v(1-u^2)}\Bigr]^{-\frac{1}{2}}
\,.
\end{equation}
$E_R$ behaves for $\epsilon\rightarrow 0$ as  

\begin{equation}\label{E+R}
E_R=\frac{\g}{\epsilon}+E+O(\epsilon)\,.
\end{equation}
Subtracting the $\tfrac{1}{\epsilon}$ term and taking $\epsilon=0$, we obtain a finite expression for the energy

\begin{equation}\label{E+}
E
=
\g\sqrt{\frac{\s}{v}}
\int^1_0\,\frac{du}{u^2}\,\biggl(\ep^{v u^2}
\Bigl(1-\cos^2{}\hspace{-1mm}\alpha\,u^4\,\ep^{2v(1-u^2)}\Bigr)^{-\frac{1}{2}}-1-u^2\biggr)
+c
=
\g\sqrt{\s}\,{\cal E}^+(v,\alpha)+c
\,.
\end{equation}
Here ${\cal E}^+$ is defined in \eqref{fE+}, and $c$ is a normalization constant. 

It is worth noting that for $\alpha=\tfrac{\pi}{2}$, the above expressions simplify to 

\begin{equation}\label{Evert}
\ell=0\,,
\qquad
E=\g\sqrt{\s}{\cal Q}(v)+c\,,
\end{equation}
where ${\cal Q}$ is defined in Appendix A. In this case, the string is stretched entirely along the radial direction. If both endpoints lie in the bulk, the second expression becomes

\begin{equation}\label{Evertb}
E=\g\sqrt{\s}\bigl({\cal Q}(v)-{\cal Q}(\bar v)\bigr)
\,.
\end{equation}
Here $\bar v=\s\rvb^2$ such that $\rv>\rvb$.

The above analysis extends straightforwardly to the case $\alpha\leq 0$. A key point, relevant for all formulas below, is that the string configuration involves two segments: one over the interval $[0,\xo]$, where $r(x)$ increases, and another over the interval $[\xo,\ell]$, where $r(x)$ decreases (see Figure \ref{ngs}). First, we define the first integral at $r=\ro$ so that $I=w(\ro)$, and then integrate the differential equation over both intervals. This yields  

\begin{equation}\label{l-}
\ell
=
\sqrt{\frac{\lambda}{\s}}\biggl[
\int^1_0 du\, u^2\, \ep^{\lambda(1-u^2)}
\Bigl(1-u^4\,\ep^{2\lambda(1-u^2)}\Bigr)^{-\frac{1}{2}}
+
\int^1_
{\sqrt{\frac{v}{\lambda}}} 
du\, u^2\, \ep^{\lambda(1-u^2)}
\Bigl(1-u^4\,\ep^{2\lambda(1-u^2)}\Bigr)^{-\frac{1}{2}}
\biggr]
=
\frac{1}{\sqrt{\s}}{\cal L}^-(\lambda,v)
\,.
\end{equation}
Here $\lambda=\s\ro^2$, and ${\cal L}^-$ is defined in \eqref{fL-}. Importantly, $\lambda$, $v$, and $\alpha$ are not independent. From the first integral it follows that 

\begin{equation}\label{v-lambda}
\frac{\ep^{\lambda}}{\lambda}=\frac{\ep^{v}}{v}\cos\alpha
\,
\end{equation}
which allows us to express $\lambda$ in terms of $v$ and $\alpha$ as

\begin{equation}\label{lambda}
\lambda=-\text{ProductLog}(-v\,\ep^{-v}/\cos\alpha)
\,.
\end{equation}
Here $\text{ProductLog}(z)$ is the principal solution for $w$ in the equation $z=w\,\ep^w$ \cite{wolf}.

As before, the string energy is computed by first rewriting the integral over $x$ in $S$ 
as an integral over $r$ and imposing the short-distance cutoff on $r$. A direct calculation gives

\begin{equation}\label{ER}
E_{R}=\g\sqrt{\frac{\s}{\lambda}}
\biggl[
\int^1_{\sqrt{\tfrac{\s}{\lambda}}\epsilon}\,\frac{du}{u^2}\,\ep^{\lambda u^2}
\Bigl[1-u^4\,\ep^{2\lambda (1-u^2)}\Bigr]^{-\frac{1}{2}}
+
\int^1_{\sqrt{\frac{v}{\lambda}}}\,\frac{du}{u^2}\,\ep^{\lambda u^2}
\Bigl(1-u^4\,\ep^{2\lambda (1-u^2)}\Bigr)^{-\oh}
\biggr] 
\,.
\end{equation}
To obtain a finite result, we subtract the $\frac{1}{\epsilon}$ term and take $\epsilon\rightarrow 0$, yielding

\begin{equation}\label{E-}
E
= 
\g\sqrt{\frac{\s}{\lambda}}
\biggl[
\int^1_0\,\frac{du}{u^2}\,
\Bigl(\ep^{\lambda u^2}\Bigl(1-u^4\,\ep^{2\lambda (1-u^2)}\Bigr)^{-\frac{1}{2}}
-1-u^2\Bigr)
+
\int^1_{\sqrt{\frac{v}{\lambda}}}\,\frac{du}{u^2}\,\ep^{\lambda u^2}
\Bigl(1-u^4\,\ep^{2\lambda (1-u^2)}\Bigr)^{-\frac{1}{2}}\biggr] 
+c
=
\g\sqrt{\s}\,{\cal E}^-(\lambda,v)
+c
\,,
\end{equation}
where $c$ is the same normalization constant introduced earlier. The function ${\cal E}^-$ is defined in Appendix A.

Finally, let us discuss the large-$\ell$ behavior. In doing so, it is convenient to start with $\alpha\leq 0$. In this case the large-$\ell$ limit corresponds to $\lambda\to 1$. From this, it follows that $v$ and $\alpha$ are not independent but subject to the constraint $v\ep^{1-v}=\cos\alpha$. Explicitly, for a given $\alpha$, the corresponding $v$ is $\text{v}_\alpha=-\text{ProductLog}(-\cos\alpha/\ep)$. The leading contributions to $\ell$ and $E$ arise from the logarithmic terms. From Eqs.\eqref{L-y=1} and \eqref{E-y=1}, we immediately find

\begin{equation}\label{Eng-large}
\ell=-\frac{1}{\sqrt{\s}}\ln(1-\lambda)+O(1)\,,
\qquad
E=-\g\ep\sqrt{\s}\,\ln(1-\lambda)+O(1)
\,.
\end{equation} 
This leads to the standard result

\begin{equation}\label{Eng-large2}
	E=\sigma\ell +O(1)
	\,,\qquad\text{with}\qquad
	\sigma=\g\ep\s
	\,.
\end{equation}
Here $\sigma$ is the string tension. To extract the subleading term, consider the difference

\begin{equation}
	E-\sigma\ell=\g\sqrt{\s}\bigl({\cal E}^-(\lambda,v)-\ep {\cal L}^-(\lambda,v)\bigr)+c
	\,.
\end{equation}
Taking $\lambda\rightarrow 1$ ($v\to\text{v}_\alpha$) and using \eqref{ILE}, we obtain

\begin{equation}\label{El-large3}
	E=\sigma\ell
	+
	c
	-\g\sqrt{\s}\,{\cal I}(\text{v}_\alpha)
	+o(1)
	\,.
\end{equation}

For $\alpha\geq 0$, the only way to reach infinite string length is to take $v\rightarrow 1$ at $\alpha=0$. This implies $\text{v}_\alpha=1$, so \eqref{El-large3} reduces to 

\begin{equation}\label{El-large4}
	E=\sigma\ell
	+c
	-\g\sqrt{\s}\,I_0
	+o(1)
	\,.
\end{equation}
Here we used the fact that ${\cal I}(1)=I_0$.

%______________________________________________________________________________
\subsubsection{A static string with endpoints in the bulk}

\begin{figure}[H]
\centering
\includegraphics[width=5.75cm]{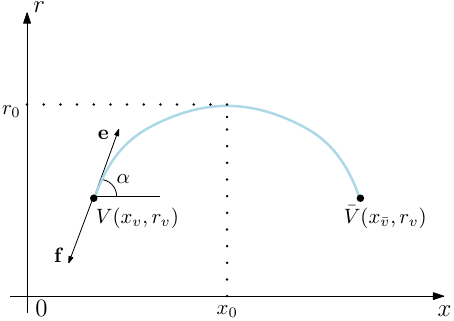}
\caption{{\small A string stretched between two symmetric points in the bulk. Here $\rv<\frac{1}{\sqrt{\s}}$, $\alpha$ denotes the tangent angle, and the arrows indicate forces acting at $V$.}}
\label{ngs2}
\end{figure}

Now consider a static string stretched between two points in the bulk at the same radial coordinate, as shown in Figure \ref{ngs2}. In this case the string profile is symmetric about $x=\ell/2$. Clearly, external forces are required to keep the  
%________________________  f - 21  __________________________________________________
%_______________________________________________________________________________
\noindent string in equilibrium. On symmetry grounds, we indicate only the forces acting at $V$. The force balance equation takes the same form as Eq.\eqref{fbeNG}, with the string tension given by

\begin{equation}\label{eb}
\mathbf{e}=\g w(\rv)(\cos\alpha,\sin\alpha)
\end{equation}
This expression for $\mathbf{e}$ follows from Eq.\eqref{e}, with $\mathbf{e}$ replaced by $-\mathbf{e}$.

The length of the string can be obtained in the standard way, by integrating the differential equation with respect to $x$ and $r$, subject to the boundary conditions $r(x_v)=r(x_{\bar v})=r_v$. Alternatively, it can be derived by considering two strings with upper endpoints at $V$ and $\bar V$, and lower endpoints on the boundary. Subtracting the length of the first string from that of the second yields 

\begin{equation}\label{l--}
\ell=x_{\bar v}-x_v
=
\frac{1}{\sqrt{\s}}\Bigl({\cal L}^-(\lambda,v)-{\cal L}^+(\alpha,v)\Bigr)
=
2\sqrt{\frac{\lambda}{\s}}
\int^1_{\sqrt{\frac{v}{\lambda}}}
 du\, u^2\, \ep^{\lambda(1-u^2)}
\Bigl(1-u^4\,\ep^{2\lambda(1-u^2)}\Bigr)^{-\frac{1}{2}}
=
\frac{1}{\sqrt{\s}}{\cal L}(\lambda,v)
\,.
\end{equation}
The energy can be computed in a similar way, giving

\begin{equation}\label{E--}
E
=
\g\sqrt{\s}\bigl({\cal E}^-(\lambda,v)-{\cal E}^+(\alpha,v)\bigr)
=
2\g\sqrt{\frac{\s}{\lambda}}
\int^1_{\sqrt{\frac{v}{\lambda}}}\,\frac{du}{u^2}\,
\ep^{\lambda u^2}\Bigl(1-u^4\,\ep^{2\lambda (1-u^2)}\Bigr)^{-\frac{1}{2}}
=
\g\sqrt{\s}\,{\cal E}(\lambda,v)
\,,
\end{equation}
where the incomplete ${\cal L}$ and ${\cal E}$ functions are defined in Appendix A. 
In our derivation, we used the relation $\cos\alpha=\frac{v}{\lambda}\ep^{\lambda-v}$. Note that no regularization is needed in \eqref{E--}, as the integral is finite.

Finally, let us discuss the large-$\ell$ limit, which corresponds to $\lambda\to 1$. As before, we consider two strings but now with upper endpoints at $\bar V$ and at the turning point $(x_0,r_0)$. From \eqref{El-large3} and \eqref{El-large4}, it follows immediately that

\begin{equation}\label{Ebulklarge}
	E=\sigma\ell
-
\g\sqrt{\s}\,{\cal J}(\text{v}_\alpha)
	+o(1)
	\,,
\end{equation}
where in the last step we used \eqref{iI}. Note that $\sigma$ is the same as in \eqref{El-large4}. This is an important feature of the soft wall model. Its physical meaning is clear: the physical string tension is determined by the expression \eqref{e} or \eqref{eb} evaluated precisely at the soft wall's position, $r=1/\sqrt{\s}$. The wall effectively prevents strings from penetrating deeper into the bulk.

%___________________________________________________________________________________
\section{Some details on the $Q\bar Q$ system}
%\label{notation}
\renewcommand{\theequation}{C.\arabic{equation}}
\setcounter{equation}{0}

This Appendix provides a brief summary of key results concerning the heavy quark-antiquark potential, which represents the ground state energy of a static quark-antiquark pair. These results are relevant to the discussions in Secs.III and IV. For standard explanations, see \cite{az1}, whose conventions we follow unless otherwise stated. 

\begin{figure}[htb]
\centering
\includegraphics[width=12cm]{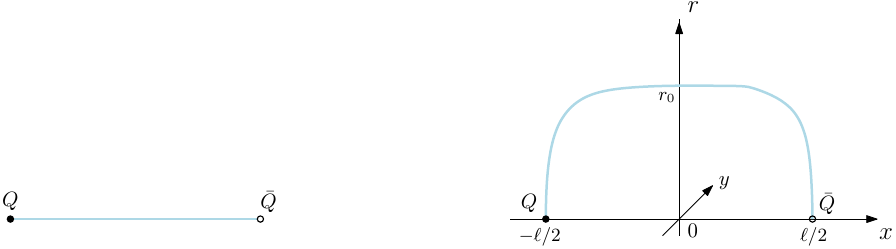}
\caption{{\small A connected string configuration in four dimensions (left) and in five dimensions (right). $\ell$ denotes the distance between the heavy quark sources, and $\ro$ is the radial coordinate of the turning point at $x=0$.}}
\label{conQQb}
\end{figure}

 In four dimensions, a static string configuration takes the form shown in Figure \ref{conQQb} on the left, which is the standard configuration 
%________________________  fig - 22 _____________
%_______________________________________________
for mesons \cite{XA}. In five dimensions, this configuration transforms into the one shown on the right. It involves a string attached to the heavy quark sources on the boundary of five-dimensional space, but in this case a gravitational force pulls the string into the interior \cite{malda}. 

For a Nambu-Goto string in the background geometry \eqref{metric}, the relation between the string energy and the quark separation distance is given parametrically as

\begin{equation}\label{EQQb}
\ell= \frac{2}{\sqrt{\s}}{\cal L}^+(0,\lambda)
 \,,
\quad
E_{\QQb}=2\g\sqrt{\s}\,{\cal E}^+(0,\lambda)+2c
\,.
\end{equation}
Here $c$ is the normalization constant, as before, and $\lambda=\s\ro^2$ is a parameter ranging from $0$ to $1$. The functions ${\cal L}^+$ and ${\cal E}^+$ are defined in Appendix A. 

The small-$\ell$ behavior of $E_{\QQb}$ is  

\begin{equation}\label{EQQb-small}
E_{\QQb}(\ell)=-\frac{\alpha_{\QQb}}{\ell}+2c+\boldsymbol{\sigma}_{\QQb}\ell +o(\ell)
\,,
\quad\text{with}\quad
\alpha_{\QQb}=(2\pi)^3\Gamma^{-4}\bigl(\tfrac{1}{4}\bigr)\g
	\,,\quad
	\boldsymbol{\sigma}_{\QQb}=\oh(2\pi)^{-2}\Gamma^{4}\bigl(\tfrac{1}{4}\bigr)\g\s
	\,,
\end{equation}
while the large-$\ell$ behavior is 

\begin{equation}\label{EQQb-large}
E_{\QQb}(\ell)=\sigma\ell+C_{\QQb}+o(1)
\,,\qquad
\text{with}
\qquad
C_{\QQb}=2c-2\g\sqrt{\s}I_0
\,.
\end{equation}
Here $\sigma$ is the string tension, and $I_0$ is defined in Appendix A. It is worth noting that the coefficients $\boldsymbol{\sigma}_{\QQb}$ and $\sigma$ are different, with their ratio given numerically by $\boldsymbol{\sigma}_{\QQb}/\sigma=0.805$. Furthermore, the difference between the constant terms in the small- and large-$\ell$ expansions is $2c-C_{\QQb}=175\,\text{MeV}$, showing that the constant term in the small-$\ell$ expansion is larger.

%_________________________________________________________________
\section{A detailed description of configuration (c)}
%\label{notation}
\renewcommand{\theequation}{D.\arabic{equation}}
\setcounter{equation}{0}

For a given $\eta$, configuration (c) can be constructed from the basic configurations of Sec.III. However, the construction becomes intricate when it involves multiple transitions between these basic configurations. In this Appendix, we describe configuration (c) and some of its properties, under the assumption that $\eta$ is not very close to $\frac{1}{\sqrt{3}}$.

%_________________________________________________________________
\subsection{The limiting values of $v$}

As explained in Sec.III, the limiting values of the parameter $v$, denoted as $\vo$, $\vt$, and $\vp$, are the solutions to Eqs.\eqref{v1}, \eqref{v1eta}, and \eqref{v1p}. Here we examine how these solutions behave as functions of $\eta$.

\begin{figure}[htbp]
\centering
\includegraphics[width=6.25cm]{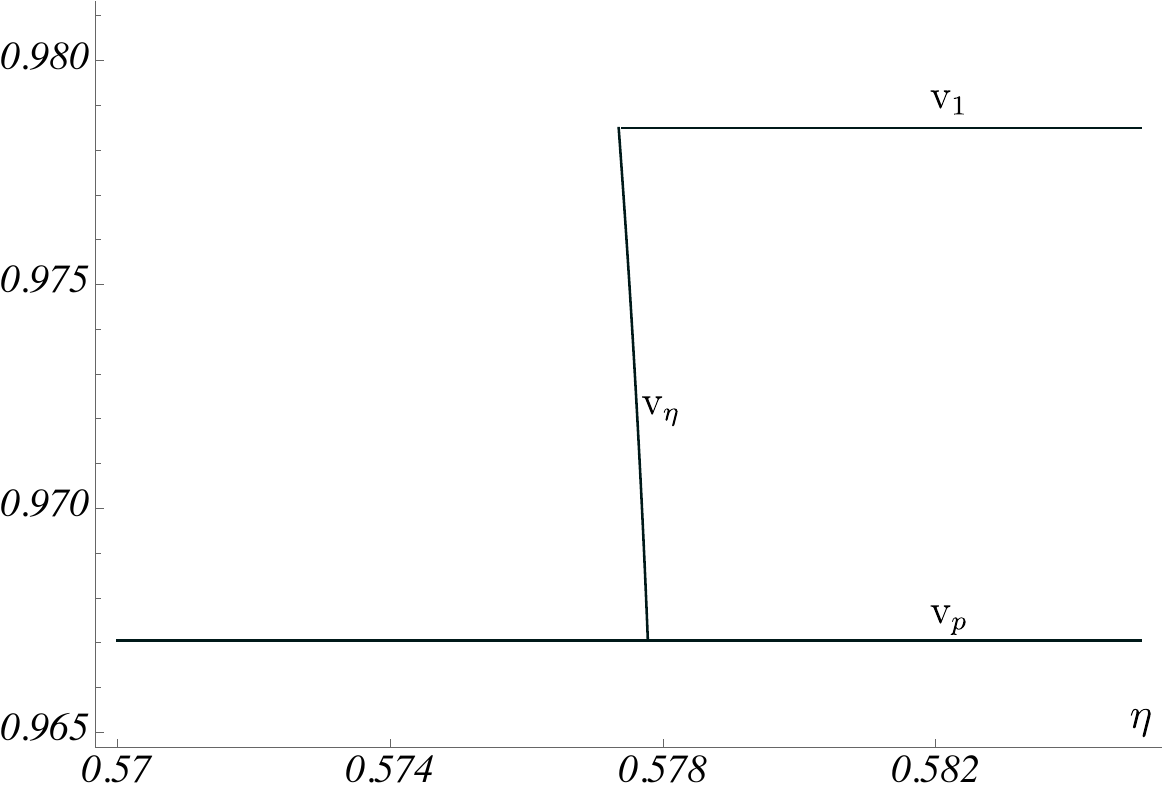}
\caption{{\small The solutions $\vo$, $\vt$, and $\vp$ as functions of $\eta$ in the vicinity of $\eta=\frac{1}{\sqrt{3}}$. Although it is true that the solution $\vo$ is independent of $\eta$, we plot it only for $\eta>\frac{1}{\sqrt{3}}$, where configuration (c) exists.}}
\label{w1}
\end{figure}

A quick way to proceed is by numerical analysis. The results are shown in Figure \ref{w1}. The solutions $\vo$ and 
%________________________  fig - 23 __________________________________
%______________________________________________________________________
\noindent $\vp$ exist for all $\eta$ and take values in the interval $[0,1]$. A simple calculation yields  $\vo=0.978$ and $\vp=0.967$. In contrast, the solution $\vt$ exists only in the very narrow interval $\frac{1}{\sqrt{3}}\leq \eta\leq \eta_{\text{\tiny p}}$.\footnote{Numerically, $\eta_{\text{\tiny  p}}-\frac{1}{\sqrt{3}}=0.0004$.} These bounds can be better understood by considering configuration (c). The lower bound corresponds to $\bar\lambda=1$, where string 3 becomes infinitely long and the second large $\ell$ limit reduces to the first one. The upper bound corresponds to $\bar\alpha=0$, where string 3 collapses to a point, transforming configuration (c) into configuration (d). It can be expressed in terms of $\vp$ as 

\begin{equation}\label{etap}
	\eta_{\text{\tiny p}}=\frac{1}{\sqrt{4\vp^2\,\ep^{2(1-\vp)}-1}}
	\,.
\end{equation}

%_________________________________________________________________
\subsection{Some preliminary comments}

As we saw in Sec.III, the large-$\ell$ limit of configuration (c) does not exist if $\eta\leq 1/\sqrt{3}$. The same is also true for the small-$\ell$ limit. A numerical analysis shows that the geometrical constraints \eqref{gcI-small} and \eqref{gcl=02}, together with the force balance equations \eqref{fbe-small}, have no solutions unless $\bar\alpha=0$, which corresponds to configuration (d). Because of these issues, no physically meaningful tetraquark configuration exists for $\eta\leq\frac{1}{\sqrt{3}}$. 

A useful tool for studying transitions between the basic configurations of Sec.III is the function 

\begin{equation}\label{fun}
f_{\eta}(v)=1-2(\eta\cos\beta-\sin\beta)\frac{{\cal L}^+(0,v)}{{\cal L}(\bar\lambda,v)}
\,,
\end{equation}
whose zeros are the solutions of the equation $\alpha(v)=0$ if $\bar\lambda(v)$ and $\beta(v)$ are defined by  

\begin{equation}\label{f-fbe}
\sin\bar\alpha=-3\k(1+4v)\ep^{-3v}
\,,\qquad
\sin\beta=\oh\cos\bar\alpha	
	\,.
\end{equation}
Clearly, at zero values, Eq.\eqref{fun} reduces to the geometric constraint \eqref{gcp}. These zeros play an important role in analyzing the tetraquark configuration, as they correspond to transitions between configurations (I) and (II).

The interval $\frac{1}{\sqrt{3}}\leq \eta\leq \eta_{\text{\tiny p}}$ is somewhat puzzling. It appears that within this range, there exists two string configurations whose large-$\ell$ expansions correspond to the different limits of configuration (II). In particular, the configuration with the linear asymptotics \eqref{EII-large5} dominates at very large $\ell$ (greater than $2000\,\text{fm}$), whereas the configuration with the linear asymptotics \eqref{EII-etalarge5} dominates at smaller $\ell$. This suggests that configuration (c) is effectively a superposition of the two. We will not pursue this issue further, as the interval is very narrow and irrelevant for determining the ground state potential of the $QQ\bar Q\bar Q$ system. It matters only for the excited states.  
%_________________________________________________________________
\subsection{The range $\eta_{\text{\tiny p}}<\eta\leq 0.7884$}

\begin{figure}[htbp]
\centering
\includegraphics[width=6.15cm]{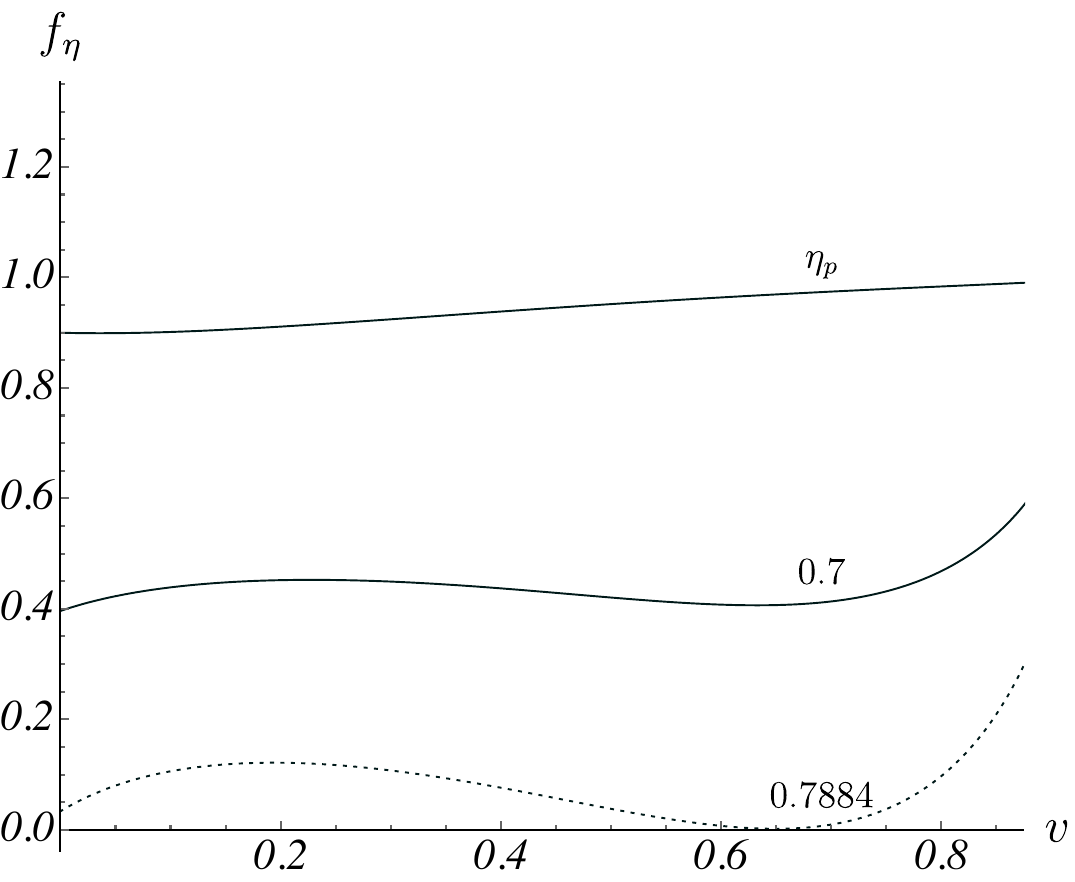}
\hspace{2.2cm}
\includegraphics[width=7.15cm]{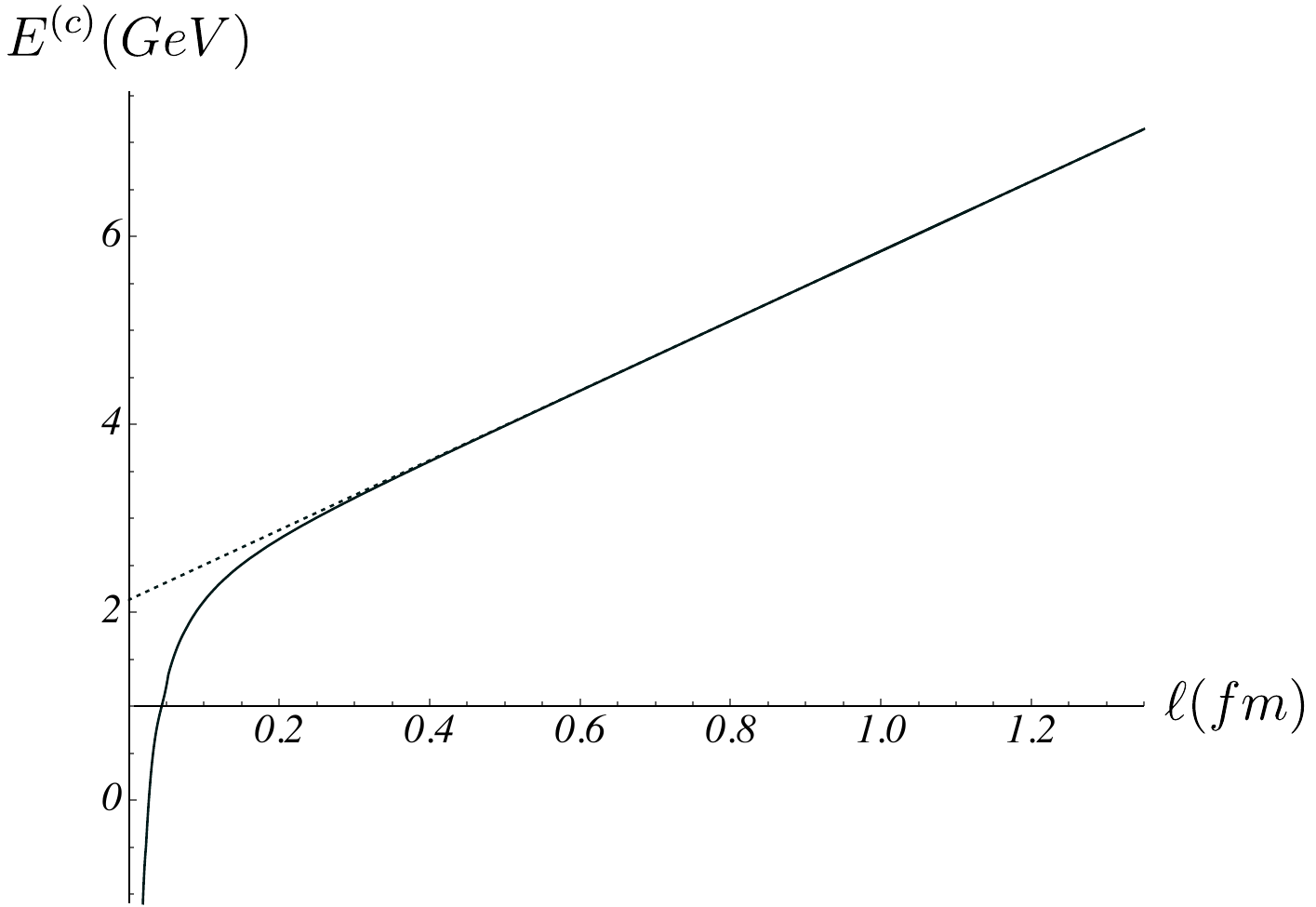}
\caption{{\small Left: The function $f_\eta$ for several values of $\eta$. Right: $E^{\text{(c)}}$ vs $\ell$ at $\eta=0.7$. The dotted line corresponds to the large-$\ell$ asymptotics \eqref{EII-large5}, here and below.}}
\label{eta1}
\end{figure}

We are interested in the zeros of $f_\eta(v)$ in the interval $0<v<\vo$. Numerical analysis shows that this function has no zeros, except for a single one at $\eta=0.7884$. For illustration, we plot $f_\eta$ as a function of $v$ in Figure \ref{eta1} on the left. However, this zero has no effect on the form of the tetraquark configuration, which is described entirely by  
%________________________  fig - 24 __________________________________
%______________________________________________________________________
configuration (II). Its energy is given parametrically by 

\begin{equation}\label{EcI}
	\ell=\ell^{\,\ii}(v)
	\,,
\quad
	E^{\text{(c)}}=E^{\,\ii}(v)
	\,,
	\quad
	0\leq v\leq\vo
	\,.
\end{equation}

We conclude with a simple example: the $\eta=0.7$ case. Figure \ref{eta1} presents both $f_\eta(v)$ and $E^{\text{(c)}}(\ell)$. A notable feature of $E^{\text{ (c)}}$ is its near linear behavior at relatively small quark separations, which the Figure shows occurs for $\ell\gtrsim 0.3\,\text{fm}$. This is significantly smaller than $\ell\gtrsim 0.6\,\text{fm}$ observed for configuration (a), where $E^{\text{(a)}}$ becomes nearly linear (see Figure \ref{Eal}).
%_________________________________________________________________
\subsection{The range $0.7884<\eta\leq\eta_{0}$}

\begin{figure}[H]
\centering
\includegraphics[width=6.75cm]{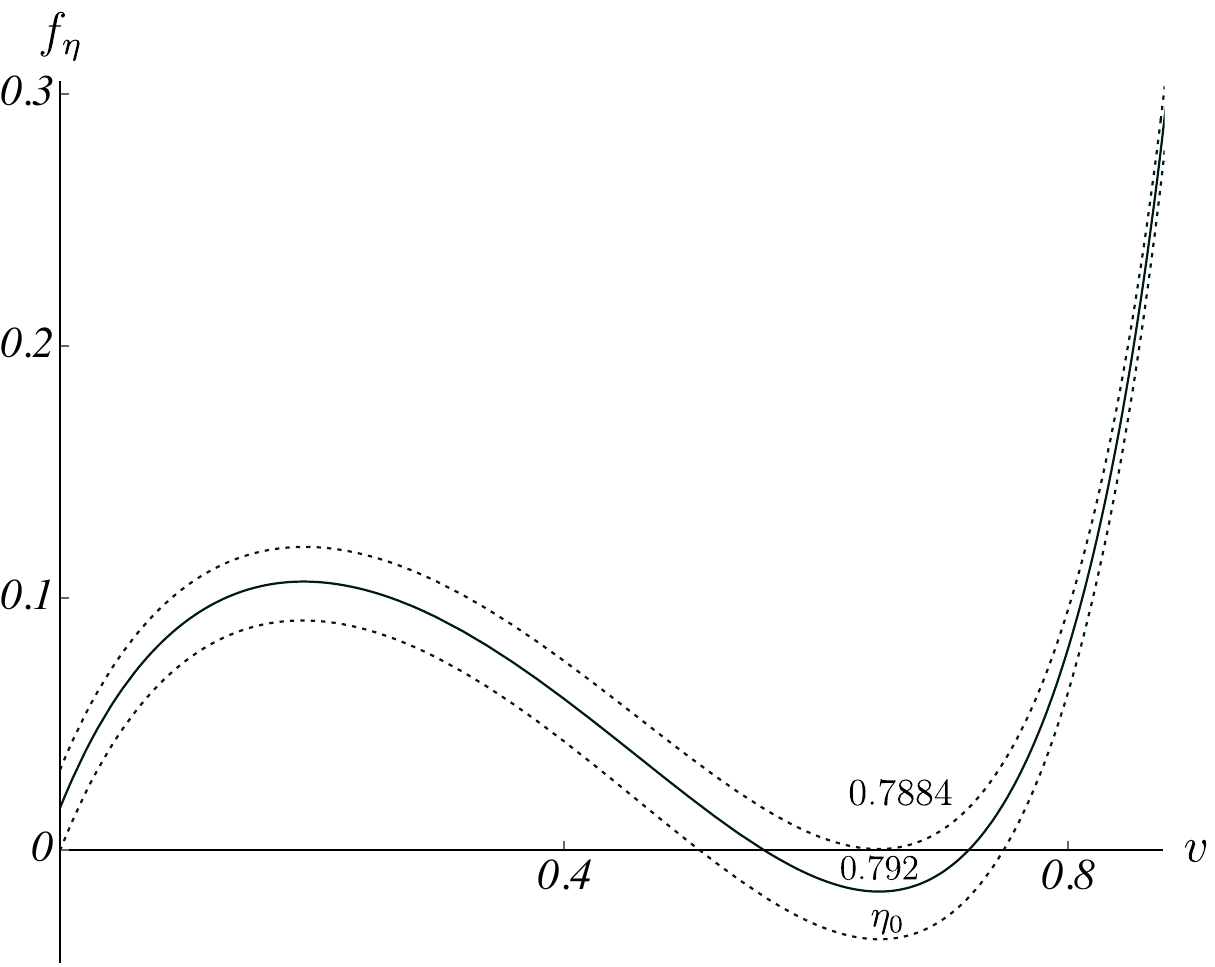}
\hspace{2cm}
\includegraphics[width=7.5cm]{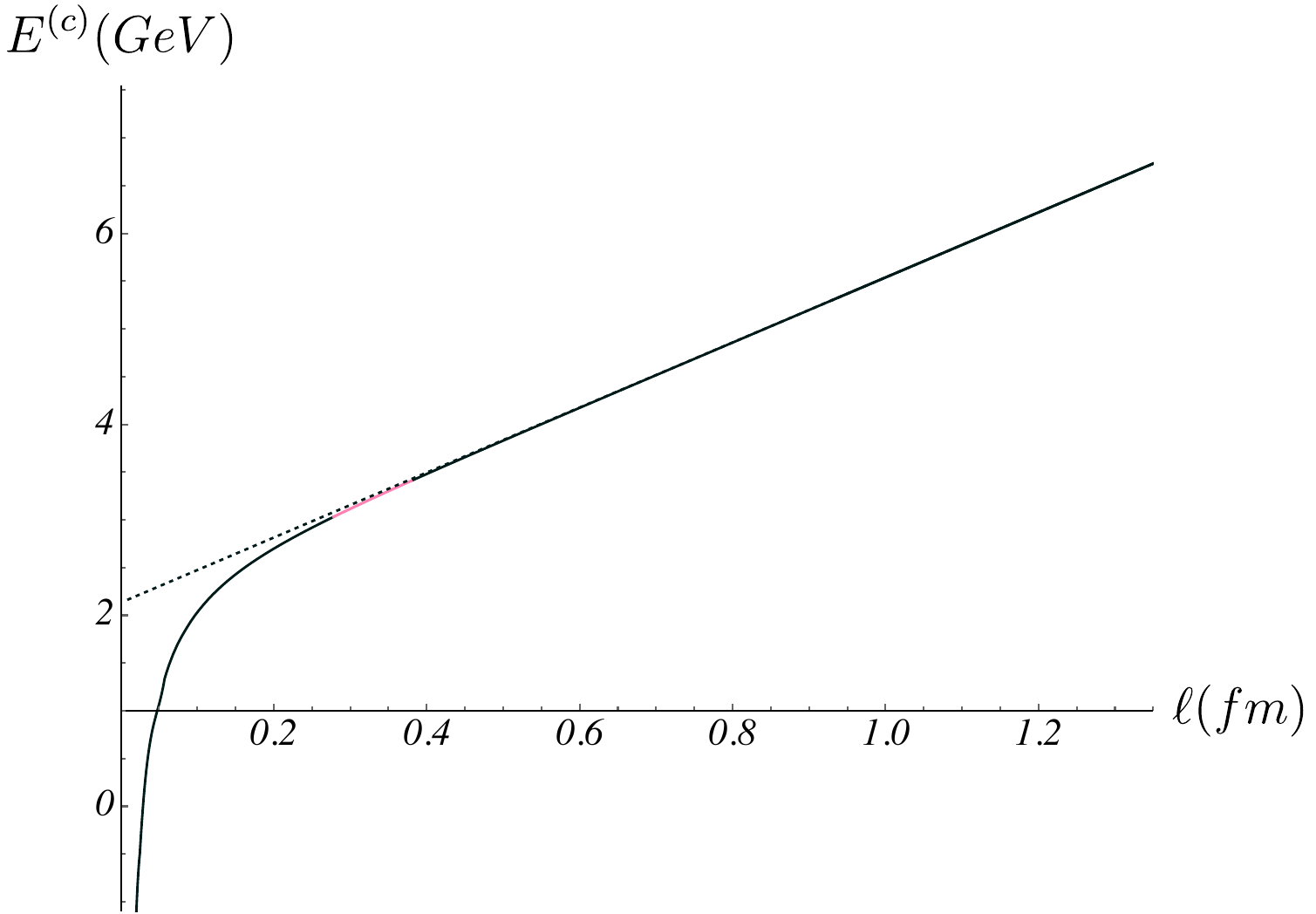}
\caption{{\small Left: The function $f_\eta$ for several values of $\eta$. Right: $E^{\text{(c)}}$ vs $\ell$ at $\eta=0.792$. The magenta and black curves correspond respectively to configurations (I) and (II), here and below.}}
\label{etaII}
\end{figure}

In this range, the function $f_\eta(v)$ has two zeros, as shown by a straightforward numerical analysis. At the upper bound, however, it has an additional zero at $v=0$. The value of $\eta_0$ can be computed analytically, with the result  

\begin{equation}\label{eta0}
\eta_{0}=
\sqrt{\frac{1-9\k^2}{3+9\k^2}}
\biggl[
1+2\frac{I\bigl(9\k^2;\tfrac{1}{2},\tfrac{3}{4}\bigr)
}{(1-9\k^2)^{\frac{3}{4}}}
\biggr]
	\,.
	\end{equation}
A simple calculation gives $\eta_{0}=0.7961$. For illustration, we plot the function $f_\eta$ in Figure \ref{etaII} on the left. 
%________________________  fig - 25 __________________________________
%______________________________________________________________________
The energy of the tetraquark configuration is now a piecewise function, described parametrically in terms of the energies of configurations (I) and (II) as 

\begin{equation}\label{EcII}
\ell=\begin{cases}
\ell^{\,\ii} (v)\,,\,\,&0\,\leq \,\, v\leq {\text v}_{\text{\tiny 01}}\,\,,\\ 
\ell^{\,\i}(v)\,,\,\,&{\text v}_{\text{\tiny 01}}\leq v\leq {\text v}_{\text{\tiny 02}}\,,\\
\ell^{\,\ii}(v)\,,\,\,&{\text v}_{\text{\tiny 02}}\leq \,v\leq \vo\,\,,
\end{cases}
\qquad
E^{\text{(c)}}=\begin{cases}
E^{\,\ii} (v)\,,\,\,&0\,\leq \,\, v\leq {\text v}_{\text{\tiny 01}}\,\,,\\  
E^{\,\i}(v)\,,\,\,&{\text v}_{\text{\tiny 01}}\leq v\leq {\text v}_{\text{\tiny 02}}\,,
\\
E^{\,\ii}(v)\,,\,\,&{\text v}_{\text{\tiny 02}}\leq \,v\leq \vo\,\,.
\end{cases}
\end{equation}
Here ${\text v}_{\text{\tiny 0\hspace{0.4pt}i}}$ are the solutions to the equation $\alpha(v)=0$.\footnote{For a single solution we simply use the notation $\vz$.}

To illustrate \eqref{EcII}, we take $\eta=0.792$. In this case, a simple calculation gives  ${\text v}_{\text{\tiny 01}}=0.5579$ and ${\text v}_{\text{\tiny 02}}=0.7213$. The functions $f_\eta$ and $E^{\text{(c)}}$ are presented in Figure \ref{etaII}. Again, $E^{\text{(c)}}$ is approximately linear for $\ell\gtrsim 0.3\,\text{fm}$.

%____________________________________________________________________________________
\subsection{The range $\eta_{0}<\eta<0.8204$}

\begin{figure}[H]
\centering
\includegraphics[width=6.75cm]{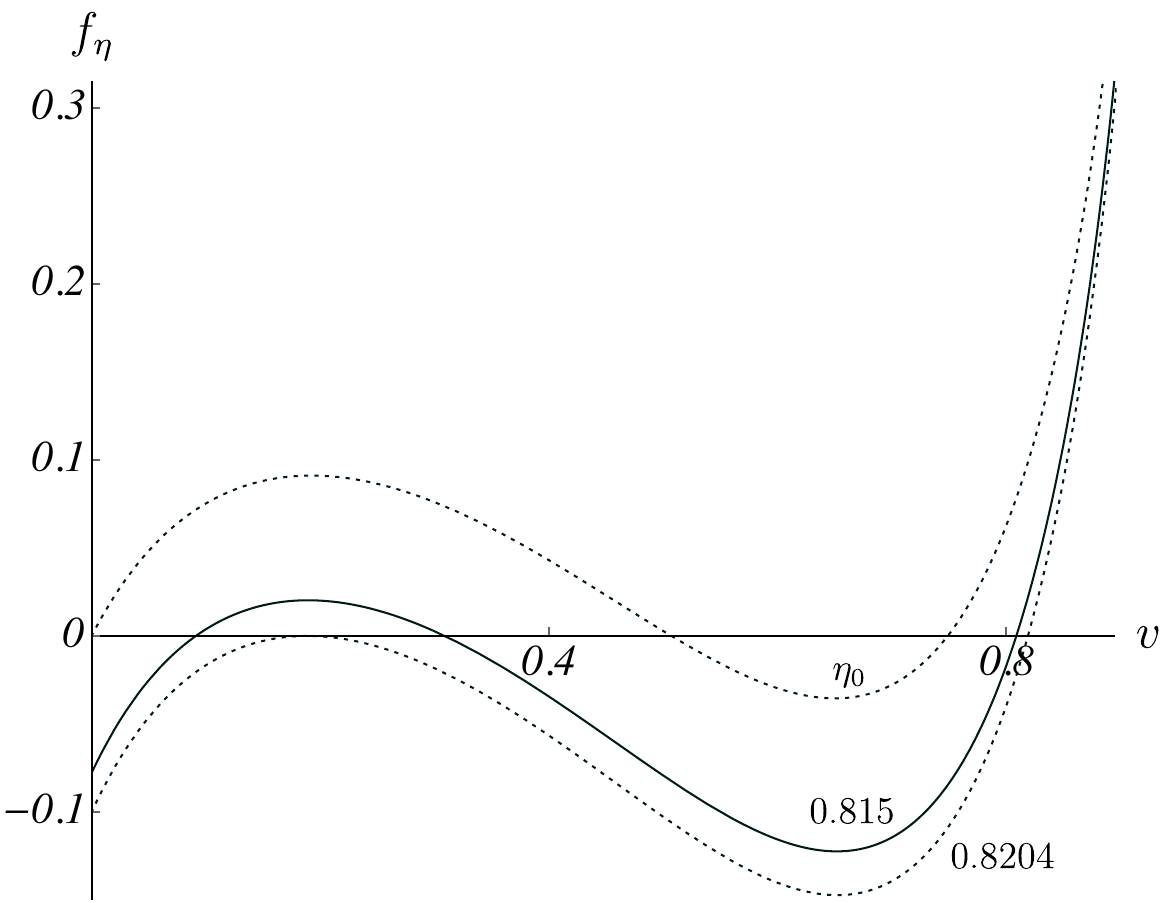}
\hspace{2cm}
\includegraphics[width=7.75cm]{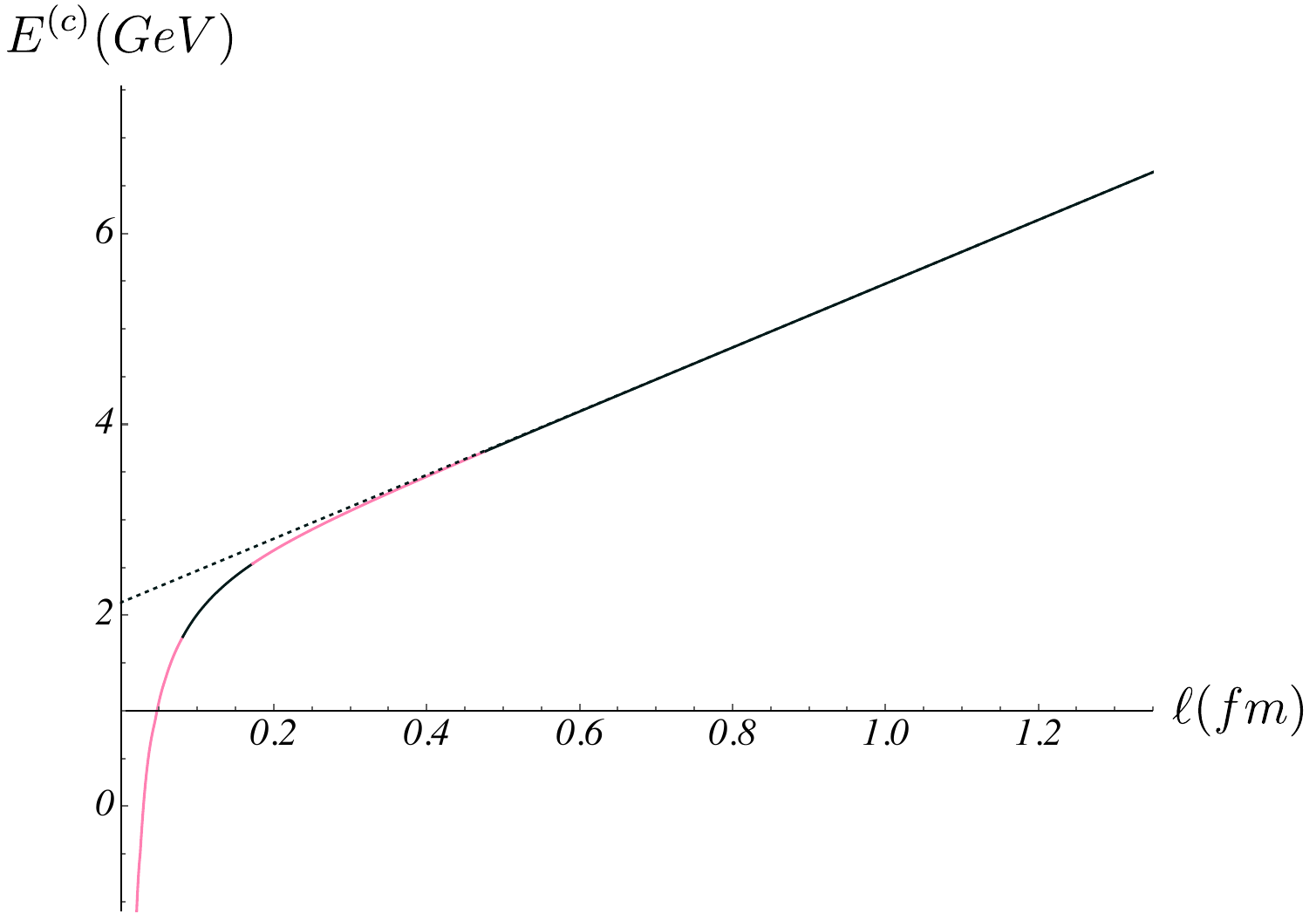}
\caption{{\small Left: The function $f_\eta$ for several values of $\eta$. Right: $E^{\text{(c)}}$ vs $\ell$ at $\eta=0.815$.}}
\label{etaIII}
\end{figure}

The function $f_\eta$ now has three zeros, as illustrated in Figure \ref{etaIII}. This implies that $E^{\text{(c)}}$ is a piecewise function 

%________________________  fig - 26  ______________________________________________
%__________________________________________________________________________________
\noindent consisting of four segments. Explicitly, 

\begin{equation}\label{EcIII}
\ell=\begin{cases}
\ell^{\,\i} (v)\,,\,\,&0\,\leq \,\, v\leq {\text v}_{\text{\tiny 01}}\,\,,\\ 
\ell^{\,\ii}(v)\,,\,\,& {\text v}_{\text{\tiny 01}}\leq v\leq {\text v}_{\text{\tiny 02}}\,,\\
\ell^{\,\i}(v)\,,\,\,&{\text v}_{\text{\tiny 02}}\leq v\leq {\text v}_{\text{\tiny 03}}\,,\\
\ell^{\,\ii}(v)\,,\,\,&{\text v}_{\text{\tiny 03}}\leq \,v\leq \vo\,\,,
\end{cases}
\qquad
E^{\text{(c)}}=\begin{cases}
E^{\,\i} (v)\,,\,\,&0\,\leq \,\, v\leq {\text v}_{\text{\tiny 01}}\,\,,\\ 
E^{\,\ii}(v)\,,\,\,&{\text v}_{\text{\tiny 01}}\leq v\leq {\text v}_{\text{\tiny 02}}\,,\\
E^{\,\i}(v)\,,\,\,&{\text v}_{\text{\tiny 02}}\leq v\leq {\text v}_{\text{\tiny 03}}\,,\\
E^{\,\ii}(v)\,,\,\,&{\text v}_{\text{\tiny 03}}^{\text{\tiny 3}}\leq \,v\leq \vo\,\,.
\end{cases}
\end{equation}
As an example, let us take $\eta=0.815$. For this value, the zeros are ${\text v}_{\text{\tiny 01}}=0.0914$, ${\text v}_{\text{\tiny 02}}=0.3084$, and ${\text v}_{\text{\tiny 03}}=0.8088$. The corresponding functions $f_\eta$ and $E^{\text{(c)}}$ are shown in Figure \eqref{etaIII}. We see that $E^{\text{(c)}}$ also becomes near linear for $\ell\gtrsim 0.3\,\text{fm}$.

%_________________________________________________________________
\subsection{The range $0.8204\leq\eta$}

\begin{figure}[htbp]
\centering
\includegraphics[width=6.5cm]{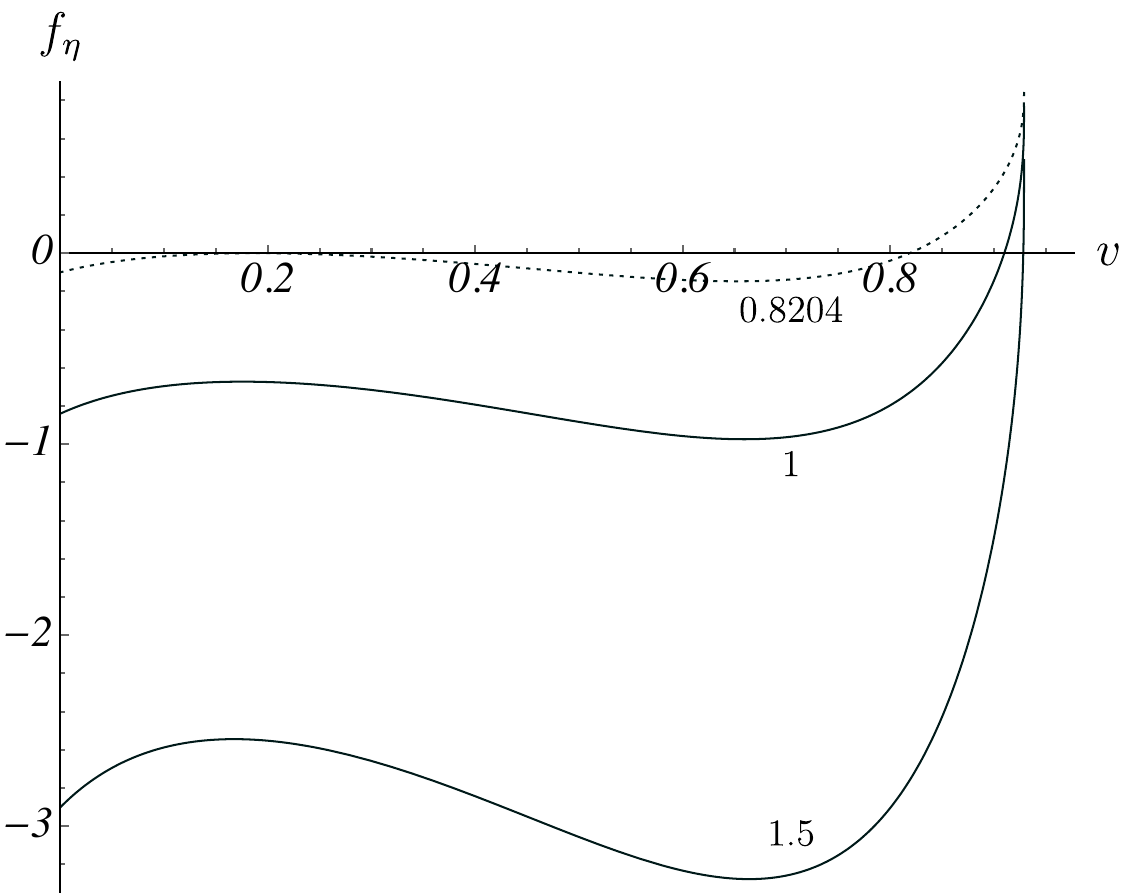}
\hspace{2cm}
\includegraphics[width=7.5cm]{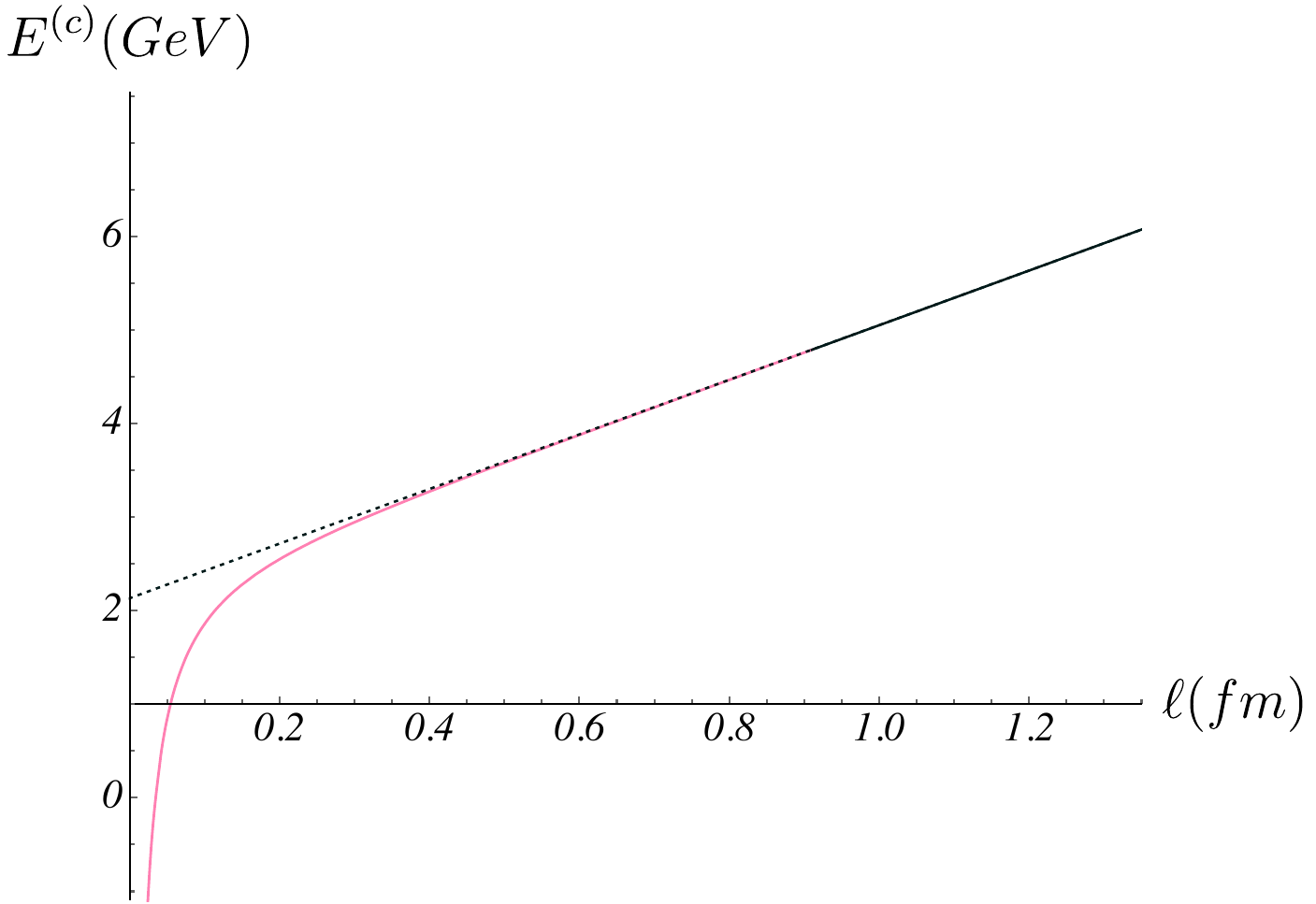}
\caption{{\small Left: The function $f$ for several values of $\eta$. Right: $E^{\text{(c)}}$ vs $\ell$ at $\eta=1$.}}
\label{etaIV}
\end{figure}

Finally, consider the range $0.8204\leq\eta$. As seen from Figure \ref{etaIV}, the function $f_{\eta}$ has a single zero, except at $\eta=0.8404$, where it has two. However, this does not affect the tetraquark configuration. 
%________________________  fig - 27 ______________________________
%__________________________________________________________________
So, the energy can be written in parametric form as

\begin{equation}\label{EcIV}
\ell=\begin{cases}
\ell^{\,\i} (v)\,,\,\,& 0\,\leq v\leq \vz\,\,,\\ 
\ell^{\,\ii}(v)\,,\,\,&\vz\leq v\leq \vo\,,
\end{cases}
\quad
E^{\text{(c)}}=\begin{cases}
E^{\,\i} (v)\,,\,\,&0\,\leq v\leq \vz\,\,,\\
E^{\,\ii}(v)\,,\,\,&\vz\leq v\leq \vo\,.
\end{cases}
\end{equation}

As a concrete example, consider $\eta=1$. In this case, the zero is given by $\vz=0.9102$. The functions $f_\eta$ and $E^{\text{(c)}}$ are plotted in Figure \ref{etaIV}. A notable feature is that $E^{\text{(c)}}$ becomes near linear for slightly larger values of $\ell$, namely for $\ell\gtrsim 0.4\,\text{fm}$. This has a natural explanation: as $\eta\to\infty$, the function $E^{\text{(c)}}$ tends to $E_{\QQb}$, where such a behavior occurs for $\ell\gtrsim 0.6\,\text{fm}$ (see Figure \ref{Eal}).

%_________________________________________________________________
\subsection{The Coulomb and quark-quark couplings}

Assuming that at sufficiently small separations the energy of the tetraquark configuration is due to pairwise quark interactions, the leading approximation to the energy takes the form

\begin{equation}\label{Ecg-small}
	E^{\text{(c)}} =-\frac{\alpha^{\text{(c)}}}{\ell}+O(1)
	=
	  -2\eta\frac{\alpha_{\QQ}}{\ell}
	-\oh \biggl(1+\frac{1}{\sqrt{1+\eta^{-2}}}\biggr)\frac{\alpha_{\QQb}}{\ell}
	+O(1)
	\,.
\end{equation} 
In general, both $\alpha_{\QQb}$ and $\alpha_{\QQ}$ may depend on $\eta$ and differ from their counterparts in the $Q\bar Q$ and $QQQ$ systems.\footnote{In the $QQQ$ system, $\alpha_{\QQ}$ is not constant but depends weakly on the angles of a triangle formed by the quarks \cite{a3Q2016,a3Q2025}.} 

 Having computed the Coulomb coefficient $\alpha^{\text{(c)}}$, we would like to estimate the quark-quark coupling that appears in \eqref{Ecg-small}. This, however, is impossible without making an additional assumption on $\alpha_{\QQb}$. There are two ways to do so. The first way is to assume the relation $\alpha_{\QQb}=2\alpha_{\QQ}$.\footnote{In the literature, it is sometimes referred to as $\frac{1}{2}$ rule.} With this, one gets  

\begin{equation}\label{alpha05}
	\alpha_{\QQ}=\frac{\alpha^{\text{(c)}}}{1+2\eta+\frac{1}{\sqrt{1+\eta^{-2}}}}
	\,.
\end{equation}
The second way is to assume that $\alpha_{\QQb}$ is the same as in the $Q\bar Q$ (meson) system. This yields

\begin{equation}\label{alphameson}
	\alpha^{\text{m}}_{\QQ}=\frac{1}{4\eta}\Bigl(2\alpha^{\text{(c)}}-\Bigl(1+\frac{1}{\sqrt{1+\eta^{-2}}}\Bigr)\alpha_{\QQb}\Bigr)
	\,,
\end{equation}
where $\alpha_{\QQb}$ is defined in \eqref{EQQb-small}. 

The results of our estimates are summarized in Table \ref{estimates4}. Importantly, for the considered values of $\eta$, the small-$\ell$ behavior of the  
%__________________________________________________________________________________________
\begin{table*}[htbp]
\renewcommand{\arraystretch}{2}
\centering 	\scriptsize
\begin{tabular}{ccccc}				
\hline
$\eta$ & \quad $\alpha^{\text{(c)}}$ & \quad $\alpha^{\text{(c)}}/\alpha^{\text{(a)}}$ & \quad $\alpha_{\QQ}/\alpha_{\QQb}$ & \quad $\alpha^{\text{m}}_{\QQ}/\alpha_{\QQb}$ 
\rule[-3mm]{0mm}{8mm}
\\
\hline \hline
1.1794         & \quad 0.5053      &  \quad 1        & \quad 0.4853  &  \quad 0.4743    \\
$\sqrt{3}$     &  \quad 0.6505     & \quad 1.2873    & \quad 0.4830  & \quad 0.4739   \\
3              &  \quad 0.9570     & \quad 1.8939    & \quad 0.4765  & \quad 0.4689   \\
6              &  \quad1.6655     &  \quad 3.2960    & \quad 0.4713  & \quad 0.4666   \\
15             &\quad  3.7854      & \quad 7.4912   & \quad 0.4679   &  \quad0.4658   \\
20             & \quad  4.9630     & \quad 9.8217    & \quad 0.4677  &  \quad 0.4635   \\
 \hline \hline
\end{tabular}
\caption{ \small Estimates for the couplings in the small-$\ell$ limit. Here $\alpha_{\QQb}$ is the Coulomb coefficient of the quark-antiquark potential (see Appendix C).}
\label{estimates4}
\end{table*}
%____________________________________________________________________________________
potential $V_0$ is determined by configuration (c) that makes our estimates also applicable to the ground state potential as well. As seen from the Table, both quark-quark couplings are close to $\oh$ and depend weakly on $\eta$. While the former behavior was observed in lattice QCD using the approximation \eqref{alpha05} \cite{sug}, the latter is our prediction, which will hopefully be testable in future numerical simulations. The expressions \eqref{alpha05} and \eqref{alphameson}, in turn, provide reasonable approximations for the $\eta$-dependent Coulomb coefficient $\alpha^{\text{(c)}}$.

%_________________________________________________________________
\section{The tetraquark configuration (c') for rhombus geometry}
%\label{notation}
\renewcommand{\theequation}{E.\arabic{equation}}
\setcounter{equation}{0}

\begin{figure}[htbp]
\centering
\includegraphics[width=5.75cm]{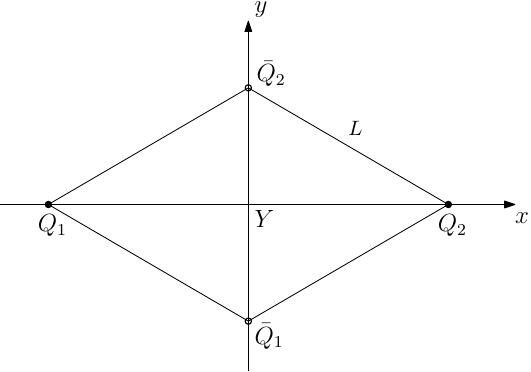}
\caption{{\small A rhombus of length $L$. The quarks and antiquarks are placed at its vertices according to type-B ordering, with point $Y$ at the origin.}}
\label{rh}
\end{figure}

Our goal here is to clarify that the tetraquark configuration exists for type-B ordering. To proceed, instead of a rectangle, we consider a rhombus in the $xy$-plane with its center at the origin, as shown in Figure \ref{rh}. We also
%________________________  fig - 26 __________________________________
%___________________________________________________________________
impose the geometrical constraint 

\begin{equation}\label{di-gc}
\vert\bar Q_1Y\vert=d\,\vert Q_1Y\vert
\,,
\end{equation}
which fixes one of the rhombus angles. A simple but useful relation is  $L=\sqrt{1+d^2}\vert Q_1Y\vert$. 

The tetraquark configuration (c') can be analyzed along the lines of Sec.IV, with the only modification arising from the above constraint. We begin by describing the basic configurations and next illustrate the construction with a concrete example. 
%__________________________________________________________________
\subsubsection{Configuration (i)}

The basic configuration (i) is similar to that shown in Figure \ref{tetrap5d}. Thus, we have $v>\bar v$, $\alpha_1=\alpha_2=\alpha$, and $\bar\alpha_1=\bar\alpha_2=\bar\alpha$, as before. The force balance equations \eqref{fbeW} remain valid, as does the expression for the energy given by the second equation in \eqref{El-cpI}. The expression for the length is easily obtained using Eq.\eqref{l-} for $\vert Q_1Y\vert$. So,
 
\begin{equation}\label{EdiI}
 L^{\text{(i)}}=\frac{\sqrt{1+d^2}}{\sqrt{\s}}\, {\cal L}^-(\lambda,v)
\,,\quad
E^{\,\text{(i)}}=\g\sqrt{\s}
\Bigl(2{\cal E}^-(\lambda,v)+2{\cal E}^+(\bar\alpha,\bar v)
+
{\cal Q}(v)-{\cal Q}(\bar v)+
3\k\frac{\ep^{-2v}}{\sqrt{v}}
+
3\k\frac{\ep^{-2\bar v}}{\sqrt{\bar v}}
\Bigr)
+4c
\,.
\end{equation}
Similarly, for the geometric constraint we get  

\begin{equation}\label{diI}
	{\cal L}^+(\bar\alpha,\bar v)=d\,{\cal L}^-(\lambda,v)
	\,,
\end{equation}
which is a light modification of \eqref{gc-pI}. The tangent angles and parameter $\bar v$ can be expressed in terms of $v$ using the force balance equation and the geometrical constraint. As a result, the energy as a function of $L$ can be written parametrically as $L=L^{\text{(i)}}(v)$ and $E=E^{\,\text{(i)}}(v)$. 

It is instructive to consider the behavior of $E^{\text{(i)}}$ at small $L$, which corresponds to the limit $v\to 0$. Taking this limit in \eqref{diI}, with the help of Eqs.\eqref{fL+smallx} and \eqref{L-y=0}, one finds $\bar v=\Bigl(d \frac{{\cal L}^-(\alpha)}{{\cal L}^+(\bar\alpha)}\Bigr)^2v$. This restricts the allowed values of $d$ to 

\begin{equation}\label{upper-d}
	d\leq \sqrt{\frac{\cos\alpha}{\cos\bar\alpha}}\,\frac{I(\cos^2\bar\alpha,\frac{3}{4},\frac{1}{2})}{1+I(\sin^2\alpha,\frac{1}{2},\frac{3}{4})}
	\,,
\end{equation}
as $v\geq \bar v$ by construction. To make a simple estimate of the upper bound, note that the values of the tangent angles are determined by Eqs.\eqref{fbeW} at $v=\bar v=0$. Using those, we get $d\leq 0.407$.\footnote{As we saw in Sec.IV, the tetraquark configuration (c') does not exist at $d=1$.} The energy behaves for $L\to 0$ as 

 \begin{equation}\label{EL-small}
	E^{\,\text{(i)}}=-\frac{\alpha^{\text{(i)}}}{L}+4c+o(1)\,,
	\quad\text{with}\quad
	\alpha^{\text{(i)}}=
	-
	\sqrt{1+d^{-2}}
\Bigl[d\,{\cal L}^-_0(\alpha)
\bigl(2{\cal E}^-_0(\alpha)-1+3\k\bigr)
+
{\cal L}^+_0(\bar\alpha)
\bigl(2{\cal E}^+_0(\bar\alpha)+1+3\k\bigr)
\Bigr]\g
\,,
\end{equation}
exhibiting the leading Coulomb term. 

%_____________________________________________________________________
\subsubsection{Configuration (ii)}

This basic configuration is similar to that shown on the right in Figure \ref{tetrap5d}, with $\bar\alpha\geq 0$. The force balance equation \eqref{fbeW-II} remains unchanged. The expressions for the length and energy follow from \eqref{EdiI} and \eqref{El-cpII}:

\begin{equation}\label{EdiII}
 L^{\text{(ii)}}=\frac{\sqrt{1+d^2}}{\sqrt{\s}}\, {\cal L}^-(\lambda,v)
\,,\quad
E^{\,\text{(ii)}}=2\g\sqrt{\s}
\Bigl({\cal E}^-(\lambda,v)+{\cal E}^+(\bar\alpha,v)
+3\k\frac{\ep^{-2v}}{\sqrt{v}}
\Bigr)
+4c
\,.
\end{equation}
The geometrical constraint now reads 

\begin{equation}\label{diII}
	{\cal L}^+(\bar\alpha,v)=d\,{\cal L}^-(\lambda,v)
	\,,
\end{equation}
which is obtained from \eqref{diI} by setting $\bar v=v$. After extracting the tangent angles from the force balance equation and the geometrical constraint, the energy can again be written in parametric form as $L=L^{\text{(ii)}}(v)$ and $E=E^{\,\text{(ii)}}(v)$. 

%___________________________________________________________________
\subsubsection{Configuration (iii)}

The only difference from configuration (ii) is that $\bar\alpha$ is negative. Therefore, all the formulas can be obtained from those of configuration (ii) by replacing ${\cal L}^+$ and ${\cal E}^+$ with ${\cal L}^-$ and ${\cal E}^-$. In this way, from \eqref{EdiII} we obtain 

\begin{equation}\label{EdiIII}
 L^{\text{(iii)}}=\frac{\sqrt{1+d^2}}{\sqrt{\s}}\, {\cal L}^-(\lambda,v)
\,,\quad
 E^{\text{(iii)}}=2\g\sqrt{\s}
\Bigl({\cal E}^-(\lambda,v)+{\cal E}^-(\bar\lambda,v)
+
3\k\frac{\ep^{-2v}}{\sqrt{v}}
\Bigr)
+4c
\,,	
\end{equation}
and from \eqref{diII}

\begin{equation}\label{diIII}
	{\cal L}^-(\bar\lambda,v)=d\,{\cal L}^-(\lambda,v)
	\,.
\end{equation}
The force balance equation \eqref{fbeW-II} remains valid. Again, the angles can be expressed in terms of $v$, and the energy can be written parametrically as $L=L^{\text{(iii)}}(v)$ and $E=E^{\,\text{(iii)}}(v)$.

Finally, let us briefly discuss the large-$L$ behavior, which corresponds to the limit $\lambda,\bar\lambda\to 1$. It follows then from Eq.\eqref{v-lambda} that $\cos\alpha=\cos\bar\alpha=v\ep^{1-v}$. Thus, in this limit configuration (c') approaches to configuration (d) of Secs.III and IV. This implies that the upper bound on $v$ is $\vp$, defined by \eqref{v1p}, and the expansions \eqref{Eld-large5} can be rewritten as 

\begin{equation}\label{EL-large}
	E^{\text{(iii)}}=2\frac{1+d}{\sqrt{1+d^2}}\,\sigma L+C^{\text{(d)}}+o(1)
	\,.
\end{equation}
Here the first term is proportional to the total length of the rhombus diagonals.

%____________________________________________________________________
\subsubsection{An example}

We now describe a concrete example, namely $d=\frac{1}{4}$, in which one can explicitly construct the tetraquark configuration from the basic configurations. As it turns out, $E^{\text{(c')}}$ is a piecewise function of $L$, given by 

\begin{equation}\label{ELc}
L=\frac{\sqrt{1+d^2}}{\sqrt{\s}}\, {\cal L}^-(\lambda,v)
\,,
\qquad
E^{\text{(c')}}=\begin{cases}
E^{\text{(i)}} (v)\,\,\,,\,\,&0\,\leq v\leq \bar{\text{v}}\,\,\,\,,\\
E^{\text{(ii)}}(v)\,\,,\,\,&\bar{\text{v}}\leq \,v\leq \vz\,\,,\\
E^{\text{(iii)}}(v)\,,\,\,&\vz\leq v\leq \vp\,.
\end{cases}
\end{equation}
Here $\bar{\text{v}}=0.399$ and $\vz=0.929$. The first value corresponds to the transition between configurations (i) and (ii), which occurs when the vertices collide, i.e., $v=\bar v$. The second value corresponds to the transition between configurations (ii) and (iii), which occurs at $\bar\alpha=0$. 

\begin{figure}[htbp]
\centering
\includegraphics[width=8cm]{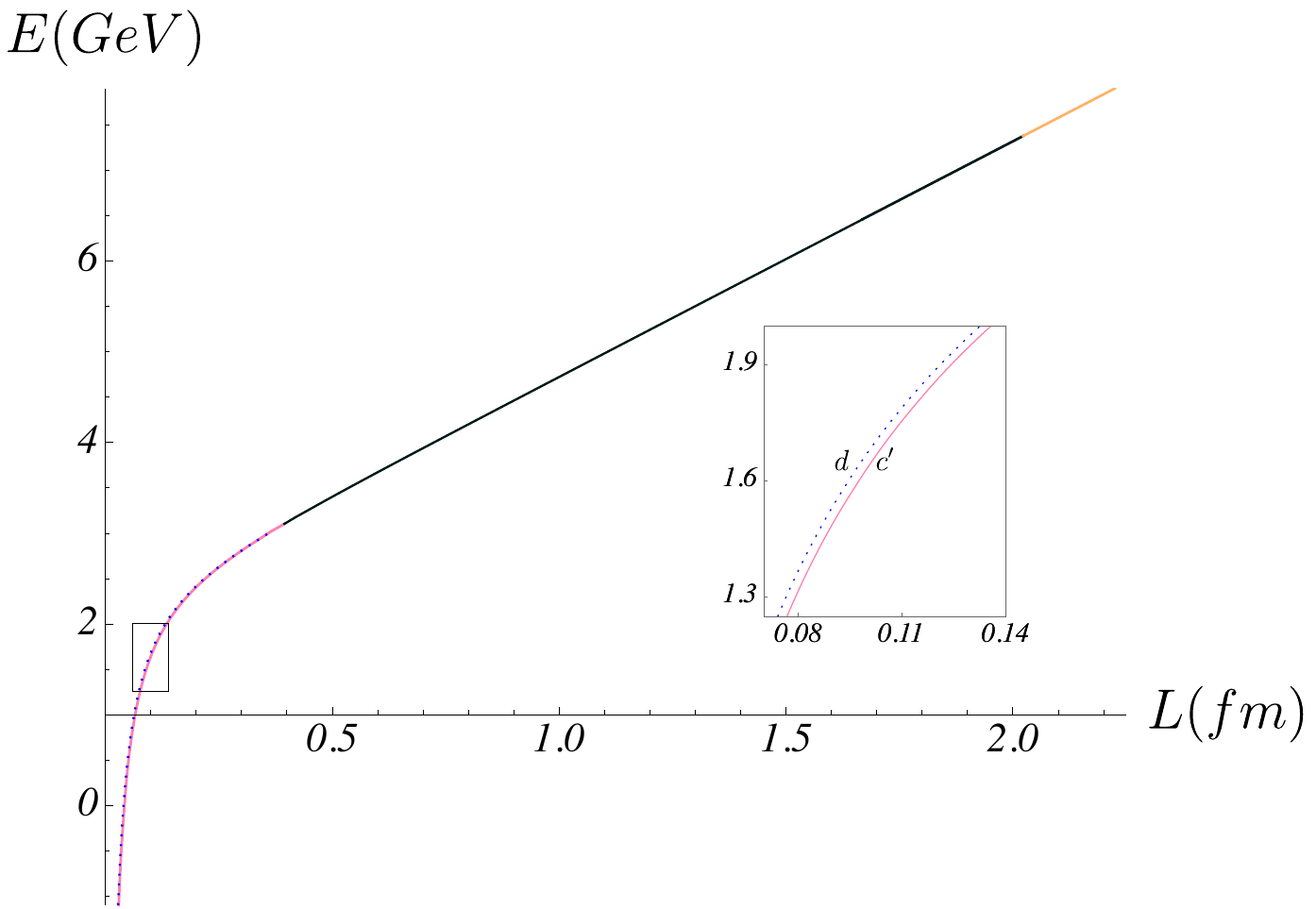}
\caption{{\small $E^{\text{(c')}}$ and $E^{\text{(d)}}$ vs $L$ at $d=\frac{1}{4}$. The magenta, black, and yellow curves correspond to the basic configurations (i)-(iii), respectively. In addition, the dotted blue curve represents configuration (ii) for $L<0.395\,\text{fm}$, which forms part of configuration (d) in this interval.}}
\label{diamond025}
\end{figure}

To complete the picture, we also present an analog of the pinched tetraquark configuration, constructed from the two basic configurations (ii) and (iii). Explicitly, 

\begin{equation}\label{ELd}
L=\frac{\sqrt{1+d^2}}{\sqrt{\s}}\, {\cal L}^-(\lambda,v)
\,,
\qquad
E^{\text{(d)}}=\begin{cases}
E^{\text{(ii)}}(v)\,\,,\,\,&0 \leq \,v\leq \vz\,\,,\\
E^{\text{(iii)}}(v)\,,\,\,&\vz\leq v\leq \vp\,.
\end{cases}
\end{equation}
Note that the configuration is antisymmetric in the sense that $\alpha\neq\bar\alpha$ unless $v=\vp$. Clearly, these configurations differ only for $L<L(\bar{\text{v}})$, due to the difference between the basic configurations (i) and (ii). In Figure \ref{diamond025} we plot $E^{\text{(c')}}(L)$ and $E^{\text{(d)}}(L)$. For $L<0.395\,\text{fm}$, the plots corresponding to configurations (i) and (ii) are indistinguishable, 
%________________________  fig - 29 __________________________________
%_______________________________________________________________________
as the difference between them is extremely small. However, a more detail analysis shows that configuration (c') has a lower energy than configuration (d). Note that configuration (i) transforms into configuration (ii) at $L=0.395\,\text{fm}$ when the baryon vertices collide. From the perspective of ten-dimensional string theory, this corresponds to the creation of a brane-antibrane bound state. A similar pinching effect has also been observed in the doubly heavy tetraquark systems \cite{a-QQqq}. 

%__________________       R E F s     ____________________________________________________

%____________________________________________________________________
\end{document}